\documentclass[jkps,preprint,fleqn,showpacs,showkeys]{revtex4}
\usepackage{graphicx}
\usepackage{amssymb}
\usepackage{amsmath}
\usepackage{bm}
\usepackage{times}
\usepackage{amsfonts}
\usepackage{amstext}
\usepackage{amsthm}
\usepackage{epsfig}
\psfigdriver{dvips}
\usepackage{latexsym}
\usepackage[matrix,curve]{xypic}
\usepackage{rotating}
\rotdriver{dvips}

\newcommand{\nc}{\newcommand}

\nc{\be}{\begin{equation}}

\nc{\ee}{\end{equation}}

\nc{\bea}{\begin{eqnarray}}

\nc{\eea}{\end{eqnarray}}

\nc{\xx}{\nonumber\\}

\nc{\ct}{\cite}

\nc{\la}{\label}

\nc{\eq}[1]{(\ref{#1})}

\def\half{\frac{1}{2}}

\def\p{\partial}

\def\CA{{\cal A}}

\def\CG{{\cal G}}
\def\CH{{\cal H}}

\def\CL{{\cal L}}

\def\CO{{\cal O}}

\def\what#1{\widehat{#1}}

\def\Tr{{\rm Tr}\,}
\def\det{{\rm det}}

\begin{document}
\setcounter{page}{0}
\title[]{Quantum Gravity from Noncommutative Spacetime}
\author{Jungjai \surname{Lee}}
\email{jjlee@daejin.ac.kr}
\affiliation{Department of Physics, Daejin University, Pocheon 487-711, Korea}
\author{Hyun Seok \surname{Yang}}
\email{hsyang@kias.re.kr}
\affiliation{School of Physics, Korea Institute for Advanced Study, Seoul 130-722, Korea}

\date[]{Received 11 September 2014}

\begin{abstract}
We review a novel and authentic way to quantize gravity.
This novel approach is based on the fact that Einstein gravity can be formulated in
terms of a symplectic geometry rather than a Riemannian geometry in the context of emergent gravity.
An essential step for emergent gravity is to realize the equivalence principle,
the most important property in the theory of gravity (general relativity),
from $U(1)$ gauge theory on a symplectic or Poisson manifold.
Through the realization of the equivalence principle, which is an intrinsic property
in symplectic geometry known as the Darboux theorem or the Moser lemma,
one can understand how diffeomorphism symmetry arises from noncommutative $U(1)$ gauge theory;
thus, gravity can emerge from the noncommutative electromagnetism, which is also an interacting theory.
As a consequence, a background-independent quantum gravity in which the prior existence of
any spacetime structure is not {\it a priori} assumed but is defined by using the fundamental ingredients
in quantum gravity theory can be formulated. This scheme for quantum gravity can be used to resolve
many notorious problems in theoretical physics, such as the cosmological constant problem,
to understand the nature of dark energy, and to explain why gravity is so weak compared to other forces.
In particular, it leads to a remarkable picture of what matter is.
A matter field, such as leptons and quarks, simply arises as a stable localized geometry,
which is a topological object in the defining algebra (noncommutative $\star$-algebra) of quantum gravity.
\end{abstract}



\maketitle

\section{INTRODUCTION}

The most important property in the theory of gravity (general
relativity) is arguably the equivalence principle. The equivalence
principle says that gravity can be interpreted as an inertial force,
so it is always possible to locally eliminate the gravitational
force by using a coordinate transformation, i.e., in a locally inertial
frame. This immediately leads to a remarkable picture in which gravity
has to describe a spacetime geometry rather than a force immanent in
spacetime. The equivalence principle also implies that gravity is
obviously a universal force such that gravity influences and is
influenced by everything that carries energy. Therefore, the
spacetime has to serve as a stage for everything supported on it, as
well as an actor for the dynamical evolution of the stage
(spacetime) itself. In order to quantize gravity, we, thus, have to
cook a frying pan and a fish altogether. How?

\subsection{What Is Quantum Gravity?}

Quantum gravity means to ``quantize" gravity. Gravity, according to
Einstein's general relativity, is the dynamics of spacetime geometry
where spacetime is a Riemannian manifold and the gravitational field
is represented by a Riemannian metric. Thus, naively quantum
gravity is meant to quantize the Riemannian manifold.
However, how to ``quantize" a Riemannian manifold is still vague.

Quantum mechanics has constituted a prominent example of ``quantization" since its foundation.
However, quantization in this case is controlled by the Planck constant
$\hbar$, whose physical dimension is length times momentum, i.e., $(x
\times p)$. Therefore, quantization in terms of $\hbar$ quantizes (or deforms in a weak sense)
a {\it particle} phase space, as we know very well. Because we consider a classical field
$\phi(x) \in C^\infty(M)$ to be a smooth function defined on a
spacetime $M$ and a many-body system describing infinitely many
particles distributed over the spacetime $M$, quantum field theory
is also defined by quantization in terms of $\hbar$ in an
infinite-dimensional particle phase space, as we clearly know.

Now consider ``quantizing" gravity. With $\hbar$ again? Because
gravity is characterized by its own intrinsic scale given by the
Newton constant $G$, where the classical gravity corresponds to the $G \to 0$
limit \cite{gravmac}, we should leave open
the possibility that the quantum gravity is defined by a deformation
in terms of $G$ instead of $\hbar$. Customarily, we have taken the
same route to the quantization of gravity as that taken by conventional
quantum field theory. Thus, conventional quantum gravity also
intends to quantize an infinite-dimensional {\it particle} phase
space associated with the metric field $g_{\mu\nu}(x)$ (or its
variants such as the Ashtekar variables or spin networks) of a
Riemannian geometry. However, we have to carefully contemplate whether
our routine approach to quantum gravity is on the right track or not
because gravity is very different from other forces, such as the
electromagnetic, weak and strong forces. For a delightful journey to
quantum gravity, it is necessary to pin down the precise object
of quantization ($\hbar$ or $G$) and clearly specify correct
variables for quantization (spacetime itself or fields associated with the spacetime).
For these reasons, we pose the following two questions \cite{nhabel}:

{\bf Q1. \; Is gravity fundamental or emergent?}

{\bf Q2. \; Which quantization defines quantum gravity: $\hbar$ or $G$?}

In {\bf Q1}, we usually refer to a physical entity (force or field) as
being {\it fundamental} when it does not have any superordinate
substructure, and {\it emergence} usually means the arising of novel
and coherent structures, patterns and properties through the
collective interactions of more fundamental entities: for example,
the superconductivity in condensed matter system or the organization
of life in biology. Regarding to question {\bf Q1}, it is quite amazing to notice
that the picture of emergent gravity was already incoded in Cartan's formulation of gravity \ct{big-book}.
In general relativity, the gravitational force is represented by a Riemannian metric of
the curved spacetime manifold $M$:
\begin{equation} \label{metric}
\Big(\frac{\partial}{\partial s}\Big)^2 = g^{\mu\nu}(y) \frac{\partial}{\partial y^\mu}
 \otimes \frac{\partial}{\partial y^\nu}.
\end{equation}
Cartan showed that the metric in Eq. (\ref{metric}) could be
defined by the tensor product of two vector fields $E_a = E_a^\mu
(y) \frac{\partial}{\partial y^\mu} \in \Gamma(TM)$ as follows:
\begin{equation} \label{cartan}
\Big(\frac{\partial}{\partial s}\Big)^2  = \eta^{ab} E_a \otimes E_b.
\end{equation}
Mathematically, a vector field $X$ on a smooth manifold $M$ is a
derivation of the algebra $C^\infty (M)$. Here, the vector fields
$E_a \in \Gamma(TM)$ are smooth sections of the tangent bundle $TM
\to M$ which are dual to the covectors $E^a = E^a_\mu (y) dy^\mu
\in \Gamma(T^* M)$; i.e., $\langle E^b, E_a \rangle = \delta^b_a$.
The expression in Eq. (\ref{cartan}) glimpses the avatar of gravity that a
spin-two graviton might arise as a composite of two spin-one {\it
vector fields}. In other words, Eq. (\ref{cartan}) can be abstracted
by using the relation $(1 \otimes 1)_S = 2 \oplus 0$. Incidentally, both
mathematician and physicist use the same word, {\it vector
field}, in spite of a bit different meaning.

Equation (\ref{cartan}) suggests that we need gauge fields that take the values
in the Lie algebra of diffeomorphisms in order to realize a
composite graviton from spin-one vector fields. To be precise, the
vector fields $E_a = E_a^\mu (y)
\frac{\partial}{\partial y^\mu} \in \Gamma(TM)$ will be identified
with ``0-dimensional" gauge fields satisfying the Lie algebra
\begin{equation} \label{lie}
[E_a, E_b] = - {f_{ab}}^c E_c.
\end{equation}
Of course, the Standard Model does not have such kind of gauge fields,
but we will see later that the desired vector fields arise from the electromagnetic fields
living in noncommutative spacetime \ct{hsy-epl,hsy-ijmpa,hsy-epjc,hsy-jhep}.
Thus, the noncommutative spacetime will allow a novel unification between electromagnetism
and Einstein gravity in a completely different context from the Kaluza-Klein unification.

Regarding question {\bf Q2}, we are willing to ponder the possibility that
the Newton constant $G$ signifies an intrinsic Poisson structure $\theta = \half \theta^{\mu\nu} (y)
\frac{\partial}{\partial y^\mu} \wedge \frac{\partial}{\partial y^\nu} \in \Gamma(\Lambda^2 TM)$
of spacetime because the gravitational constant $G$ carries the physical dimension
of $({\rm length})^2$ in natural units. Recall that the particle
phase space $P$ has its intrinsic Poisson structure $\eta = \hbar
\frac{\partial}{\partial x^i} \wedge \frac{\partial}{\partial p_i}$ \ct{mechanics}
so that the Poisson bracket on the vector space $C^\infty (P)$ is given by
\be \la{p-poisson}
\{f, g\}_\hbar (x,p) = \hbar \Big( \frac{\partial f}{\partial x^i} \frac{\partial g}{\partial p_i}
- \frac{\partial f}{\partial p_i} \frac{\partial g}{\partial x^i} \Big),
\ee
where $f, g \in C^\infty(P)$. Here, we have intentionally inserted the
Planck constant $\hbar$ into $\eta$ to make it dimensionless like $\theta$.
Then, the canonical quantization can be done by association
with a commutative algebra $C^\infty(P)$ of physical observables of a
classical mechanical system, a noncommutative algebra $\CA_\hbar$ of
linear operators on a suitable Hilbert space $\CH$. That is, the
physical observables $f, g \in C^\infty(P)$ are replaced by
self-adjoint operators $\widehat{f}, \widehat{g} \in \CA_\hbar$
acting on $\CH$, and the Poisson bracket in Eq. \eq{p-poisson} is replaced
by a quantum bracket
\be \la{qm-bracket}
\{f, g\}_\hbar \; \rightarrow \; -i [\widehat{f},
\widehat{g}].
\ee
This completes the quantization of the mechanics of a particle system
whose phase space $P$ is now noncommutative.

In the same way, one can define a Poisson bracket $\{\cdot,\cdot\}_\theta: C^\infty(M) \times C^\infty(M)
\to C^\infty(M)$ by using the Poisson structure $\theta\in \Gamma(\Lambda^2 TM)$
of the spacetime manifold $M$:
\be \la{p-bracket}
\{ f, g\}_\theta (y) = \theta (df, dg) =
\theta^{\mu\nu} (y) \Big(
\frac{\partial f}{\partial y^\mu} \frac{\partial g}{\partial y^\nu}
- \frac{\partial f}{\partial y^\nu} \frac{\partial g}{\partial
y^\mu} \Big)
\ee
where $f,g \in C^\infty(M)$. In the case where $\theta^{\mu\nu}$ is
a constant cosymplectic matrix of rank $2n$, one can apply the same
canonical quantization to the Poisson manifold $(M, \theta)$. One
can associate to a commutative algebra $(C^\infty(M),\{\cdot,\cdot\}_\theta)$
of smooth functions defined on the spacetime $M$, a noncommutative algebra $\CA_\theta$ of
linear operators on a suitable Hilbert space $\CH$. That is, the
smooth functions $f, g \in C^\infty(M)$ become noncommutative
operators (fields) $\widehat{f}, \widehat{g} \in \CA_\theta$ acting
on $\CH$, and the Poisson bracket in Eq. \eq{p-bracket} is replaced by a
noncommutative bracket \ct{nc-review,szabo-cqg}
\be \la{q-bracket}
\{f, g\}_\theta \; \rightarrow \; -i [\widehat{f},
\widehat{g}].
\ee
This completes the quantization of the Poisson algebra $(C^\infty(M),\{\cdot,\cdot\}_\theta)$
where spacetime $M$ is now noncommutative, i.e.,
\be \la{nc-spacetime}
[y^\mu, y^\nu] = i \theta^{\mu\nu}.
\ee
Throughout the paper, we will omit the hat for noncommutative
coordinates $y^\mu \in \CA_\theta$ for notational convenience.

The question still remains. What is the relation between the Poisson
algebra $(C^\infty(M),\{\cdot,\cdot\}_\theta)$ and (quantum) gravity?
We will not try to answer the question right now. Instead, we
want to point out that the vector fields in Eq. \eq{cartan} can be
derived from the Poisson algebra \ct{hsy-epl,hsy-ijmpa,hsy-epjc,hsy-jhep}.
It is well-known \ct{mechanics} that, for a given Poisson algebra $(C^\infty(M),
\{\cdot,\cdot\}_\theta)$, there exists a natural map $C^\infty (M) \to
TM: f \mapsto X_f$ between smooth functions in $C^\infty(M)$ and
vector fields in $TM$ such that
\be \la{duality}
X_f (g) = \{f, g \}_{\theta}
\ee
for any $g \in C^\infty(M)$. Indeed, the assignment in Eq. \eq{duality}
between a Hamiltonian function $f$ and the corresponding Hamiltonian
vector field $X_f$ is the Lie algebra homomorphism in the sense
\be \la{poisson-lie}
X_{\{f,g\}_{\theta}} = [X_f,X_g],
\ee
where the right-hand side represents the Lie bracket between the
Hamiltonian vector fields.

Motivated by the homomorphism between the Poisson algebra
$(C^\infty(M), \{\cdot,\cdot\}_\theta)$ and the Lie algebra of
vector fields \ct{mechanics}, one may venture to formulate Einstein
gravity in terms of a symplectic or a Poisson geometry rather than
a Riemannian geometry. Suppose that there is a set of fields defined
on a symplectic manifold $M$
\be \la{p-field}
\{D_a(y) \in C^\infty(M)| y \in M, \; a = 1, \cdots, 2n \}.
\ee
According to the map in Eq. \eq{duality}, the smooth functions in
Eq. \eq{p-field} can be mapped to vector fields as follows:
\be \la{v-field}
V_a[f](y) \equiv \{D_a (y), f(y)\}_{\theta} = - \theta^{\mu\nu}
\frac{\partial D_a(y)}{\partial y^\nu} \frac{\partial f(y)}{\partial y^\mu}.
\ee
The vector fields $V_a (y) = V_a^\mu (y) \frac{\partial}{\partial
y^\mu} \in \Gamma(TM)$ take values in the Lie algebra of
volume-preserving diffeomorphisms because $\p_\mu V_a^\mu = 0$ by
definition. Because the vector fields $V_a$ need not be orthonormal
though they are orthogonal frames, it will be possible to relate the
vector fields $V_a \in \Gamma(TM)$ to the orthonormal frames
(vielbeins) $E_a$ in Eq. \eq{cartan} by $V_a = \lambda E_a$,
with $\lambda \in C^\infty(M)$ to be determined \ct{hsy-jhep}.

Our above reasoning implies that a field theory equipped with the
fields in Eq. \eq{p-field} on a symplectic or a Poisson manifold may
give rise to Einstein gravity. If this is the case, quantum gravity
will be much more accessible because there is a natural symplectic or
Poisson structure, so it is obvious how to quantize the
underlying system, as was already done in Eqs. \eq{q-bracket} and \eq{nc-spacetime}.
Following this line of thought, we will aim to answer the question what quantum gravity
is by carefully addressing the issues in ${\bf Q1}$ and ${\bf Q2}$.

\subsection{Quartet of Physical Constants}

The physical constants defining a theory prescribe all the essential
properties of the theory. Planck realized that the union
of three ``fundamental" constants in Nature, the Planck constant
$\hbar$ in quantum mechanics, the universal velocity $c$ in
relativity, and the Newton constant $G$ in gravity, uniquely fixed
the characteristic scales for quantum gravity:
\bea \la{trio}
&& M_{pl} = \sqrt{\frac{c \hbar}{G}} = 2.2 \times 10^{-5} g, \xx
&& L_{pl} = \sqrt{\frac{G \hbar}{c^3}} = 1.6 \times 10^{-33} cm, \\
&& T_{pl} = \sqrt{\frac{G \hbar}{c^5}} = 5.4 \times 10^{-44} s.
\nonumber
\eea
The expression in Eq. \eq{trio} holds in four dimensions, and it has a
different expression in other dimensions. Interestingly, one
cannot construct a dimensionless quantity out of the three constants
except in two dimensions. In two dimensions, the combination $G
\hbar /c^3$ is a dimensionless quantity. This may be understood by
noticing that pure two-dimensional gravity is topological, so it
should not trigger any dynamical scale. Therefore, except for two
dimensions, each constant must play an independent role in the
theory of quantum gravity.

Recent developments in theoretical physics have revealed in many ways that
gravity may not be a fundamental, but rather an emergent, phenomenon,
as string theory has demonstrated in the last decade \ct{string-book}.
This means that the Newton constant $G$ can be determined by
using some of the quantities in a theory defining emergent gravity. Because we
want to derive gravity from some gauge theory, it is proper to consider the quartet of
physical constants by adding a coupling constant $e$, which is the electric charge.
A general gauge coupling constant sometimes will be denoted by $g_{YM}$.
Using the symbol $L$ for length, $T$ for time, and $M$ for mass and
writing $[X]$ for the dimension of some physical quantity $X$, we
have the following in $D$ dimensions:
\bea \la{quartet}
&& [\hbar] = M L^2 T^{-1} , \xx && [c] = L T^{-1}, \xx && [e^2] = M
L^{D-1} T^{-2}, \\
&& [G] = M^{-1} L^{D-1} T^{-2}. \nonumber
\eea
A remarkable point of the system in Eq. \eq{quartet} is that it specifies
the following intrinsic scales independently of dimensions:
\bea \la{3-quartet}
&& M^2 = \Big[ \frac{e^2}{G} \Big], \xx
&& L^2 = \Big[ \frac{G \hbar^2}{e^2 c^2} \Big], \\
&& T^2 = \Big[ \frac{G \hbar^2}{e^2 c^4} \Big]. \nonumber
\eea
From the four-dimensional case where $e^2/\hbar c
\approx 1/137$, one can see that the scales in Eq. \eq{3-quartet} are not
so different from the Planck scales in Eq. \eq{trio}.

Note that the first relation $GM^2 = e^2$ in Eq. \eq{3-quartet} implies
that at the mass scale $M$, the gravitational and the electromagnetic
interactions become of equal strength. This is a desirable property
for our purpose because we want to derive gravity from gauge theory!
Also, the system in Eq. \eq{quartet} should be expected to admit a
dimensionless quantity because the four quantities are determined by
three variables. That quantity is given by the following combination:
\be \la{dless}
\Big( \frac{ec}{\sqrt{G} \hbar} \Big)^D \cdot \frac{\hbar^3 G^2}{c^5
e^2} \; = \; {\rm dimensionless}.
\ee
One can see from Eq. \eq{dless} that in lower dimensions, it is
possible to construct a dimensionless quantity out of only three parameters:
$\{\hbar, c, G\}$ in $D=2$, $\{c, e, G\}$ in $D=3$, $\{\hbar, c,
e\}$ in $D=4$, and $\{\hbar, e, G\}$ in $D=5$. In $D \geq 6$
dimensions, we need all of the four constants in Eq. \eq{quartet} to
have a dimensionless quantity. This smells of interesting hidden
physics, but we have not yet figured out what it is \cite{6d-mphys}.
However, we will clearly see what the scales in Eq. \eq{3-quartet} set by the system
in Eq. \eq{quartet} mean. Notice that the length $L$ in
Eq. \eq{3-quartet} is the Compton wavelength of mass $M$ for which the
gravitational and the electromagnetic interactions have the same
strength. It turns out to be the scale of spacetime noncommutativity where
the conspiracy between gravity and gauge theory takes place.

Equation \eq{3-quartet} implies that, if gauge theory whose coupling
constant is given by $e$ is equipped with an intrinsic length scale
$L$, the Newton constant $G$ can be determined by using field theory
parameters only, i.e., $\frac{G\hbar^2}{c^2} \sim e^2 L^2$, hinting
an intimate correspondence between gravity and gauge theory
\ct{hsy-jhep}. For example, a noncommutative gauge theory and a field
theory on a D-brane are cases where $L^2$ can be identified with
$|\theta|$, the noncommutativity of spacetime, for the former and with
$2\pi \alpha'$, the size of a string, for the latter \cite{dbi}.
As we will discuss later, a theory of quantum gravity
must be background independent; thus, the dynamical scale for
quantum gravity will be generated by a vacuum condensate that is
exactly one in Eq. \eq{3-quartet}.

\subsection{Noncommutative Spacetime as Quantum Geometry}

We have discussed some reasons the gravitational constant $G$ dictates
a symplectic or a Poisson structure to spacetime $M$. Thereby, a field
theory will be defined on a symplectic manifold, and, as we argued
before, Einstein gravity may arise from the field theory. If this is
the case, quantum gravity will be defined by quantizing the field
theory in terms of the underlying Poisson structure, which is simply
a Dirac quantization such as Eq. \eq{q-bracket} for a canonical symplectic structure.
Therefore, if all these are smoothly working, quantum geometry can be defined by
using a field theory on a noncommutative space such as Eq. \eq{nc-spacetime}.
Let us briefly sketch how that is possible.

A symplectic structure $B = \half B_{\mu\nu} dy^\mu \wedge dy^\nu$
of spacetime $M$ defines a Poisson structure $\theta^{\mu\nu}
\equiv (B^{-1})^{\mu\nu}$ on $M$, where $\mu, \nu = 1, \dots, 2n$. The Dirac
quantization with respect to the Poisson structure $\theta^{\mu\nu}$
then leads to the quantum phase space in Eq. \eq{nc-spacetime}. Because the
Poisson structure $\theta^{\mu\nu}$ defines canonical pairs between
noncommutative coordinates $y^\mu \in \CA_\theta$, one can introduce
annihilation and creation operators, $a_i$ and $a_j^\dagger, \;
i,j = 1, \cdots, n$, by using those pairs. For example,
$a_1 = (y^1 + i y^2) / \sqrt{2\theta^{12}}, \; a_1^\dagger = (y^1 - i y^2) /\sqrt{2\theta^{12}}$, etc.
Then, the Moyal algebra in Eq. \eq{nc-spacetime} is equal to the Heisenberg algebra of an
$n$-dimensional harmonic oscillator, i.e.,
\begin{equation} \label{vacuum-spacetime}
[y^\mu, y^\nu] = i \theta^{\mu\nu} \;\; \Leftrightarrow \;\; [a_i,
a_j^\dagger] = \delta_{ij}.
\end{equation}

From quantum mechanics, the representation space of noncommutative $\mathbb{R}^{2n}$ is
well-known to be given by an infinite-dimensional, separable Hilbert space
\be \la{fock}
\CH = \{ |\vec{n} \rangle \equiv |n_1,\cdots,n_n \rangle,
\; n_i=0,1, \cdots \}
\ee
that is orthonormal, i.e., $\langle \vec{n}| \vec{m} \rangle = \delta_{\vec{n}\vec{m}}$,
and complete, i.e., $\sum_{\vec{n} = 0}^\infty |\vec{n} \rangle \langle \vec{n}| = 1$.
Note that any smooth function on a noncommutative space can be represented
by an operator acting on an appropriate Hilbert space ${\cal H}$, which
consists of a noncommutative $\star$-algebra ${\cal A}_\theta$, like
a set of observables in quantum mechanics. Therefore, any field
$\widehat{\Phi} \in \CA_\theta $ in the noncommutative space in Eq. \eq{vacuum-spacetime}
becomes an operator acting on the Fock space ${\cal H}$ and
can be expanded in terms of the complete operator basis
\begin{equation}\label{matrix-basis}
\CA_\theta = \{ |\vec{n} \rangle \langle \vec{m}|, \; n_i, m_j = 0,1, \cdots \},
\end{equation}
that is,
\begin{equation}\label{op-matrix}
    \widehat{\Phi}(y) = \sum_{\vec{n},\vec{m}} \Phi_{\vec{n}\vec{m}}
    |\vec{n} \rangle \langle \vec{m}|.
\end{equation}
One may use the `Cantor diagonal method' to put the $n$-dimensional
non-negative integer lattice in $\CH$ into a one-to-one correspondence
with the infinite set of natural numbers (i.e., $1$-dimensional
positive integer lattice): $|\vec{n} \rangle
\leftrightarrow |n \rangle, \; n=1,\cdots,N
\to \infty $. In this one-dimensional basis, Eq. \eq{op-matrix} can
be relabeled in the following form:
\be \la{matrix-expansion}
\widehat{\Phi}(y) = \sum_{n,m=1}^\infty M_{nm} \;|n \rangle \langle m |.
\ee
One can regard $M_{nm}$ in Eq. \eq{matrix-expansion} as elements of
an $N \times N$ matrix $M$ in the $N \to \infty$ limit. We then get
the following important relation \ct{nc-review,szabo-cqg}:
\be \la{ncft-matrix}
\mathrm{Any \; field \; on \; noncommutative} \; \mathbb{R}^{2n} \; \cong \; N \times N \;
\mathrm{matrix} \; \mathrm{at} \; N \to \infty.
\ee
If the field $\Phi(y) \in C^\infty(M)$ was originally a real field,
then $M$ should be a Hermitian matrix. The relation in Eq. \eq{ncft-matrix}
means that a field in the noncommutative $\star$-algebra
$\CA_\theta$ can be regarded as a master field of a large $N$
matrix.

Now the very notion of points in a noncommutative space, such as the
Moyal space in Eq. \eq{nc-spacetime}, is doomed; instead the points are replaced by
states in $\CH$. Thus, the usual concept of geometry based on a smooth
manifold will be replaced by a theory of operator algebra, e.g.,
the noncommutative geometry {\it \`a la} Connes \ct{connes} or the
theory of deformation quantization {\it \`a la} Kontsevich
\ct{kontsevich}. Furthermore, through the matrix representation in Eq. \eq{ncft-matrix}
of the operator algebra, one can achieve a coordinate-free description of quantum field theory.
Therefore, it should be possible to achieve a {\it background-independent}
formulation of quantum gravity in which the interactions between
fundamental ingredients can be defined without introducing any
spacetime structure \ct{hsy-jhep}. The most natural objects for that purpose
are algebra-valued fields, which can be identified with elements in a noncommutative $\star$-algebra
such as noncommutative gauge fields \cite{SW-ncft}.

Several matrix models \ct{bfss,ikkt,mst} have appeared in string
theory. They have illustrated how the matrix model can be regarded as a
nonperturbative formulation of gravity or string theory in the sense
that it describes a quantized geometry with an arbitrary topology.
Matrix theory contains multiple branes with arbitrary topologies
as its spectrum and allows a topology change of the spacetime manifold
as a sequence of the change of matrix data \ct{taylor}.

\subsection{Outline of the Paper}

In this review, we will not survey other approaches to emergent gravity
because good expositions \ct{analogue} already exist and our
approach is quite different, although the underlying philosophy may be
the same. Our unique clue is based on the fact that Einstein gravity
can be formulated in terms of a symplectic geometry \ct{hsy-jhep}.
Basically, we are considering a symplectic geometry as a commutative
limit of a noncommutative geometry that is regarded as a microscopic
structure of spacetime, just as classical mechanics on a mundane
scale is simply a coarse graining of quantum mechanics in the atomic world.
A Riemannian geometry, thus, appears in a macroscopic world as a coarse graining
of the noncommutative geometry, as we already briefly outlined.

Our line of thought has been motivated by several similar ideas,
mostly by the AdS/CFT correspondence \ct{ads-cft} and matrix models
\ct{bfss,ikkt,mst} in string theory. Also, the work by Rivelles \ct{rivelles}
and the following works \ct{hsy-before} triggered thoughts about a
noncommutative field theory as a theory of gravity. A series of
interesting works along this line has recently been conducted by
Steinacker and his collaborators \ct{harold1,harold2,harold3}. See his recent review
\ct{harold-review} and references therein. Also, there are many closely related
works \ct{rel-old,rel-work1,rel-work2,rel-work3,bubbling,berenstein,rel-matrix,compact-matrix,schenkel}.
However, the emergent gravity based on noncommutative field
theories is relatively new, so it would be premature to have an
extensive review about this subject because it is still in an early
stage of development. Therefore we will focus on ours.
Although this review is basically a coherent survey of recent
works \ct{hsy-epl,hsy-ijmpa,hsy-epjc,hsy-jhep,hsy-inst,hsy-mpla,hsy-bjp,hsy-cc,hsy-siva}
of the second author, it also contains several new results and
many clarifications, together with important pictures about quantum
gravity. Early reviews can be found in Refs. \ct{hsy-mpla} and \ct{hsy-bjp}.

In Section II we elucidate the reason Einstein gravity can emerge
from electromagnetism as long as spacetime admits a symplectic
structure and explicitly show how Einstein's equations can be
derived from the equations of motion for electromagnetic fields on a
symplectic spacetime. The emergent gravity we propose here actually
corresponds to a field theory realization of open-closed string
duality or large-$N$ duality in string theory and to a generalization
of homological mirror symmetry \ct{kont-mirror} in the sense that the deformation of
the symplectic structure $\omega$ is isomorphic to the deformation of
the Riemannian metric $g$ in the triple $(M, g, \omega)$.
In Section II.A, we explain the equivalence principle for the
electromagnetic force, stating that there always exists a coordinate
transformation to locally eliminate the electromagnetic force if
spacetime supports a symplectic structure. This important property
for emergent gravity originates from the Darboux theorem
\ct{darboux} or the Moser lemma \ct{moser} in symplectic geometry,
which also explains how diffeomorphism symmetry in general
relativity arises from such a gauge theory. We explain how the
equations of motion for $U(1)$ gauge fields are mapped to those for
vector fields defined by Eq. \eq{v-field}.
In Section II.B, we first initiate the emergent gravity by showing \ct{hsy-epl} that
self-dual electromagnetism on a symplectic 4-manifold is
equivalent to self-dual Einstein gravity. Although it was previously
proved in Ref. \ct{hsy-jhep}, we newly prove it again in a more geometric
way to illustrate how elegant and beautiful the emergent gravity is.
In Section II.C, we derive Einstein's equations from the
electromagnetism on a symplectic manifold, rigorously confirming the
speculation in Section I.A. We find \ct{hsy-jhep} that the emergent gravity from
electromagnetism predicts some exotic energy whose physical
nature will be identified in Section V.B.

In Section III, we discuss how to quantize Einstein gravity in the
context of emergent gravity by using the canonical quantization
in Eq. \eq{q-bracket} of a spacetime Poisson manifold.
We argue in Section III.A that the equivalence principle for the
emergent gravity can be lifted to a noncommutative spacetime such as
Eq. \eq{nc-spacetime}. The equivalence principle in a noncommutative spacetime is
realized as a gauge equivalence between star products in the context
of deformation quantization \ct{kontsevich}, and so dubbed the ``quantum equivalence
principle" \ct{hsy-ijmpa}. This implies that quantum gravity can consistently be
derived from the quantum equivalence principle and that matter fields can
arise from the quantized spacetime.
In Section III.B, it is shown \ct{hsy-epjc} that the emergent gravity from a
noncommutative $\star$-algebra $\CA_\theta$ can be understood as a
large-$N$ duality such as the AdS/CFT correspondence and the matrix
models in string theory. The gravitational metric determined by
large-$N$ matrices or noncommutative gauge fields is explicitly derived.
We clarify in Section III.C how emergent gravity achieves the background-independent formulation
in which any kind of spacetime structure is not {\it a priori} assumed, but is defined by the theory.
An important picture of emergent gravity is identified. If a classical
geometry is to be described from a background-independent theory, it is necessary to
have a nontrivial vacuum defined by a ¡°coherent¡± condensation of
gauge fields, e.g., the vacuum defined by Eq. \eq{vacuum-spacetime},
which is also the origin of a spacetime symplectic or a Poisson
structure such as Eq. \eq{p-bracket}.
In Section III.D, the emergent gravity is generalized to a generic
noncommutative spacetime such as the case with $\theta^{\mu\nu}=$
nonconstant \ct{hsy-ijmpa} and a general Poisson manifold \ct{hsy-siva}.

In Section IV, we speculate what particles and matter fields are and
how they arise from a noncommutative $\star$-algebra $\CA_\theta$.
We claim that a matter field, such as quarks and leptons, is defined
by a stable localized geometry, which is a topological object in the
defining algebra (noncommutative $\star$-algebra) of quantum gravity \ct{hsy-jhep}.
First we review in Section IV.A Feynman's view \ct{feynman,crlee,feynman-gen}
of the electrodynamics of a charged particle to understand why an extra
internal space is necessary to introduce the weak and the strong forces.
The extra dimensions appear with a Poisson structure of Lie algebra
type implemented with some localizability condition to stabilize the internal space.
In Section IV.B, we understand the Feynman's derivation of gauge
forces as the Darboux transformation in Eq. \eq{darboux-tr} between two
symplectic structures where one of them is a deformation of the
other in terms of external gauge fields. This beautiful idea is not
ours, but was noticed at Ref. \ct{sternberg} long ago.
In Section IV.C, we define a stable state in a large-$N$ gauge theory
and relate it to the K-theory \ct{karoubi,k-theory1,k-theory2,k-theory3}.
With the correspondence in Eq. \eq{ncft-matrix}, the K-theory class is
mapped to the K-theory of a noncommutative $\star$-algebra
$\CA_\theta$. We argue, using the well-known, but rather mysterious,
math, the Atiyah-Bott-Shapiro construction of K-theory generators in
terms of the Clifford module \ct{abs}, that the topological object
defined by large-$N$ matrices or the noncommutative $\star$-algebra
$\CA_\theta$ describes fermions such as quarks and leptons. It turns
out that an extra noncommutative space is essential to realize the weak and the strong forces.

In Section V, we discuss the most beautiful aspects of emergent gravity.
Remarkably, emergent gravity reveals a novel picture
about the origin of spacetime, dubbed as emergent spacetime, which
is radically different from any previous physical theories, all of
which describe what happens in a given spacetime.
In Section V.A, we point out that the concept of emergent time is
naturally defined as long as spacetime admits an intrinsic
symplectic or Poisson structure. The time evolution of a spacetime
geometry is defined by Hamilton's equation defined by the
spacetime Poisson structure,  Eq. \eq{p-bracket}. Because the symplectic
structure triggered by the vacuum condensate in Eq. \eq{vacuum-spacetime} not only causes
the emergence of spacetime but also specifies an orientation of spacetime manifold,
we are tempted to conceive that the emergent gravity may explain the ¡°arrow of time¡± in the
cosmic evolution of our Universe - the most notoriously difficult problem in quantum gravity.
We analyze in Section V.B the anatomy of spacetime derived from a
noncommutative gauge theory or large-$N$ matrix model. We explain
why there is no cosmological constant problem in emergent gravity
\ct{hsy-cc}. We point out that a vacuum energy of a Planck scale does not
gravitate, unlike Einstein gravity and that a flat spacetime is emergent from
the Planck energy condensation in vacuum.
Finally, we try to identify the physical nature of the exotic
energy-momentum tensor whose existence was predicted in Section II.C.
Surprisingly, it mimics all the properties of dark energy, so we
suggest the energy as a plausible candidate for dark energy
\ct{hsy-jhep}. If so, the quantum gravity defined by noncommutative
gauge theory seems to resolve many notorious problems in theoretical
physics: for example, the cosmological constant problem,
the nature of dark energy and the reason gravity is so weak compared to other forces.

In the final section, we try to understand why the emergent gravity
defined by a noncommutative geometry resembles string theory.
We conclude with several remarks about important open issues
and speculate on a proper mathematical framework for emergent gravity and quantum gravity.

\section{EMERGENT GRAVITY}

In order to argue that gravity can emerge from some field theory, it
is important to identify how the essential properties of gravity can be
realized in the underlying field theory. If not, the emergent
gravity cannot physically be viable. Therefore, we will reasonably
argue how the equivalence principle, the most important property in
the theory of gravity (general relativity), can be realized from
$U(1)$ gauge theory on a symplectic manifold $M$. Through the
realization of the equivalence principle in the context of
symplectic geometry, we can understand how diffeomorphism symmetry
arises from noncommutative $U(1)$ gauge theory and gravity can
emerge from noncommutative electromagnetism, which is also an
interacting theory.

\subsection{The Equivalence Principle from Symplectic Geometry}

Consider a $U(1)$ bundle supported on a symplectic manifold $(M,B)$.
Physically, we are considering open strings moving on a D-brane whose
data are given by $(M, g, B)$, where $M$ is a smooth manifold
equipped with a metric $g$ and a symplectic structure $B$. The
worldsheet action of open strings, with a compact notation, reads as
\be \la{open-action}
S = \frac{1}{4 \pi \alpha^\prime} \int_{\Sigma} |d X|^2  -
\int_{\Sigma} B -
\int_{\p \Sigma} A,
\ee
where $X: \Sigma \to M$ is a map from an open string worldsheet
$\Sigma$ to a target spacetime $M$ and $B(\Sigma) = X^*B(M)$ and
$A(\p \Sigma) = X^* A(M)$ are pull-backs of spacetime gauge fields to the
worldsheet $\Sigma$ and the worldsheet boundary $\p
\Sigma$, respectively.

From the compact notation in Eq. \eq{open-action}, it is obvious that
the string action in Eq. \eq{open-action} respects the following local gauge symmetries:\\
(I) Diff($M$)-symmetry
\be \la{diff-symm}
X \to X^\prime = X^\prime (X) \; \in \; {\rm Diff}(M),
\ee
(II) $\Lambda$-symmetry
\be \la{lambda-symm}
(B,\; A) \to (B - d\Lambda, \; A + \Lambda),
\ee
where the gauge parameter $\Lambda$ is a one-form in $M$. Note that
the $\Lambda$-symmetry is present only when $B \neq 0$, so it is a stringy
symmetry by nature. When $B=0$, the symmetry in Eq. \eq{lambda-symm} is
reduced to $A \to A + d\lambda$, which is the ordinary $U(1)$ gauge
symmetry because $A$ is a connection of the $U(1)$ bundle.

The $\Lambda$-symmetry then predicts a very important result. The
presence of a $U(1)$ bundle on a symplectic manifold $(M, B)$ should
appear only with the combination $\Omega = B + F$, where $F=dA$ because
$\Omega$ is the only gauge-invariant 2-form under the transformation
in Eq. \eq{lambda-symm}. Because we regard $B \in \Gamma(\Lambda^2 T^*M)$ as a
symplectic structure over $M$, the electromagnetic force $F=dA$ appears only
as the local deformation of the symplectic structure $\Omega (x) = (B + F) (x)$.

Another important result derived from the open string action,
Eq. \eq{open-action}, is that the triple $(M, g, B)$
comes only in the combination $(M, g + \kappa B)$, where $\kappa
\equiv 2 \pi \alpha' = 2 \pi l_s^2$ denotes the string scale.
Note that the Riemannian metric $g$ and the symplectic structure $B$
in the triple $(M, g, B)$ can be regarded as an bundle isomorphism
from a tangent bundle $TM$ to a cotangent bundle $T^*M$ because
both are nondegenerate bilinear maps on $TM$, i.e., $(g, B) : TM \to
T^*M$. Therefore, the so-called DBI ``metric" $g +
\kappa B: TM \to T^*M$, which maps $X \in TM$ to $\xi = (g + \kappa
B)(X) \in T^*M$, embodies a generalized geometry \ct{general-geom}
continuously interpolating between a symplectic geometry $(|\kappa
Bg^{-1}| \gg 1)$ and a Riemannian geometry $(|\kappa Bg^{-1}| \ll 1)$.
Including the excitation of open strings, one can combine the two
results to conclude that the data of `D-manifold' are given by $(M,
g, \Omega)= (M, g + \kappa \Omega)$.

Consider another D-brane whose `D-manifold' is described by
different data $(N, G, B) = (N, G + \kappa B)$. A question is
whether there exists a diffeomorphism $\phi : N \to M$ such that
$\phi^* (g + \kappa \Omega) = G + \kappa B$ on $N$. In order to
answer the question, let us shortly digress to some important
aspects of simplectic geometry. The symplectic geometry respects an important property, known as the
Darboux theorem \ct{darboux}, stating that every symplectic manifold
of the same dimension is locally indistinguishable \ct{mechanics}.
To be precise, let $(M, \omega)$ be a symplectic manifold.
Consider a smooth family $\omega_t = \omega_0 + t (\omega_1 -
\omega_0)$ of symplectic forms joining $\omega_0$ to $\omega_1$, where
$[\omega_0] = [\omega_1] \in H^2(M)$ and $\omega_t$ is symplectic
$\forall t \in [0, 1]$. A remarkable point (due to the Moser lemma
\ct{moser}) is that there exists a one-parameter family of diffeomorphisms
$\phi: M \times \mathbb{R} \to M$ such that $\phi^*_t(\omega_t) =
\omega_0, \; 0 \leq t \leq 1$. If there exist such diffeomorphisms as a flow generated
by time-dependent vector fields $X_t \equiv \frac{d \phi_t}{dt}
\circ \phi_t^{-1}$, one would have for all $0 \leq t \leq 1$ that
\be \la{symplectic-evolution}
{\cal L}_{X_t} \omega_t + \frac{d \omega_t}{dt} = 0
\ee
because, by the Lie derivative formula, one has
\bea \la{time-flow}
0 = \frac{d}{dt} \big( \phi_t^* \omega_t \big) &=&
\phi_t^* \big( \mathcal{L}_{X_t} \omega_t \big) + \phi_t^* \frac{d \omega_t}{dt} \xx
&=& \phi_t^* \Big( \mathcal{L}_{X_t} \omega_t + \frac{d
\omega_t}{dt} \Big).
\eea
Using Cartan's magic formula, $\CL_X = d \iota_X + \iota_X d$,
for the Lie derivative along the flow of a vector field $X$, one can
see that Eq. \eq{symplectic-evolution} can be reduced to Moser's equation
\be \la{moser-eq}
\iota_{X_t} \omega_t + A = 0,
\ee
where $\omega_1 - \omega_0 = dA$.

In summary, there always exists a one-parameter family of
diffeomorphisms as a flow generated by a smooth family of
time-dependent vector fields $X_t$ satisfying Eq. \eq{moser-eq} for
the change of the symplectic structure within the same cohomology
class from $\omega_0$ to $\omega_1$, where $\omega_1 - \omega_0 = dA$.
The evolution of the symplectic structure is locally described
by the flow $\phi_t$ of $X_t$ satisfying $\frac{d
\phi_t}{dt} = X_t \circ \phi_t$ and starting at $\phi_0$ = identity.
Thus, one has $\phi_1^* \omega_1 =  \phi_0^* \omega_0 =
\omega_0$, so $\phi_1$ provides a chart describing the evolution
from $\omega_0$ to $\omega_1 = \omega_0 + dA$. In terms of local
coordinates, there always exists a coordinate transformation
$\phi_1$ whose pullback maps $\omega_1 = \omega_0 + dA$ to
$\omega_0$, i.e., $\phi_1 : y
\mapsto x = x(y)$ so that
\be \la{darboux-tr}
\frac{\p x^a}{\p y^\mu}\frac{\p x^b}{\p y^\nu} {\omega_1}_{ab}(x)
= {\omega_0}_{\mu\nu}(y).
\ee

This can directly be applied to the open string case,
Eq. \eq{open-action}, by considering a local Darboux chart $(U;
y^1, \cdots, y^{2n})$ centered at $p \in U$ and valid on an open
neighborhood $U \subset M$ such that $\omega_0|_U =
\frac{1}{2} B_{\mu\nu} dy^\mu \wedge dy^\nu$, where $B_{\mu\nu}$ is a constant
symplectic matrix of rank $2n$. Now, consider a flow $\phi_t: U
\times [0,1] \to M$ generated by the vector field $X_t$ satisfying
Eq. \eq{moser-eq}. Under the action of $\phi_\epsilon$ with an
infinitesimal $\epsilon$, one finds that the point $p \in U$ whose
coordinates are $y^\mu$ is mapped to $\phi_\epsilon (y) \equiv
x^\mu(y) = y^\mu + \epsilon X^\mu(y)$. Using the inverse map
$\phi^{-1}_\epsilon: x^\mu \mapsto y^\mu (x)= x^\mu - \epsilon
X^\mu(x)$, the symplectic structure $\omega_0|_U = \frac{1}{2}
B_{\mu\nu} (y) dy^\mu \wedge dy^\nu$ can be expressed as
\bea \la{moser}
(\phi^{-1}_\epsilon)^* (\omega_0|_y) &=& \frac{1}{2} B_{\mu\nu}(x -
\epsilon X) d(x^\mu - \epsilon X^\mu)
\wedge d(x^\nu - \epsilon X^\nu) \xx & \approx & \frac{1}{2} \Big[ B_{\mu\nu} -
\epsilon X^a (\partial_a B_{\mu\nu} + \partial_\nu B_{a \mu}
+ \partial_\mu B_{\nu a}) + \epsilon \Big( \partial_\mu (B_{\nu a}
X^a) - \partial_\nu (B_{\mu a} X^a) \Big) \Big] dx^\mu \wedge dx^\nu
\xx &\equiv& B + \epsilon F,
\eea
where $A_\mu(x) = B_{\mu a}(x) X^a(x)$ or $\iota_X B + A = 0$ and
$dB = 0$ was used so that the second term vanished.
Equation \eq{moser} can be rewritten as $\phi_\epsilon^* (B + \epsilon F)
= B$, which means that the electromagnetic force $F = dA$ can always
be eliminated by a local coordinate transformation generated by the
vector field $X$ satisfying Eq. \eq{moser-eq}.

Now, let us go back to the previous question. We considered a
symmetry transformation which is a combination of the
$\Lambda$-transformation, Eq. \eq{lambda-symm}, followed by a
diffeomorphism $\phi : N \to M$. It transforms the DBI metric $g +
\kappa B$ on $M$ according to $g + \kappa B \to \phi^* (g + \kappa
\Omega)$. The crux is that there exists a diffeomorphism
$\phi : N \to M$ such that $\phi^* (\Omega) = B$, which is precisely
the Darboux transformation in Eq. \eq{darboux-tr} in symplectic geometry.
Then, we arrive at a remarkable fact \ct{hsy-jhep} that two different
DBI metrics, $g + \kappa \Omega$ and $G + \kappa B$, are diffeomorphic to each other, i.e.,
\be \la{diff-dbi}
\phi^* (g + \kappa \Omega) = G + \kappa B,
\ee
where $G = \phi^* (g)$. Because the open string theory, Eq. \eq{open-action}, respects
the diffeomorphism symmetry, Eq. \eq{diff-symm}, the D-manifolds described by $(M, g, \Omega)$ and
$(N, G, B)$ must be physically equivalent. Note that this property
holds for any pair $(g,B)$ of a Riemannian metric $g$ and a symplectic structure $B$.

The above argument reveals superb physics in gauge theory.
There ``always" exists a coordinate transformation to locally
eliminate the electromagnetic force $F=dA$ as long as a manifold $M$
supports a symplectic structure $B$; i.e., $(M, B)$ defines a
symplectic manifold. That is, a symplectic structure on a spacetime
manifold $M$ admits a novel form of the equivalence principle, known
as the Darboux theorem, for the geometrization of the
electromagnetic force. Because it is always possible to find a coordinate
transformation $\phi \in {\rm Diff}(M)$ such that $\phi^* (B+F) =
B$, the relationship $\phi^*
\big(g + \kappa (B+F) \big) = G +
\kappa B$ clearly shows that the electromagnetic fields in the DBI
metric $g + \kappa (B+F)$ now appear as a new metric $G = \phi^*
(g)$ after the Darboux transformation in Eq. \eq{diff-dbi}. One may also
consider the inverse relationship $\phi_*(G +
\kappa B)= g + \kappa (B+F)$, which implies that a nontrivial metric
$G$ in the background $B$ can be interpreted as an inhomogeneous
condensation of gauge fields on a `D-manifold' with metric $g$.
We might point out that the relationship in the case of $\kappa =
2 \pi \alpha' = 0$ is the familiar diffeomorphism in a Riemannian
geometry, so it says nothing marvelous. See footnote \cite{dbi}.

We observed that the Darboux transformation between symplectic
structures immediately leads to the diffeomorphism between two
different DBI metrics. In terms of local coordinates $\phi: y
\mapsto x = x(y)$, the diffeomorphism in Eq. \eq{diff-dbi} explicitly reads as
\be \la{dbi-iso}
(g + \kappa \Omega)_{ab} (x) = \frac{\p y^\mu}{\p x^a}
\Bigl(G_{\mu\nu}(y)+  \kappa B_{\mu\nu}(y) \Bigr) \frac{\p y^\nu}{\p x^b},
\ee
where $\Omega = B + F$ and
\be \la{gauge-metric}
G_{\mu\nu}(y) = \frac{\p x^a}{\p y^\mu} \frac{\p x^b}{\p y^\nu}
g_{ab}(x).
\ee
Equation \eq{dbi-iso} conclusively shows how gauge fields on a symplectic
manifold manifest themselves as a spacetime geometry. To expose the
intrinsic connection between gauge fields and spacetime geometry,
let us represent the coordinate transformation in Eq. \eq{dbi-iso} as
\be \la{cov-cod}
x^a(y) = y^a + \theta^{ab} \widehat{A}_b(y),
\ee
with $\theta^{ab} = (B^{-1})^{ab}$. Note that the coordinate
transformation in Eq. \eq{cov-cod} for the case $F(x) = 0$ is trivial, i.e.,
$\widehat{A}_a(y) = 0$ and $G_{ab} = g_{ab}$ as it should be, while
it is nontrivial for $F(x) \neq 0$. The nontrivial coordinate transformation
can be encoded into smooth functions $\widehat{A}_a(y)$, which will be identified with
noncommutative gauge fields. Clearly, Eq. \eq{gauge-metric} embodies
how the metric on $M$ is deformed by the presence of noncommutative
gauge fields.

We showed how the diffeomorphism symmetry in Eq. \eq{diff-dbi} between two
different DBI metrics arises from $U(1)$ gauge theory on a
symplectic manifold. Surprisingly (at least to us), the
diffeomorphism symmetry in Eq. \eq{diff-dbi} is realized as a novel form of
the equivalence principle for the electromagnetic force
\ct{hsy-ijmpa}. Therefore, one may expect electromagnetism on a
symplectic manifold to be a theory of gravity; in other words,
gravity can emerge from electromagnetism on a symplectic or Poisson manifold.

A low-energy effective field theory deduced from the open string
action in Eq. \eq{open-action} describes an open string dynamics on a
$(p+1)$-dimensional D-brane worldvolume \ct{string-book}. The
dynamics of D-branes is described by open string field theory whose
low-energy effective action is obtained by integrating out all the
massive modes and keeping only massless fields that are slowly varying
at the string scale $\kappa = 2 \pi l_s^2$. For a $Dp$-brane in
closed string background fields, the action describing the resulting
low-energy dynamics is given by
\begin{equation} \label{dbi-general}
S = \frac{2\pi}{g_s (2\pi \kappa)^{\frac{p+1}{2}}}
\int d^{p+1} x \sqrt{\det(g + \kappa (B + F))}
+ {\cal O}(\sqrt{\kappa} \partial F, \cdots),
\end{equation}
where $F=dA$ is the field strength of $U(1)$ gauge fields. The DBI
action in Eq. \eq{dbi-general} respects the two local symmetries,
Eqs. \eq{diff-symm} and \eq{lambda-symm}, as expected.

Note that ordinary $U(1)$ gauge symmetry is a special case of
Eq. \eq{lambda-symm} where the gauge parameter $\Lambda$ is exact,
namely, $\Lambda = d \lambda$, so that $B \to B$ and $A \to A +
d\lambda$. One can see from Eq. \eq{moser-eq} that the $U(1)$ gauge
transformation is generated by a Hamiltonian vector field
$X_\lambda$ satisfying $\iota_{X_\lambda} B + d \lambda = 0$.
Therefore, the gauge symmetry acting on $U(1)$ gauge fields as $A
\to A + d\lambda$ is a diffeomorphism symmetry generated by the
vector field $X_\lambda$ satisfying $\CL_{X_\lambda} B = 0$, which is
known to be a symplectomorphism. We see here that the $U(1)$ gauge
symmetry on the symplectic manifold $(M,B)$ turns into a
``spacetime" symmetry rather than an ``internal" symmetry. This fact
already implies an intimate connection between gauge fields on a
symplectic manifold and spacetime geometry.

It was shown in Eq. \eq{dbi-iso} that the strong isotopy in Eq. \eq{darboux-tr}
between symplectic structures brings in the diffeomorphic equivalence,
Eq. \eq{dbi-iso}, between two different DBI metrics, which, in turn, leads
to a remarkable identity \ct{cornalba} between DBI actions:
\begin{equation} \label{mirror}
\int_M d^{p+1} x \sqrt{\det \bigl(g(x) + \kappa (B + F)(x)\bigr)}
= \int_N d^{p+1} y \sqrt{\det\bigl(G(y) + \kappa B(y)\bigr)}.
\end{equation}
The property in Eq. \eq{mirror} appearing in the geometrization of
electromagnetism may be summarized in the context of a derived
category. More closely, if $M$ is a complex manifold whose complex
structure is given by $J$, we see that dynamical fields on the
left-hand side of Eq. \eq{mirror} act only as the deformation of
the symplectic structure $\Omega (x) = (B + F)(x)$ in the triple $(M, J,
\Omega)$ while those on the right-hand side appear only as
the deformation of the complex structure $K = \phi^*(J)$ in the triple
$(N, K, B)$ through the metric in Eq. \eq{gauge-metric}. In this notation,
the identity in Eq. \eq{mirror} can, thus, be written as follows:
\be \la{h-mirror}
(M, J, \Omega) \cong  (N, K, B).
\ee
The equivalence, Eq. \eq{h-mirror}, is very reminiscent of the homological
mirror symmetry \ct{kont-mirror}, which states the equivalence between
the category of A-branes (derived Fukaya category corresponding to
the triple $(M, J, \Omega)$) and the category of B-branes (derived
category of coherent sheaves corresponding to the triple $(N, K, B)$).

Because the open string action in Eq. \eq{open-action} basically describes a
$U(1)$-bundle (the Chan-Paton bundle) on a D-brane whose data are
given by $(M, g, B)$, $U(1)$ gauge fields, the connections of the
$U(1)$ bundle are regarded as dynamical fields while the metric $g$
and the two-form $B$ are considered as background fields. However,
Eq. \eq{mirror} clearly shows that gauge field fluctuations can be
interpreted as a {\it dynamical} metric on the brane given by
Eq. \eq{gauge-metric}. In all, one may wonder whether the right-hand
side of Eq. \eq{mirror} can be rewritten as a theory of gravity.
Remarkably, it is the case, as will be shown soon.

Here, we will use the background-independent prescription \ct{SW-ncft} in which the open
string metric $\widehat{g}_{ab}$, the noncommutativity $\theta^{ab}$
and the open string coupling constant $\widehat{g}_s$ are determined by
\be \la{open}
\theta^{ab} = \Bigl(\frac{1}{B}\Bigr)^{ab}, \quad
\widehat{g}_{ab} = - \kappa^2 \Bigl( B \frac{1}{g} B \Bigr)_{ab},
\quad \widehat{g}_s = g_s \sqrt{\det (\kappa B g^{-1})}.
\ee
In terms of these parameters, the couplings are related by
\bea \la{coupling1}
&& \frac{1}{g_{YM}^2} =
\frac{\kappa^{2-n}}{(2\pi)^{n-1}
  \widehat{g}_s}, \\
  \la{coupling2}
&& \frac{\sqrt{\det \widehat{g}}}{\widehat{g}_s} =
\frac{\kappa^n}{g_s |{\rm Pf} \theta|},
\eea
where $p+1 \equiv 2n$. For constant $g$ and $B$, one can
rewrite the right-hand side of Eq. \eq{mirror} by using the open string variables
and defining new covariant fields $D_a (y) \equiv B_{ab} x^b (y)$ as
\be \label{mirror-open}
\int d^{p+1} y \sqrt{\det\bigl(G(y) + \kappa B \bigr)} =
\int d^{p+1} y \sqrt{\det\bigl(\kappa B + \kappa^2 \widehat{G}(y)\bigr)},
\ee
where
\be \la{opengauge-metric}
\widehat{G}_{\mu\nu}(y) = \widehat{g}^{ab} \frac{\p D_a}{\p y^\mu} \frac{\p D_b}{\p y^\nu}.
\ee
One can expand the right-hand side of Eq. \eq{mirror-open} around the
background $B$ in powers of $\kappa$, arriving at the following result:
\bea \label{semi-expansion}
&& \int d^{p+1} y \sqrt{\det\bigl(\kappa B + \kappa^2 \widehat{G}(y)\bigr)} \xx &&
\qquad = \int d^{p+1} y \sqrt{\det\bigl(\kappa B \bigr)}
\Bigl(1 + \frac{\kappa^2}{4}  \widehat{g}^{ac} \widehat{g}^{bd}
\{D_a, D_b\}_{\theta} \{D_c, D_d\}_{\theta} + \cdots \Bigr),
\eea
where $\{D_a, D_b\}_{\theta}$ is the Poisson bracket defined by Eq. \eq{p-bracket}.
The second part of Eq. \eq{semi-expansion} can then be written in a form with a constant metric
$\widehat{g}^{ab} = - \frac{1}{\kappa^2} (\theta g \theta)^{ab}$ as
\be \la{semi-matrix}
S_D = \frac{1}{4 g_{YM}^2} \int d^{p+1} y \sqrt{\det
\widehat{g}} \; \widehat{g}^{ac} \widehat{g}^{bd}
\{D_a, D_b\}_{\theta} \{D_c, D_d\}_{\theta},
\ee
where the gauge coupling constant $g_{YM}$ was recovered after collecting all factors including
the one in Eq. \eq{dbi-general}. From now on, let us set the metric $\widehat{g}^{ab} = \delta^{ab}$
for simplicity.

Note that
\bea \la{symp-fs}
\{D_a, D_b\}_{\theta} &=& -B_{ab} + \p_a \widehat{A}_b - \p_b
\widehat{A}_a + \{\widehat{A}_a, \widehat{A}_b\}_{\theta} \xx
&\equiv& - B_{ab} + \widehat{F}_{ab}
\eea
and
\bea \la{symp-cfs}
\{D_a, \{D_b, D_c \}_{\theta} \}_{\theta} &=& \p_a \widehat{F}_{bc} + \{\widehat{A}_a,
\widehat{F}_{bc} \}_{\theta} \xx
&\equiv& \widehat{D}_a \widehat{F}_{bc}.
\eea
Therefore, the Jacobi identity for the Poisson bracket in Eq. \eq{p-bracket} can be written as
\bea \la{poisson-jacobi}
0 &=& \{D_a, \{D_b, D_c \}_{\theta} \}_{\theta} + \{D_b, \{D_c, D_a
\}_{\theta} \}_{\theta} + \{D_c, \{D_a, D_b \}_{\theta} \}_{\theta} \xx
&=& \widehat{D}_a \widehat{F}_{bc} + \widehat{D}_b \widehat{F}_{ca}
+ \widehat{D}_c \widehat{F}_{ab}.
\eea
Similarly, the equations of motion derived from the action
in Eq. \eq{semi-matrix} read as
\be \la{poisson-eom}
\{D^a, \{D_a, D_b \}_{\theta} \}_{\theta} = \widehat{D}^a \widehat{F}_{ab} = 0.
\ee
Note that electromagnetism on a symplectic manifold is a
nonlinear interacting theory as the self-interaction in
Eq. \eq{symp-fs} clearly shows.

Going from the left-hand side of Eq. \eq{mirror} to the right-hand side,
we have eliminated the $U(1)$ gauge field in terms of the local coordinate
transformation in Eq. \eq{cov-cod}. Nevertheless, if one looks at the action in Eq. \eq{semi-matrix},
which was obtained by expanding the DBI action free from gauge fields,
gauge fields seem to appear again on first look. One may thus suspect that
the action in Eq. \eq{semi-matrix} does not satisfy the equivalence principle we have justified before.
However, one has to notice that the gauge fields in Eq. \eq{cov-cod} should be regarded as dynamical
coordinates describing a fluctuating metric as in Eq. \eq{gauge-metric}.
Rather an interesting point is that the fluctuation of the emergent metric, Eq. \eq{gauge-metric},
can be written in the form of gauge theory on a symplectic spacetime.
This highlights a key feature in realizing the gauge/gravity duality in noncommutative spacetime.

One can identify the defining fields $D_a(y) \in C^\infty(M), \; a =
1, \cdots, p+1 = 2n$, in the action in Eq. \eq{semi-matrix} with the set
in Eq. \eq{p-field} and, according to the map in Eq. \eq{v-field}, the fields $D_a(y)$
can be mapped to vector fields $V_a \in \Gamma(TM)$. One can
immediately see by identifying $f = D_a$ and $g = D_b$ and by using the
relation in Eq. \eq{symp-fs} that the Lie algebra homomorphism
in Eq. \eq{poisson-lie} leads to the following identity:
\be \la{map-fvv}
X_{\widehat{F}_{ab}} = [V_a, V_b],
\ee
where $V_a \equiv X_{D_a}$ and $V_b \equiv X_{D_b}$. Similarly,
using Eq. \eq{symp-cfs}, one can further deduce that
\be \la{map-vvv}
X_{\widehat{D}_a \widehat{F}_{bc}} = [V_a, [V_b, V_c]].
\ee
Then, the map in Eq. \eq{map-vvv} translates the Jacobi identity
in Eq. \eq{poisson-jacobi} and the equations of motion
in Eq. \eq{poisson-eom} into some relations between the vector fields $V_a$
defined by Eq. \eq{v-field}. That is, we have the following correspondence \ct{hsy-jhep}:
\bea \la{map-jacobi}
&& \widehat{D}_{[a} \widehat{F}_{bc]} = 0 \;\;
\Leftrightarrow \;\; [V_{[a}, [V_b, V_{c]}]] = 0, \\
\la{map-eom}
&& \widehat{D}^a \widehat{F}_{ab} = 0 \;\;
\Leftrightarrow \;\; [V^a, [V_a, V_b]] = 0,
\eea
where $[a,b,c]$ denotes a fully antisymmetrization of indices
$(a,b,c)$ and the bracket between the vector fields on the
right-hand side is defined by the Lie bracket.

As we remarked before, the vector fields $V_a \in \Gamma(TM)$ can be related to
the vielbeins $E_a \in \Gamma(TM)$ in Eq. \eq{cartan} by $V_a = \lambda E_a$,
with $\lambda \in C^\infty(M)$ to be determined, so the double Lie brackets
in Eqs. \eq{map-jacobi} and \eq{map-eom} will be related to Riemann curvature tensors because
they are involved with the second-order derivatives of the metric
in Eq. \eq{metric} or the vielbein in Eq. \eq{cartan}. It will be rather
straightforward to derive Einstein gravity from the set of equations,
Eqs. \eq{map-jacobi} and \eq{map-eom}, which is the subject of the following two subsections.

\subsection{Warm-up with a Beautiful Example}

First, let us briefly summarize some aspects in differential geometry \ct{big-book} that
are useful in understanding some concepts of emergent gravity.
Let $E^a = E^a_\mu dy^\mu$ be the basis of 1-forms dual to a given frame $E_a = E_a^\mu \p_\mu$.
If we define the local matrix of connection 1-forms by
\be \la{spin-conn}
{\omega^a}_b = {{\omega_\mu}^a}_b dy^\mu = {{\omega_c}^a}_b E^c,
\ee
the first Cartan's structure equation
\be \la{torsion}
T^a = dE^a + {\omega^a}_b \wedge E^b
\ee
describes the torsion as a 2-form in terms of the vielbein basis, and
the second structure equation
\be \la{curvature}
{R^a}_b = d{\omega^a}_b + {\omega^a}_c \wedge {\omega^c}_b
\ee
allows to compute the matrix-valued curvature 2-form by using the connection.

The metric compatibility leads to the symmetry $\omega_{ab} =
-\omega_{ba}$, which, together with the torsion-free condition
$T^a=0$, uniquely determines the connection 1-form, Eq. \eq{spin-conn},
which is nothing other than the Levi-Civita connection in the vielbein formalism
\begin{eqnarray} \label{spin-connection}
\omega_{abc} &=& \frac{1}{2} (f_{abc} - f_{bca} + f_{cab})
\nonumber \\
&=& - \omega_{acb},
\end{eqnarray}
where
\begin{equation} \label{f}
f_{abc} = E_a^\mu E_b^\nu (\partial_\mu E_{\nu c} - \partial_\nu
E_{\mu c})
\end{equation}
are the structure functions of the vectors $E_a
\in \Gamma(TM)$ defined by Eq. \eq{lie}.

Now, let us specialize to a Riemannian four-manifold $M$. Because the
spin connection $\omega_{\mu a b}$ and the curvature tensor
$R_{\mu\nu ab}$ are antisymmetric on the $ab$ index pair, one can
decompose them into a self-dual part and an anti-self-dual part as
follows \ct{hsy-jhep}:
\begin{eqnarray} \label{spin-sd-asd}
&& \omega_{\mu ab} = A_\mu^{(+)i} \eta^i_{ab} + A_\mu^{(-)i}
\bar{\eta}^i_{ab}, \\
\label{curvature-sd-asd}
&& R_{\mu\nu ab} = F_{\mu\nu}^{(+)i} \eta^i_{ab} + F_{\mu\nu}^{(-)i}
\bar{\eta}^i_{ab},
\end{eqnarray}
where the $4 \times 4$ matrices $\eta^i_{ab} \equiv
\eta^{(+)i}_{ab}$ and ${\bar \eta}^i_{ab} \equiv \eta^{(-)i}_{ab}$
for $i=1,2,3$ are 't Hooft symbols defined by
\begin{eqnarray} \label{tHooft-symbol}
&& {\bar \eta}^i_{jk} = {\eta}^i_{jk} = {\varepsilon}_{ijk}, \qquad
j,k \in
\{1,2,3\}, \nonumber\\
&& {\bar \eta}^i_{4j} = {\eta}^i_{j4} = \delta_{ij}.
\end{eqnarray}
Note that the 't Hooft matrices intertwine the group structure of
index $i$ with the spacetime structure of indices $a, b$.
See appendix A in Ref. \ct{hsy-jhep} for some useful identities for
the 't Hooft tensors.

Any $SO(4) = SU(2)_L \times SU(2)_R/\mathbf{Z}_2$ rotations can be
decomposed into self-dual and anti-self-dual rotations. Let us
introduce two families of $4
\times 4$ matrices defined by
\begin{equation} \label{thooft-matrix}
(T^i_+)_{ab} \equiv \eta^i_{ab}, \qquad (T^i_-)_{ab} \equiv {\bar
\eta}^i_{ab}.
\end{equation}
Then, one can show that $T^i_\pm$ satisfy $SU(2)$ Lie algebras, i.e.,
\begin{equation} \label{thooft-su2}
[T^i_\pm, T^j_\pm] = - 2 \varepsilon^{ijk} T^k_\pm,
\qquad [T^i_\pm, T^j_\mp] = 0.
\end{equation}
Indeed the 't Hooft tensors in Eq. (\ref{thooft-matrix}) are two
independent spin $s = \frac{3}{2}$ representations of the $SU(2)$ Lie
algebra. A deep geometric meaning of the 't Hooft tensors is to
specify the triple $(I,J,K)$ of complex structures of $\mathbb{R}^4
\cong \mathbb{C}^2$ for a given orientation as the simplest hyper-K\"ahler manifold.
The triple complex structures $(I,J,K)$ form a quaternion, that can
be identified with the $SU(2)$ generators $T^i_\pm$ in Eq. (\ref{thooft-matrix}).

Using the representation $\omega_{\mu ab}^{(\pm)} = A_\mu^{(\pm)i}
(T^i_\pm)_{ab} = (A_\mu^{(\pm)})_{ab}$ in Eq. \eq{spin-sd-asd}, we can write the
(anti-)self-dual curvature tensors in Eq. \eq{curvature-sd-asd} in the form
\begin{equation} \label{su2-curvature}
F_{\mu\nu}^{(\pm)} = \partial_\mu A_\nu^{(\pm)} - \partial_\nu
A_\mu^{(\pm)} + [A_\mu^{(\pm)}, A_\nu^{(\pm)}].
\end{equation}
Therefore, we see that $A_\mu^{(\pm)}$ can be identified with
$SU(2)_{L,R}$ gauge fields and $F_{\mu\nu}^{(\pm)}$ with their field
strengths. Indeed, one can also show that the local $SO(4)$ rotations
\begin{equation} \label{spin-so4}
\omega_\mu \to \Lambda \omega_\mu \Lambda^{-1} + \Lambda \partial_\mu \Lambda^{-1}
\end{equation}
with ${\Lambda^a}_b(y) \in SO(4)$ can be represented as the gauge transformations
of the $SU(2)$ gauge field $A_\mu^{(\pm)}$
\begin{equation} \label{spin-su2}
A^{(\pm)}_\mu \to \Lambda_\pm A^{(\pm)}_\mu
\Lambda^{-1}_\pm + \Lambda_\pm \partial_\mu \Lambda^{-1}_\pm
\end{equation}
where $\Lambda_\pm \in SU(2)_{L,R}$ \cite{emg-gauge}.

With the form language where $d = dy^\mu \partial_\mu = E^a E_a$
and $A = A_\mu dy^\mu = A_a E^a$, the field strength, Eq. (\ref{su2-curvature}),
of $SU(2)$ gauge fields in a coordinate basis is given by
\begin{eqnarray} \label{form-field-cod}
F^{(\pm)} &=& dA^{(\pm)} + A^{(\pm)} \wedge A^{(\pm)} \nonumber \\
&=& \frac{1}{2} F^{(\pm)}_{\mu\nu} dy^\mu \wedge dy^\nu \nonumber \\
&=&  \frac{1}{2} \Big( \partial_\mu A_\nu^{(\pm)} - \partial_\nu
A_\mu^{(\pm)} + [A_\mu^{(\pm)}, A_\nu^{(\pm)}] \Big) dy^\mu \wedge
dy^\nu
\end{eqnarray}
or in a non-coordinate basis by
\begin{eqnarray} \label{form-field-noncod}
F^{(\pm)} &=& \frac{1}{2} F^{(\pm)}_{ab} E^a \wedge E^b \nonumber \\
&=&  \frac{1}{2} \Big( E_a A_b^{(\pm)} - E_b A_a^{(\pm)} +
[A_a^{(\pm)}, A_b^{(\pm)}] + {f_{ab}}^c A^{(\pm)}_c \Big)E^a \wedge
E^b
\end{eqnarray}
where we used in Eq. (\ref{form-field-noncod}) the structure equation
\begin{equation} \label{structure-eq}
dE^a = \frac{1}{2} {f_{bc}}^a  E^b \wedge E^c.
\end{equation}

As we remarked, the 't Hooft tensors $\{\eta^{(\pm) i}_{ab}\}$
specify the triple of complex structures of the simplest
hyper-K\"ahler manifold $\mathbb{R}^4$ satisfying the quaternion algebra.
Because any Riemannian metric can be written as $g_{\mu\nu} (y) =
E_\mu^a (y) E_\nu^b (y) \delta_{ab}$, one can introduce the
corresponding triple of local complex structures on the Riemannian manifold $M$
\begin{equation} \label{comp-str-e}
J^{(\pm) i} = \frac{1}{2} \eta^{(\pm) i}_{ab} E^a \wedge E^b,
\end{equation}
which are inherited from $\mathbb{R}^4$, or in terms of local coordinates
\begin{equation} \label{comp-str-x}
J^{(\pm) i} = \frac{1}{2} \eta^{(\pm) i}_{ab} E_\mu^a E_\nu^b \;
dy^\mu
\wedge dy^\nu \equiv \frac{1}{2} J^{(\pm) i}_{\mu\nu} dy^\mu
\wedge dy^\nu.
\end{equation}
One can easily check that the local complex structures $J^{(\pm) i}$
still satisfy the quaternion algebra
\begin{eqnarray} \label{local-quaternion}
[ J^{(\pm) i}, J^{(\pm) j} ]_{\mu\nu}  &\equiv&  J^{(\pm) i}_{\mu
\lambda}
{J^{(\pm) j \lambda}}_\nu - J^{(\pm) j}_{\mu\lambda} {J^{(\pm) i \lambda}}_\nu \nonumber \\
&=& - 2 \varepsilon^{ijk}  J^{(\pm) k}_{\mu\nu}, \nonumber \\
{[J^{(\pm) i}, J^{(\mp) j}]}_{\mu\nu}  &=&  0.
\end{eqnarray}
Now, it is easy to see that the torsion-free condition $T^a =0$ is
equivalent to the one in which the complex structures $J^{(\pm) i}$ are
covariantly constant, i.e.,
\begin{eqnarray}
dJ^{(\pm) i} &=& \frac{1}{2} \eta^{(\pm) i}_{ab} d E^a \wedge E^b -
\frac{1}{2} \eta^{(\pm) i}_{ab} E^a \wedge d E^b \nonumber \\
&=& - [\eta^{(\pm) i} \eta^{(\mp) j}]_{ab} A^{(\mp) j} \wedge E^a
\wedge E^b + 2 \varepsilon^{ijk} A^{(\pm) j} \wedge J^{(\pm) k} \nonumber \\
&=& 2 \varepsilon^{ijk} A^{(\pm) j} \wedge J^{(\pm) k},
\end{eqnarray}
where we used the fact that $[\eta^{(\pm) i} \eta^{(\mp) j}]_{ab}$
is symmetric, i.e., $[\eta^{(\pm) i} \eta^{(\mp) j}]_{ab} =
[\eta^{(\pm) i}
\eta^{(\mp) j}]_{ba}$. Therefore, we get the relation
\begin{equation} \label{cov-comp-str}
d_A J^{(\pm) i} \equiv d J^{(\pm) i} - 2 \varepsilon^{ijk} A^{(\pm) j}
\wedge J^{(\pm) k} = 0.
\end{equation}

All these properties can be beautifully summarized using the
Palatini formalism of Einstein gravity, in which the spin connection and
the vielbein are regarded as independent variables. The
Einstein-Hilbert action in the Palatini formalism is given by
\begin{equation} \label{palatini}
S_{P} = \frac{1}{4} \int_M {\varepsilon_{ab}}^{cd} E^a \wedge E^b
\wedge R_{cd}.
\end{equation}
By varying the action in Eq. \eq{palatini} with respect to the vierbein and
the spin connection, one can get the torsion-free condition $T^a = dE^a
+ {\omega^a}_b \wedge E^b = 0$, as well as the Einstein equation
$R_{ab} - \frac{1}{2} \delta_{ab} R = 0$, thus recovering the Einstein gravity.

If the decomposition in Eq. \eq{curvature-sd-asd} is used, the Palatini action
in Eq. (\ref{palatini}) can be recast into the beautiful form
\begin{eqnarray} \label{palatini-su2}
S_{P} &=& \frac{1}{4} \int_M {\varepsilon_{ab}}^{cd} E^a \wedge E^b
\wedge R_{cd} \nonumber \\
&=& \frac{1}{4} \int_M {\varepsilon_{ab}}^{cd} E^a \wedge E^b
\wedge \big(F^{(+)i} \eta^i_{cd} + F^{(-)i} \bar{\eta}^i_{cd} \big)
\nonumber \\
&=& \frac{1}{2} \int_M E^a \wedge E^b
\wedge \big(F^{(+)i} \eta^i_{ab} - F^{(-)i} \bar{\eta}^i_{ab} \big)
\nonumber \\
&=& \int_M   \big( J^{(+)i} \wedge F^{(+)i} -  J^{(-)i} \wedge
F^{(-)i}
\big).
\end{eqnarray}
The action in Eq. (\ref{palatini-su2}) immediately shows that the condition
in Eq. (\ref{cov-comp-str}) is, indeed, the equations of motion for $SU(2)$
gauge fields $A^{(\pm)i}$. Interestingly, the Palatini action in Eq. (\ref{palatini-su2})
is invariant under a local deformation given by
\begin{equation} \label{palatini-def}
A^{(\pm) i} \to A^{(\pm) i}, \qquad J^{(\pm)i} \to J^{(\pm)i} + d_A
\Lambda^{(\pm) i},
\end{equation}
with an arbitrary one-form $\Lambda^{(\pm)} \in SU(2)$. The deformation
symmetry in Eq. (\ref{palatini-def}) should be true due to the Bianchi
identity $d_A F^{(\pm)} = 0$.

The gravitational instantons \ct{grav-instanton} are defined by the
self-dual solution to the Einstein equation
\begin{equation} \label{g-instanton}
R_{efab} = \pm \frac{1}{2} {\varepsilon_{ab}}^{cd} R_{efcd}.
\end{equation}
Note that a metric satisfying the self-duality equation, Eq. (\ref{g-instanton}), is necessarily
Ricci-flat because $R_{ab} = {R_{acb}}^c = \pm \frac{1}{6} {\varepsilon_{b}}^{cde} R_{a[cde]} =
0$ and thus automatically satisfies the vacuum Einstein equations.
If the decomposition in Eq. (\ref{curvature-sd-asd}) is used, Eq. \eq{g-instanton} can be written as
\bea \la{half-flat}
F_{ef}^{(+)i} \eta^i_{ab} + F_{ef}^{(-)i}
\bar{\eta}^i_{ab} &=& \pm \frac{1}{2} {\varepsilon_{ab}}^{cd} (F_{ef}^{(+)i} \eta^i_{cd} + F_{ef}^{(-)i}
\bar{\eta}^i_{cd}) \xx
&=&\pm (F_{ef}^{(+)i} \eta^i_{ab} - F_{ef}^{(-)i}
\bar{\eta}^i_{ab}).
\eea
Therefore, we should have $F_{ab}^{(-)i} = 0$ for the self-dual case
with a plus sign in Eq. \eq{g-instanton} and $F_{ab}^{(+)i} = 0$ for
the anti-self-dual case with a minus sign, thus imposing the
self-duality equation, Eq. \eq{g-instanton}, is equivalent to the
half-flat equation $F^{(\pm)i} = 0$. Because the Riemann curvature
tensors satisfy the symmetry property
\be \la{riem-symm}
R_{abcd} = R_{cdab},
\ee
one can rewrite the self-duality equation, Eq. \eq{g-instanton}, as follows:
\begin{equation} \label{gr-instanton}
R_{abef} = \pm \frac{1}{2} {\varepsilon_{ab}}^{cd} R_{cdef}.
\end{equation}
Then, using the decomposition in Eq. (\ref{curvature-sd-asd}) again, one
can similarly show \ct{our-note} that the gravitational instanton
in Eq. (\ref{gr-instanton}) can be understood as an $SU(2)$ Yang-Mills
instanton, i.e.,
\begin{equation} \label{ym-instanton}
F^{(\pm)}_{ab} = \pm \frac{1}{2} {\varepsilon_{ab}}^{cd} F^{(\pm)}_{cd},
\end{equation}
where $F^{(\pm)}_{ab} = F^{(\pm) i}_{ab} T^i_\pm = E_a^\mu E_b^\nu
F_{\mu\nu}^{(\pm)}$ are defined by Eq. (\ref{form-field-noncod}).

A solution to the half-flat equation $F^{(\pm)} = 0$ is given by $A^{(\pm)} = \Lambda_\pm d \Lambda^{-1}_\pm$;
then, Eq. \eq{spin-su2} shows that it is always possible to choose a
self-dual gauge $A^{(\pm)i} = 0$. In this gauge, Eq. (\ref{cov-comp-str}) reduces
to the property $d J^{(\pm)i} = 0$; that is, the triple complex structures in one of
the $(\pm)$-sectors are all closed. In other words, there is the triple $\{ J^{(\pm)i}\}$ of globally
well-defined complex structures. This means that the metric $ds^2 = E^a \otimes E^a$ describes
a hyper-K\"ahler manifold. In the end, the gravitational instantons defined by Eq. (\ref{g-instanton})
can be characterized by the following property:
\begin{equation} \label{gi-hyper-k}
F^{(\pm)i} = 0 \quad \Leftrightarrow \quad d J^{(\pm)i} = 0, \qquad
\forall \, i = 1,2,3.
\end{equation}

In order to solve the equations in Eq. (\ref{gi-hyper-k}), let us introduce
linearly-independent four-vector fields $V_a$ and a volume form
$\nu$ on $M$. Then, one can easily check \ct{osaka} that the
(anti-)self-dual ansatz \cite{vol-form}
\begin{equation} \label{sd-ansatz}
J^{(\pm)i} = \frac{1}{2} \eta^{(\pm)i}_{ab} \iota_a \iota_b \nu,
\end{equation}
where $\iota_a$ denotes the inner derivation with respect to $V_a$,
immediately solves the equations $d J^{(\pm)i} = 0$ if and only if
the vector fields satisfy the following equations \ct{g-instanton}
\begin{eqnarray} \label{sde-u1}
&& \frac{1}{2} \eta^{(\pm)i}_{ab} [V_a,V_b] = 0, \\
\label{sde-vol}
&& {\cal L}_{V_a} \nu = 0, \qquad \forall \, a = 1, \cdots, 4.
\end{eqnarray}
This can simply be seen by applying the formula \cite{cartan-homo}
\begin{equation} \label{math-formula}
d(\iota_X \iota_Y \alpha) = \iota_{[X,Y]} \alpha + \iota_Y {\cal
L}_X \alpha - \iota_X {\cal L}_Y \alpha + \iota_X \iota_Y d\alpha
\end{equation}
for vector fields $X,Y$ and a $p$-form $\alpha$.

Now go back to the action in Eq. \eq{semi-matrix} and consider the
self-duality equation of $U(1)$ gauge fields defined by
\be \la{self-dual-gauge}
\widehat{F}_{ab} = \pm \frac{1}{2} {\varepsilon_{ab}}^{cd}
\widehat{F}_{cd}.
\ee
Note that the self-duality equation, Eq. \eq{self-dual-gauge}, is
nonlinear due to the Poisson commutator term in Eq. \eq{symp-fs}, so
there exist nontrivial solutions \ct{hsy-inst}. After quantization, Eq. \eq{q-bracket},
they become noncommutative $U(1)$ instantons \ct{nc-instanton}.
One can translate the self-duality equation, Eq. \eq{self-dual-gauge}, to the
self-duality equation between vector fields according to the map in Eq. \eq{map-fvv}:
\be \la{iso-map}
\widehat{F}_{ab} = \pm \frac{1}{2} {\varepsilon_{ab}}^{cd}
\widehat{F}_{cd} \quad \Leftrightarrow \quad
[V_a, V_b] = \pm \frac{1}{2} {\varepsilon_{ab}}^{cd} [V_c, V_d].
\ee
Recall that the vector fields $V_a$ are all divergence-free, i.e.,
$\p_\mu V_a^\mu = 0$; in other words, ${\cal L}_{V_a} \nu = 0$.
Therefore, we see that the self-duality equation, Eq. \eq{self-dual-gauge},
for gauge fields is certainly equivalent to Eqs. (\ref{sde-u1}) and (\ref{sde-vol}).
In conclusion, we finally proved \ct{hsy-epl,hsy-ijmpa,hsy-epjc,hsy-jhep} the equivalence
between $U(1)$ instantons defined by Eq. \eq{self-dual-gauge} and gravitational instantons
defined by Eq. \eq{g-instanton}.

\subsection{Einstein Gravity from Electromagnetism on a Symplectic Manifold}

As a warm-up, we have illustrated with self-dual gauge fields how the
Darboux theorem in symplectic geometry implements a deep principle
to realize a Riemannian manifold as an emergent geometry from gauge
fields on a symplectic manifold through the correspondence in Eq. \eq{v-field}
whose metric is given by
\be \la{emergent-metric}
ds^2 = \delta_{ab} E^a \otimes E^b = \lambda^2 \delta_{ab} V^a_\mu
V^b_\nu dy^\mu \otimes dy^\nu,
\ee
where $E^a = \lambda V^a \in \Gamma(T^*M)$ are dual one-forms. Now,
we will generalize the emergent gravity to arbitrary gauge fields on
a $2n$-dimensional symplectic manifold $(M, B)$ and derive Einstein
equations from Eqs. \eq{map-jacobi} and \eq{map-eom}.

First let us determine what $\lambda \in C^\infty(M)$ in Eq. \eq{emergent-metric} is.
Introduce the structure equation of the vector fields $V_a = \lambda E_a \in \Gamma(TM)$
\be \la{lie-v}
[V_a, V_b] = - {\mathbf{f}_{ab}}^c V_c.
\ee
By comparing with Eq. \eq{lie}, one can get the relation between the two
structure functions
\be \la{2-structure}
{\mathbf{f}_{ab}}^c = \lambda {f_{ab}}^c - V_a \log \lambda
\delta^c_b + V_b \log \lambda \delta^c_a.
\ee
With the definition in Eq. \eq{lie-v}, the self-duality equation,
Eq. \eq{iso-map}, may be written in a compact form
\be \la{self-sc}
\eta^{(\pm)i}_{ab} {\mathbf{f}_{ab}}^c = 0.
\ee

Suppose that the $2n$-dimensional volume form whose four-dimensional
example was introduced in Eq. \eq{sd-ansatz} is given by
\be \la{vol-form}
\nu = \lambda^2 V^1 \wedge \cdots \wedge V^{2n};
\ee
in other words,
\be \la{lambda}
\lambda^2 = \nu(V_1, \cdots, V_{2n}).
\ee
The volume form in Eq. \eq{vol-form} can be related to the Riemannian one
$\nu_g = E^1 \wedge \cdots \wedge E^{2n}$ as
\be \la{rel-vol}
\nu = \lambda^{2 - 2n} \nu_g.
\ee
Acting ${\cal L}_{V_a}$ on both sides of Eq. \eq{lambda}, we get
\bea \la{div-vol}
{\cal L}_{V_a} \Bigl( \nu (V_1, \cdots, V_{2n}) \Bigr) &=& ({\cal
L}_{V_a} \nu) (V_1, \cdots, V_{2n}) + \sum_{b=1}^{2n}
\nu (V_1, \cdots, {\cal L}_{V_a} V_b, \cdots,  V_{2n}) \xx
&=& ({\cal L}_{V_a} \nu)(V_1, \cdots, V_{2n}) +
\sum_{b=1}^{2n} \nu (V_1, \cdots, [V_a, V_b], \cdots,  V_{2n}) \xx
&=& (\nabla \cdot V_a + 2(1-n) V_a \log \lambda + {\mathbf{f}_{ba}}^b ) \nu (V_1,
\cdots,  V_{2n}) \xx
&=& (2 V_a \log \lambda) \nu (V_1, \cdots, V_{2n}).
\eea
Because ${\cal L}_{V_a} \nu  = (\nabla \cdot V_a + 2(1-n) V_a \log \lambda) \nu = 0$,
Eq. \eq{div-vol} leads to the relation
\be \la{sc-lambda}
\rho_a \equiv {\mathbf{f}_{ba}}^b = 2 V_a \log \lambda;
\ee
then, from Eq. \eq{2-structure},
\be \la{lie-lambda}
{f_{ba}}^b = (3-2n) E_a \log \lambda.
\ee
Conversely, if ${\mathbf{f}_{ba}}^b = 2 V_a \log \lambda$, the vector fields $V_a$ preserve
the volume form $\nu$, i.e., ${\cal L}_{V_a} \nu = (\nabla \cdot V_a + 2(1-n) V_a \log \lambda) \nu = 0 \;
\forall a = 1, \cdots, 2n$. Equation \eq{lie-lambda} implies that the vector fields $E_a$
preserve the volume form $\widetilde{\nu} = \lambda^{3-2n} \nu_g$,
which can be proven as follows:
\be \la{vol-pres}
\mathcal{L}_{E_a} (\lambda^{3-2n} \nu_g) =
d \bigl(\iota_{E_a} (\lambda^{3-2n} \nu_g) \bigr) =
d\bigl(\iota_{\lambda E_a} (\lambda^{2-2n} \nu_g) \bigr) =
d\bigl(\iota_{V_a} \nu \bigr) = \mathcal{L}_{V_a} \nu = 0.
\ee

In a non-coordinate (anholonomic) basis $\{E_a\}$ satisfying the
commutation relation in Eq. \eq{lie}, the spin connections are defined by
\be \la{D-spin1}
\nabla_a E_c = {{\omega_a}^b}_c  E_b,
\ee
where $\nabla_a \equiv \nabla_{E_a}$ is the covariant derivative in
the direction of a vector field $E_a$. Acting on the dual basis
$\{E^a\}$, they are given by
\be \la{D-spin2}
\nabla_a E^b = - {{\omega_a}^b}_c E^c.
\ee
Because we will impose the torsion-free condition, i.e.,
\be \la{torsion-free}
T(a,b) = \nabla_{[a} E_{b]} - [E_a,E_b]= 0,
\ee
the spin connections are related to the structure functions
\be \la{spin-structure}
f_{abc} = -\omega_{acb} + \omega_{bca}.
\ee
The Riemann curvature tensors in the basis $\{E_a\}$ are defined by
\be \la{riemann-def}
R(a,b) = [\nabla_a, \nabla_b] - \nabla_{[a,b]}
\ee
or in component form, by
\bea \la{D-riemann}
{{R_{ab}}^c}_d &=& \langle E^c, R(E_a, E_b) E_d \rangle \xx &=& E_a
{{\omega_b}^c}_d - E_b {{\omega_a}^c}_d + {{\omega_a}^c}_e
{{\omega_b}^e}_d - {{\omega_b}^c}_e {{\omega_a}^e}_d  + {f_{ab}}^e
{{\omega_e}^c}_d.
\eea

Imposing the condition that the metric in Eq. \eq{emergent-metric} is
covariantly constant, i.e.,
\be \la{metric-condition}
\nabla_c \Bigl( \delta_{ab} E^a \otimes E^b \Bigr) = 0,
\ee
or, equivalently,
\be \la{spin-condition}
\omega_{cab}= - \omega_{cba},
\ee
we see that the spin connections $\omega_{cab}$ have the same number of
components as $f_{abc}$. Thus, Eq. \eq{spin-structure} has a unique
solution, and it is precisely given by Eq. \eq{spin-connection}. The
definition in Eq. \eq{riemann-def}, together with the metricity condition
in Eq. \eq{spin-condition}, immediately leads to the following symmetry property:
\be \la{curvature-anti-symm}
R_{abcd} = - R_{abdc} = -R_{bacd}.
\ee
If the relation in Eq. \eq{2-structure} is used, the spin connections in
Eq. \eq{spin-connection} are now determined by the gauge theory bases
\be \la{spin-conformal}
\lambda \omega_{abc} = \half(\mathbf{f}_{abc} - \mathbf{f}_{bca}
+ \mathbf{f}_{cab}) - V_b \log \lambda \delta_{ca} + V_c \log
\lambda \delta_{ab}.
\ee

The spacetime geometry described by the metric in Eq. \eq{emergent-metric}
is an emergent gravity arising from gauge fields whose underlying
theory is defined by the action in Eq. \eq{semi-matrix}. The fundamental
variables in our approach are, of course, gauge fields that are
subject to Eqs. \eq{map-jacobi} and \eq{map-eom}. A
spacetime metric is now regarded as a collective variable defined by
a composite or bilinear of gauge fields. Therefore, we are going to get a
viable realization of the idea we speculated about with Eq. \eq{cartan};
is it possible to show that the equations of motion, Eq. \eq{map-eom}, for gauge fields
together with the Bianchi identity in Eq. \eq{map-jacobi} can be rewritten as the Einstein equations
for the metric in Eq. \eq{emergent-metric}? For this purpose, we first want to represent
the Riemann curvature tensors in Eq. \eq{D-riemann}, originally
expressed with the orthonormal basis $E_a$, in terms of the gauge
theory basis $V_a$. That representation will be useful because we
will eventually impose Eqs. \eq{map-jacobi} and \eq{map-eom} on them.

Indeed, everything is prepared because all calculations can straightforwardly be done using
Eqs. \eq{2-structure} and \eq{spin-conformal}. All the details can be found in Ref. \ct{hsy-jhep}.
Here, we will briefly sketch essential steps.
One can easily derive the following identity:
\bea \la{1-Bianchi}
&& R(E_a, E_b) E_c + R(E_b, E_c) E_a + R(E_c, E_a) E_b \xx &=&
 [E_a, [E_b, E_c]] + [E_b, [E_c, E_a]] + [E_c, [E_a, E_b]]
\eea
by using the torsion-free condition in Eq. \eq{torsion-free}. The Jacobi
identity then immediately leads to $R_{[abc]d} = 0$. Because $V_a =
\lambda E_a$, we have the relation
\be \la{2-jacobi}
[V_{[a}, [V_b, V_{c]}]] = \lambda ^3 [E_{[a}, [E_b, E_{c]}]],
\ee
where all the terms containing the derivatives of $\lambda$ cancel each other.
As we promised, the first Bianchi identity $R_{[abc]d} = 0$ follows from the Jacobi identity
in Eq. \eq{map-jacobi}. Thus, we pleasingly confirm that
\be \la{bianchi-true}
\widehat{D}_{[a} \widehat{F}_{bc]} = 0 \quad \Leftrightarrow \quad
R_{[abc]d} = 0.
\ee
The Bianchi identity in Eq. \eq{bianchi-true}, together with
Eq. \eq{curvature-anti-symm}, leads to the symmetry in Eq. \eq{riem-symm}. It
should be emphasized that the equivalence in Eq. \eq{bianchi-true} holds for
arbitrary gauge fields in any even dimensions.

From the above derivation, we have to notice that, if torsion-free
condition in Eq. \eq{torsion-free} are not imposed, the equivalence
in Eq. \eq{bianchi-true} must be corrected. This can be seen from the Bianchi
identities \ct{big-book}
\bea \la{bianchi-1}
&& DT^a \equiv dT^a + {\omega^a}_b \wedge T^b = {R^a}_b \wedge E^b,
\\
&& D{R^a}_b \equiv d{R^a}_b + {\omega^a}_c \wedge {R^c}_b - {R^a}_c
\wedge {\omega^c}_b = 0,
\eea
which are integrability conditions derived from Eqs. \eq{torsion} and
\eq{curvature}. In general, the equivalence in Eq. \eq{bianchi-true} holds only if $DT^d = 0$,
where ${R_c}^d \wedge E^c = \half {R_{abc}}^d E^a
\wedge E^b \wedge E^c = \frac{1}{6} {R_{[abc]}}^d E^a
\wedge E^b \wedge E^c = 0$.
Note that Eq. \eq{map-jacobi} is simply a consequence of the Jocobi
identity for the Poisson bracket in Eq. \eq{p-bracket}, which should be
true for a general Poisson structure (not necessarily nondegenerate)
as long as the Schouten-Nijenhuis bracket $[\theta, \theta]_{SN} \in
\Gamma(\Lambda^3 TM)$ vanishes. Thus, Eq. \eq{map-jacobi}
will still be true for a generic Poisson manifold. Also, one may get
the relation ${R_{[abc]}}^d - D_{[a} T^d_{bc]} = 0$, instead of
Eq. \eq{2-jacobi}, for a nonzero torsion. Nevertheless, we conjecture that the
torsion will identically vanish even for a general Poisson manifold
because the existence of a Poisson structure implies the equivalence
principle in the theory of emergent gravity.

The mission for the equations of motion, Eq. \eq{map-eom}, is more
nontrivial. After some technical manipulation, a remarkably
simple form for the Ricci tensor can be obtained in four dimensions
\ct{hsy-jhep}:
\bea \la{ricci-final}
R_{ab} &=& - \frac{1}{\lambda^2}  \Big[ \mathbf{f}_d^{(+)i}
\eta^i_{ac} \mathbf{f}_d^{(-)j} \bar{\eta}^j_{bc} + \mathbf{f}_d^{(+)i}
\eta^i_{bc} \mathbf{f}_d^{(-)j} \bar{\eta}^j_{ac} \xx
&& - \big(\mathbf{f}_c^{(+)i}
\eta^i_{ac} \mathbf{f}_d^{(-)j} \bar{\eta}^j_{bd} + \mathbf{f}_c^{(+)i}
\eta^i_{bc} \mathbf{f}_d^{(-)j} \bar{\eta}^j_{ad} \big) \Big].
\eea
Here, we also decomposed $\mathbf{f}_{abc}$ into
self-dual and anti-self-dual parts as in Eq. \eq{spin-sd-asd}:
\be \la{f-sd-asd}
\mathbf{f}_{abc} = \mathbf{f}_c^{(+)i} \eta^i_{ab}
+ \mathbf{f}_c^{(-)i} \bar{\eta}^i_{ab},
\ee
where
\be \la{f-decomposition}
\mathbf{f}_c^{(\pm)i} \eta^{(\pm)i}_{ab} = \half \Big(
\mathbf{f}_{abc} \pm \half {\varepsilon_{ab}}^{de}
\mathbf{f}_{dec} \Big).
\ee
Recall that we want to relate the equations of motion, Eq. \eq{map-eom}, together with the Bianchi
identity in Eq. \eq{map-jacobi}, to the Einstein equations for the emergent metric \eq{emergent-metric}.
For later use, let us also introduce a completely antisymmetric tensor defined by
\be \la{3-form}
\Psi_{abc} \equiv \mathbf{f}_{abc} +
\mathbf{f}_{bca} + \mathbf{f}_{cab} \equiv \varepsilon_{abcd} \Psi_d.
\ee
Using the decomposition in Eq. \eq{f-sd-asd}, one can easily see that
\be \la{psi-decomp}
\Psi_a = - \frac{1}{3 !} \varepsilon_{abcd} \Psi_{bcd} =
- \Big( \mathbf{f}_b^{(+)i} \eta^i_{ab} - \mathbf{f}_b^{(-)i}
\bar{\eta}^i_{ab} \Big)
\ee
while Eq. \eq{sc-lambda} leads to
\be \la{rho-decomp}
\rho_a = \mathbf{f}_{bab} = - \Big( \mathbf{f}_b^{(+)i} \eta^i_{ab}
+ \mathbf{f}_b^{(-)i} \bar{\eta}^i_{ab} \Big).
\ee
Note that the right-hand side of Eq. \eq{ricci-final} is purely
interaction terms between the self-dual and the anti-self-dual parts in
Eq. \eq{f-sd-asd}. Therefore, if gauge fields satisfy the
self-duality equation, Eq. \eq{self-sc}, i.e., $\mathbf{f}_a^{(\pm)i} =0$,
they describe a Ricci-flat manifold, i.e., $R_{ab}=0$.
Of course, this result is completely consistent with the previous self-dual case.

The next step is to calculate the Einstein tensor to identify the
form of the energy-momentum tensor
\bea \la{einstein-tensor}
E_{ab} &=& R_{ab} - \half \delta_{ab} R \xx &=& -
\frac{1}{\lambda^2}
\Big( \mathbf{f}_d^{(+)i}
\eta^i_{ac} \mathbf{f}_d^{(-)j} \bar{\eta}^j_{bc} + \mathbf{f}_d^{(+)i}
\eta^i_{bc} \mathbf{f}_d^{(-)j} \bar{\eta}^j_{ac} \Big) \xx
&& + \frac{1}{\lambda^2} \Big( \mathbf{f}_c^{(+)i}
\eta^i_{ac} \mathbf{f}_d^{(-)j} \bar{\eta}^j_{bd} + \mathbf{f}_c^{(+)i}
\eta^i_{bc} \mathbf{f}_d^{(-)j} \bar{\eta}^j_{ad} - \delta_{ab} \mathbf{f}_d^{(+)i}
\eta^i_{cd} \mathbf{f}_e^{(-)j} \bar{\eta}^j_{ce} \Big)
\eea
where the Ricci scalar $R$ is given by
\be \la{final-ricci-scalar}
R = \frac{2}{\lambda^2} \mathbf{f}_b^{(+)i}
\eta^i_{ab} \mathbf{f}_c^{(-)j} \bar{\eta}^j_{ac}.
\ee
We have adopted the conventional view that the gravitational field
is represented by the spacetime metric itself. The problem, thus,
reduces to finding field equations to relate the metric
in Eq. \eq{emergent-metric} to the energy-momentum distribution. According to our
scheme, Eq. \eq{einstein-tensor}, therefore, corresponds to such field
equations, i.e., the Einstein equations. First, notice that
the right-hand side of Eq. \eq{einstein-tensor} identically vanishes for self-dual
gauge fields, satisfying $\mathbf{f}_a^{(\pm)i} = 0$, whose energy-momentum tensor
also identically vanishes because their action is topological, i.e., metric independent.
Howecer, for general gauge fields for which
$\mathbf{f}_a^{(\pm)i} \neq 0$, the right-hand side of Eq. \eq{einstein-tensor} no longer
vanishes, which, in turn, enforces $E_{ab} =  R_{ab} - \half \delta_{ab} R
\neq 0$. Because the Einstein tensor $E_{ab}$ equals some energy-momentum tensor for matter fields,
the non-vanishing Einstein tensor implies that there is a nontrivial energy-momentum tensor
coming from $U(1)$ gauge fields. In other words, the presence of $U(1)$ gauge fields
on a symplectic spacetime not only deforms spacetime geometry according to the correspondence
in Eq. \eq{v-field} but also plays a role of matter fields contributing to the energy-momentum tensor.
Indeed, this should be the case because the action in Eq. \eq{semi-matrix} reduces
to the ordinary Maxwell theory in the commutative limit and thus has a nontrivial energy-momentum tensor.
Therefore, it is natural to identify the right-hand side of Eq. \eq{einstein-tensor}
with some energy-momentum tensor determined by $U(1)$ gauge fields.

We intentionally make the following separation into two kinds of
energy-momentum tensors denoted by $T_{ab}^{(M)}$ and $T_{ab}^{(L)}$:
\bea \la{em-tensor-maxwell}
\frac{8 \pi G}{c^4} T_{ab}^{(M)} &=& - \frac{1}{\lambda^2}
\Big( \mathbf{f}_d^{(+)i}
\eta^i_{ac} \mathbf{f}_d^{(-)j} \bar{\eta}^j_{bc} + \mathbf{f}_d^{(+)i}
\eta^i_{bc} \mathbf{f}_d^{(-)j} \bar{\eta}^j_{ac} \Big) \xx
&=& - \frac{1}{\lambda^2}
\mathbf{f}_d^{(+)i}  \mathbf{f}_d^{(-)j} \Big(
\eta^i_{ac} \bar{\eta}^j_{bc} + \eta^i_{bc} \bar{\eta}^j_{ac} \Big) \xx
&=& - \frac{1}{\lambda^2}
\Big(\mathbf{f}_{acd} \mathbf{f}_{bcd} - \frac{1}{4}
\delta_{ab} \mathbf{f}_{cde} \mathbf{f}_{cde} \Big), \\
\la{dark-energy}
\frac{8 \pi G}{c^4} T_{ab}^{(L)} &=& \frac{1}{\lambda^2}
\Big(\mathbf{f}_c^{(+)i}
\eta^i_{ac} \mathbf{f}_d^{(-)j} \bar{\eta}^j_{bd} + \mathbf{f}_c^{(+)i}
\eta^i_{bc} \mathbf{f}_d^{(-)j} \bar{\eta}^j_{ad} - \delta_{ab} \mathbf{f}_d^{(+)i}
\eta^i_{cd} \mathbf{f}_e^{(-)j} \bar{\eta}^j_{ce} \Big) \xx
&=& \frac{1}{\lambda^2} \mathbf{f}_c^{(+)i} \mathbf{f}_d^{(-)j}
\Big(\eta^i_{ac} \bar{\eta}^j_{bd} + \eta^i_{bc} \bar{\eta}^j_{ad}
- \delta_{ab} \eta^i_{ec} \bar{\eta}^j_{ed} \Big) \xx &=&
\frac{1}{2\lambda^2} \Big( \rho_a \rho_b - \Psi_a
\Psi_b -
\frac{1}{2} \delta_{ab}(\rho_c^2 - \Psi_c^2) \Big)
\eea
where we have used the decomposition in Eq. \eq{f-decomposition} and the relations
$$ \mathbf{f}_b^{(+)i} \eta^i_{ab} = - \half(\rho_a + \Psi_a), \qquad
\mathbf{f}_b^{(-)i} \bar{\eta}^i_{ab} = - \half(\rho_a - \Psi_a).$$
With this notation, the Einstein equations, Eq. \eq{einstein-tensor}, can be written as
\bea \la{emergent-eq}
E_{ab} &=& R_{ab} - \half \delta_{ab} R \xx &=&
\frac{8 \pi G}{c^4} \Big( T_{ab}^{(M)} + T_{ab}^{(L)} \Big).
\eea
The main motivation of the above separation was the fact that the energy-momentum tensor $T_{ab}^{(M)}$
is traceless, i.e., $T_{aa}^{(M)} = 0$, because of the property
$\eta^{(\pm)i}_{ab} \eta^{(\mp)j}_{ab} = 0$, so the Ricci scalar
in Eq. \eq{final-ricci-scalar} is determined by the second
energy-momentum tensor $T_{ab}^{(L)}$ only.

First, let us identify the real character of the energy-momentum
tensor in Eq. \eq{em-tensor-maxwell}. When one stares at the
energy-momentum tensor in Eq. \eq{em-tensor-maxwell}, one may find that
it is very reminiscent of the Maxwell energy-momentum tensor given
by
\be \la{em-maxwell}
T_{ab}^{(em)} = \frac{\hbar^2 c^2}{g^2_{YM}}\Big( F_{ac} F_{bc} -
\frac{1}{4} \delta_{ab} F_{cd} F_{cd} \Big),
\ee
which is also traceless, i.e., $T_{aa}^{(em)} = 0$. Indeed, it was
argued in Ref. \ct{hsy-jhep} that the energy-momentum tensor
in Eq. \eq{em-tensor-maxwell} can be mapped to Eq. \eq{em-maxwell} by reversely
applying the map in Eq. \eq{v-field}, so to speak, by translating the map
$\Gamma(TM) \to C^\infty(M)$ \cite{sign-prob}.

There is another reason the energy-momentum tensor
in Eq. \eq{em-tensor-maxwell} should be mapped to Eq. \eq{em-maxwell}.
Consider a commutative limit in which $|\theta|^2 \equiv
\widehat{g}_{ac} \widehat{g}_{bd} \theta^{ab} \theta^{cd} =
\kappa^2 |\kappa B g^{-1}|^2 \to 0$. In this limit, we should recover
the ordinary Maxwell theory from the action in Eq. \eq{semi-matrix}, which
may be more obvious from the left-hand side of Eq. \eq{mirror}. Because
the action in Eq. \eq{semi-matrix} is defined in the commutative limit,
which reduces to the Maxwell theory at $|\theta| = 0$, the Maxwell
theory should play a role in the Einstein equation in Eq. \eq{emergent-eq}.
Of course, it would be most natural for it to appear on the right-hand side of the Einstein
equation, Eq. \eq{emergent-eq}, as an energy-momentum tensor, as we explained above.

If so, it is still necessary to understand how the gravitational
constant $G$ in Eq. \eq{emergent-eq} arose from the gauge theory
in Eq. \eq{semi-matrix} because it did not contain $G$ from the outset.
Recall that both Eqs. \eq{emergent-eq} and \eq{em-maxwell} are valid even
in $D$-dimensions. Because the energy-momentum tensor carries the
physical dimension of energy density, i.e., $[T^{(em)}_{ab}] =
\frac{M L^2 T^{-2}}{L^{D-1}}$ and $[R_{ab}] = L^{-2}$, we need some
physical constant carrying the physical dimension of $M^{-1} L^{D-1} T^{-2}$ in Eq. \eq{emergent-eq}.
Of course, it is the Newton constant $G$. See Eq. \eq{quartet}.
However, we pointed out that, if a field theory is equipped with an intrinsic length scale,
which is precisely the case for the action in Eq. \eq{semi-matrix} with $L^2 =
|\theta|$, the gravitational constant $G$ can arise purely from the
field theory. In our case, this means \ct{hsy-jhep} that the
gravitational constant $G$ can be determined from only the field theory
parameters in Eq. \eq{semi-matrix}:
\be \la{newton-constant}
\frac{G \hbar^2}{c^2} \sim g_{YM}^2 |\theta|.
\ee
We will wait to Section V to pose an important question on
what the physical implications of Eq. \eq{newton-constant} are because
we are not yet prepared for that question.

Now, it is in order to ask about the real character of the energy-momentum
tensor in Eq. \eq{dark-energy}. As we pointed out before, $T_{ab}^{(M)}$ in
Eq. \eq{em-tensor-maxwell} is traceless, so the Ricci scalar
in Eq. \eq{final-ricci-scalar} should genuinely be determined by the
second energy-momentum tensor in Eq. \eq{dark-energy}. For example, let us
consider a maximally symmetric space in which the curvature and
the Ricci tensors are given by
\be \la{hom-curvature}
R_{abcd} = \frac{R}{D(D-1)} (\delta_{ac} \delta_{bd} - \delta_{ad}
\delta_{bc}), \quad R_{ab} = \frac{R}{D} \delta_{ab}.
\ee
Then, let us simply assume that the Einstein equations, Eq. \eq{emergent-eq},
allow a nearly maximally symmetric spacetime. In this case, the
energy-momentum tensor in Eq. \eq{dark-energy} will be dominant, and the
global structure of spactime will be determined by $T^{(L)}_{ab}$ only.

To descry closer aspects of the energy-momentum tensor
in Eq. \eq{dark-energy}, let us consider the following decomposition:
\bea \la{dec-emde}
&& \rho_a \rho_b = \frac{1}{4} \delta_{ab} \rho_c^2  +
\Big( \rho_a \rho_b - \frac{1}{4} \delta_{ab} \rho_c^2 \Big), \xx
&& \Psi_a \Psi_b = \frac{1}{4} \delta_{ab} \Psi_c^2  +
\Big( \Psi_a \Psi_b - \frac{1}{4} \delta_{ab} \Psi_c^2 \Big).
\eea
In the above decomposition, the first terms correspond to scalar
modes and will be a source of the expansion/contraction of spacetime
while the second terms correspond to quadruple modes and will give
rise to the shear distortion of spacetime, which can be seen via
Raychauduri's equation \eq{raychaudhuri}. For a nearly maximally
symmetric spacetime, the second terms can, thus, be neglected. In this
case, the energy-momentum tensor in Eq. \eq{dark-energy} behaves as a
cosmological constant for a (nearly) constant-curvature spacetime,
i.e.,
\be \la{em-cc}
T_{ab}^{(L)} = - \frac{c^4 R}{32 \pi G} \delta_{ab}.
\ee
In Section V, we will consider the Wick rotation, $y^4 = iy^0$, of
the energy-momentum tensor in Eq. \eq{dark-energy} and discuss a very
surprising aspect of it in Minkowski spacetime.

\section{QUANTUM GRAVITY}

Riemannian geometry has been charged with a primary role in
describing the theory of gravity, but many astronomical phenomena
involved with very strongly gravitating systems, e.g., the Big-Bang,
black holes, etc., disclose that a Riemannian geometry describing
a smooth spacetime manifold is not enough. Instead, it turns out that
a ``quantum geometry" is necessary to describe such extremely
gravitating systems. Unfortunately, we still do not know how to
quantize a Riemannian manifold in order to define the quantum
geometry.

In the previous section, we showed that a Riemannian geometry could
emerge from a Poisson geometry in the context of emergent gravity.
The underlying Poisson geometry has been defined by a $U(1)$ gauge
theory on a Poisson spacetime. Therefore, we may quantize the Poisson
geometry to define quantum gravity. Now, we want to explore how the
Poisson geometry defined by the $U(1)$ gauge theory on the Poisson
spacetime can be quantized to describe quantum geometries.

\subsection{Quantum Equivalence Principle}

In Section I, we have suggested that the quantization of
gravity might be defined by the spacetime deformation in terms of
$G$ rather than $\hbar$. If spin-two graviton were really a
fundamental particle, it could be physically viable to quantize
gravity in terms of the Planck constant $\hbar$ which will quantize
the particle phase space of gravitons. However, recent developments in
string theory \ct{string-book}, known as the AdS/CFT
duality, open-closed string duality, matrix models, etc., imply that
gravity may be a collective phenomenon emergent from gauge fields.
That is, the spin-two graviton might arise as a composite of two
spin-one gauge bosons. Presumably, this composite nature of gravitons
is already immanent in the vielbein formalism as the metric
expression in Eq. \eq{cartan} politely insinuates.

In Section II, we showed that Einstein gravity can be
formulated in terms of a symplectic geometry rather than a Riemannian
geometry in the context of emergent gravity. An essential step for
emergent gravity was to realize the equivalence principle, the most
important property in the theory of gravity (general relativity),
from $U(1)$ gauge theory on a symplectic or a Poisson manifold.
Through the realization of the equivalence principle, which is an
intrinsic property in symplectic geometry \ct{mechanics}, known
as the Darboux theorem or the Moser lemma, we can understand how
diffeomorphism symmetry arises from symplectic $U(1)$ gauge theory
and how gravity can emerge from symplectic electromagnetism,
which is also an interacting theory.

A unique feature of gravity disparate from other physical
interactions is that it is characterized by the Newton constant $G$
whose physical dimension is $({\rm length})^2$ in natural units.
We have to deeply ruminate about its physical origin. Our
proposal is that it is inherited from a Poisson structure of
spacetime. In order to support that, we have elucidated how gravity
can emerge from a field theory on such a spacetime. Also, we have
realized such an idea in Eq. \eq{newton-constant} that the gravitational
constant $G$ can be purely determined by the gauge theory parameters,
signaling the emergence of gravity from the field theory. Remarkably,
it turns out that $U(1)$ gauge theory defined with an intrinsic
length scale set by the Poisson structure in Eq. \eq{p-bracket} should be a
theory of gravity.

Therefore, it is now obvious how to quantize gravity if gravity is
emergent from a gauge theory defined on a symplectic or Poisson
manifold. We already briefly speculated in subsection I.C how
quantum geometry can arise from the quantization of spacetime, i.e.,
a noncommutative spacetime. We will clarify more how the essential
properties of emergent gravity can be lifted to the noncommutative
spacetime. In particular, we want to clarify how the Darboux theorem,
as the equivalence principle for emergent gravity, can be realized in
a full noncommutative geometry. It was already convincingly argued
in Ref. \ct{hsy-ijmpa} that such a kind of ``quantum equivalence principle"
exists in the context of deformation quantization {\it \`a la}
Kontsevich as a gauge equivalence between star products. Actually,
this gauge equivalence between star products reduces in the commutative
limit to the usual Darboux theorem and was the basis of the
Seiberg-Witten map between commutative and noncommutative gauge
fields \ct{SW-ncft}, as was also discussed in Ref. \ct{hsy-ijmpa}. Therefore,
a general noncommutative deformation of emergent gravity would be possible
because Kontsevich already proved \ct{kontsevich} that any Poisson manifold can always
be quantized at least in the context of deformation quantization.

As we argued in Section II, the Darboux theorem, or more
precisely the Moser lemma, in symplectic geometry is enough to derive
Einstein gravity because the latter arises from a $U(1)$ gauge theory on
a symplectic manifold. Now, we want to quantize the $U(1)$ gauge
theory defined by the action in Eq. \eq{semi-matrix} {\it \`a la} Dirac, i.e., by
adopting the quantization rule in Eq. \eq{q-bracket}:
\bea \la{nc-photon}
&& D_a \in C^\infty(M) \;\; \rightarrow \;\;
\widehat{D}_a = B_{ab} y^b + \widehat{A}_a \in  \CA_\theta, \xx
&& \{D_a, D_b\}_{\theta} \;\; \rightarrow
\;\; -i [\widehat{D}_a,
\widehat{D}_b] \equiv - B_{ab} + \widehat{F}_{ab},
\eea
where $\widehat{F}_{ab} \in \CA_\theta$ is the noncommutative field
strength defined by
\be \la{nc-field}
\widehat{F}_{ab} = \p_a \widehat{A}_b - \p_b \widehat{A}_a -i
[\widehat{A}_a, \widehat{A}_b].
\ee
Here, we understand the noncommutative fields in Eq. \eq{nc-photon}
as being self-adjoint operators acting on the Hilbert space $\CH$. Then, we
get the $U(1)$ gauge theory defined on the noncommutative spacetime in Eq. \eq{nc-spacetime} as
\be \la{nc-matrix}
\widehat{S} = - \frac{1}{4 G_s} \Tr_{\CH}
[\widehat{D}_a, \widehat{D}_b] [\widehat{D}^a, \widehat{D}^b],
\ee
where $G_s \equiv g_s/2 \pi \kappa^2$ and the trace $\Tr_{\CH}$ is
defined over the Fock space in Eq. \eq{fock} and can be identified with
\be \la{trace}
 \Tr_{\CH} \equiv \int \frac{d^{2n}y}{(2\pi)^n |{\rm Pf} \theta|}.
\ee
The Jacobi identity for the operator algebra $\CA_\theta$ leads to
the Bianchi identity
\be \la{nc-jacobi}
[\widehat{D}_{[a}, [\widehat{D}_b, \widehat{D}_{c]}]] = -
\widehat{D}_{[a} \widehat{F}_{bc]} = 0,
\ee
and the equations of motion derived from the action in Eq. \eq{nc-matrix} read as
\be \la{nc-eom}
[\widehat{D}^a, [\widehat{D}_a, \widehat{D}_b]] = - \widehat{D}^a
\widehat{F}_{ab} = 0,
\ee
where
\be \la{nccov-f}
\widehat{D}_a \widehat{F}_{bc} = \p_a \widehat{F}_{bc}
- i [\widehat{A}_a, \widehat{F}_{bc}].
\ee

In classical mechanics, the set of possible states of a particle
system forms a Poisson manifold $P$. The observables that we want to
measure are smooth functions in $C^\infty(P)$, forming a commutative
(Poisson) algebra. In quantum mechanics, the set of possible states
is represented by a Hilbert space $\CH$. The observables are
self-adjoint operators acting on $\CH$, forming a noncommutative
$\star$-algebra. The change from a Poisson manifold to a Hilbert
space is a pretty big one.

A natural question is whether a quantization, such as
Eq. \eq{q-bracket}, for a spacetime manifold $M$ with a general
Poisson structure $\pi = \half \pi^{\mu\nu} (x)
\frac{\partial}{\partial x^\mu} \wedge \frac{\partial}{\partial x^\nu} \in \Gamma(\Lambda^2 TM)$
is always possible with a radical change in the nature of the
observables. The problem is how to construct the Hilbert space for a
general Poisson manifold, which is, in general, highly nontrivial.
Deformation quantization was proposed in Ref. \ct{d-quantization} as an alternative, where
the quantization is understood to be a deformation of the algebra $\CA
= C^\infty(M)$ of classical observables. Instead of building a
Hilbert space from a Poisson manifold and associating an algebra of
operators to it, we are only concerned with the algebra $\CA$ to
deform the commutative product in $C^\infty(M)$ to a noncommutative
and associative product. In a canonical phase space where $\pi =
\theta$ such as the case we have considered so far, it is easy to
show that the two approaches have a one-to-one correspondence through
the Weyl-Moyal map
\ct{nc-review,szabo-cqg}:
\begin{equation}\label{moyal-product}
\what{f} \cdot \what{g} \cong (f \star g)(y) = \left.\exp\left(\frac{i}{2}
\theta^{\mu\nu} \partial_{\mu}^{y}\partial_{\nu}^{z}\right)f(y)g(z)\right|_{y=z}.
\end{equation}

Recently Kontsevich answered the above question in the context of
deformation quantization \ct{kontsevich}. He proved that every
finite-dimensional Poisson manifold $M$ admits a canonical
deformation quantization and that changing coordinates leads to
gauge-equivalent star products. We briefly recapitulate his results,
which will be useful for our later discussions.
Let $\CA$ be the algebra over $\mathbb{R}$ of smooth functions on a
finite-dimensional $C^\infty$-manifold $M$. A star product on $\CA$
is an associative $\mathbb{R}[[\hbar]]$--bilinear product on the algebra
$\CA[[\hbar]]$, a formal power series in $\hbar$ with coefficients
in $C^\infty (M) = \CA$, given by the following formula for $f,g \in
\CA \subset \CA[[\hbar]]$ \cite{d-para}:
\be \la{d-star}
(f,g) \mapsto f \star g = fg + \hbar{B_1(f,g)}+ \hbar^2{B_2(f,g)}+ \cdots
\in \CA[[\hbar]],
\ee
where $B_i(f,g)$ are bidifferential operators. There is a natural
gauge group which acts on star products. This group consists of
automorphisms of $\CA[[\hbar]]$ considered as an
$\mathbb{R}[[\hbar]]$--module (i.e., linear transformations $\CA \to \CA$
parameterized by $\hbar$). If $D(\hbar) = 1+ \sum_{n\geq{1}}{\hbar^{n}D_n}$ is such an automorphism,
where $D_n: \CA \to \CA$ are differential operators, it acts on the set of star
products as
\be \la{star-equiv}
\star \rightarrow \star^\prime, \quad
f(\hbar) \star^\prime g(\hbar) = D(\hbar)
\Bigl( D(\hbar)^{-1} (f(\hbar)) \star  D(\hbar)^{-1} (g(\hbar)) \Bigr)
\ee
for $f(\hbar), g(\hbar) \in \CA[[\hbar]]$. This is evident from the
commutativity of the diagram
$$
\xymatrix@+0.5cm{ \CA[[\hbar]] \times \CA[[\hbar]]
\ar[r]^-\star \ar[d]_{D(\hbar) \times D(\hbar)}\ & \CA[[\hbar]]
\ar[d]^{D(\hbar)} \cr
\CA[[\hbar]] \times \CA[[\hbar]] \ar[r]^-{\star^\prime} & \CA[[\hbar]]. \cr } \
$$

Two star products $\star$ and $\star^\prime$ are called equivalent
if there exists an automorphism $D(\hbar)$, a formal power series of
differential operators, satisfying Eq. \eq{star-equiv}. We are
interested in star products up to the gauge equivalence. This
equivalence relation is closely related to the cohomological
Hochschild complex of algebra $\CA$ \ct{kontsevich}, i.e.,
the algebra of smooth polyvector fields on $M$.
For example, it follows from the associativity of the product
in Eq. \eq{d-star} that the symmetric part of $B_1$ can be killed by a
gauge transformation that is a coboundary in the Hochschild
complex, and that the antisymmetric part of $B_1$, denoted as
$B^-_1$, comes from a bivector field $\pi \in \Gamma(\Lambda^2 TM)$
on $M$:
\be \la{bi-vector}
B_1^-(f,g) = \langle \pi, df \otimes dg \rangle.
\ee
In fact, any Hochschild coboundary can be removed by using a gauge
transformation $D(\hbar)$, thus leading to the gauge equivalent star
product in Eq. \eq{star-equiv}. The associativity at $\CO(\hbar^2)$ further
constrains that $\pi$ must be a Poisson structure on $M$; in other
words, $[\pi, \pi]_{SN} = 0$, where the bracket is the
Schouten-Nijenhuis bracket on polyvector fields (see
Ref. \ct{kontsevich} for the definition of this bracket and the
Hochschild cohomology). Thus, gauge equivalence classes of star
products modulo $\CO(\hbar^2)$ are classified by Poisson structures
on $M$. It was shown
\ct{kontsevich} that there are no other obstructions to deforming
the algebra $\CA$ up to arbitrary higher orders in $\hbar$.

For an equivalence class of star products for any Poisson manifold,
Kontsevich arrived at the following general result \ct{kontsevich}:

{\it The set of gauge equivalence classes of star products on a
smooth manifold $M$ can be naturally identified with the set of
equivalence classes of Poisson structures depending formally on
$\hbar$
\be \la{theorem-1.1}
\pi = \pi(\hbar)= \pi_1 \hbar + \pi_2 \hbar^2 + \cdots
\in \Gamma(\Lambda^2 TM)[[\hbar]], \quad
[\pi,\pi]_{SN} = 0 \in \Gamma(\Lambda^3 TM)[[\hbar]]
\ee
modulo the action of the group of formal paths in the diffeomorphism
group of $M$, starting at the identity diffeomorphism. Also, if we
change coordinates in Eq. \eq{d-star}, we obtain a gauge-equivalent
star product.}

This theorem means that the set of equivalence classes of
associative algebras close to algebras of functions on manifolds is
in one-to-one correspondence with the set of equivalence classes of
Poisson manifolds module diffeomorphisms.

Suppose that the Poisson tensor $\pi = \half
\pi^{\mu\nu} (x) \frac{\partial}{\partial x^\mu}
\wedge \frac{\partial}{\partial x^\nu} \in \Gamma(\Lambda^2 TM)$ is a
nondegenerate constant bi-vector and denote it with $\theta$ again.
In this case, the star product is given by Eq. \eq{moyal-product}, the
so-called Moyal product. If we make an arbitrary change of
coordinates, $y^\mu \mapsto x^a(y)$, in the Moyal $\star$-product
in Eq. \eq{moyal-product}, which is nothing but the Kontsevich's star product
in Eq. \eq{d-star} with the constant Poisson bi-vector $\theta$,
we will get a new star product, Eq. \eq{d-star}, defined by a Poisson
bi-vector $\pi(\hbar)$. However, the resulting star product has to be
gauge equivalent to the Moyal product in Eq. \eq{moyal-product} and
$\pi(\hbar)$ belongs to the same equivalence class of Poisson
structures and so could be determined by using the formal power series
in Eq. \eq{theorem-1.1} for the original Poisson bi-vector $\theta$.
Conversely, if two star products $\star$ and $\star^\prime$ are
gauge equivalent in the sense that there exists an automorphism
$D(\hbar)$ satisfying Eq. \eq{star-equiv}, the Poisson structures
$\theta$ and $\pi$ defining the star products $\star$ and
$\star^\prime$, respectively, must belong to the same gauge
equivalence class. This is the general statement of the above theorem.

Actually, it is easy to show that the gauge equivalence relation
in Eq. \eq{star-equiv} between star products reduces to the Darboux
transformation \eq{darboux-tr} in the commutative limit where $D(\hbar)
= 1$. After identifying $\pi^{-1} = B + F$ and $\theta^{-1} = B$, we
get in this limit
\be \la{star-darboux}
\{x^a, x^b \}_\theta (y) = \theta^{\mu\nu} \frac{\p x^a}{\p y^\mu}
\frac{\p x^b}{\p y^\nu} = \pi^{ab}(x) = \Big( \frac{1}{B+F}
\Big)^{ab}(x),
\ee
which is precisely the inverse of Eq. \eq{darboux-tr} if $\omega_1 =
\pi^{-1}$ and $\omega_0 = \theta^{-1}$.
Therefore, we propose \ct{hsy-ijmpa} the ``quantum equivalence
principle" as the gauge equivalence in Eq. \eq{star-equiv} between star
products in the sense that the Darboux theorem as the equivalence
principle for emergent gravity is lifted to a noncommutative geometry.
Furthermore, the isomorphism in Eq. \eq{cov-cod} from the Lie algebra of
Poisson vector fields to the Lie algebra of vector fields as
derivations of $C^\infty(M)$ can be lifted to the noncommutative
spacetime in Eq. \eq{nc-spacetime} as follows: Consider an adjoint
operation of noncommutative gauge fields $\widehat{D}_a(y) \in
\CA_\theta$ in Eq. \eq{nc-photon},
\be \la{nc-vector}
\widehat{V}_a[\widehat{f}](y) \equiv - i [
\widehat{D}_a(y), \widehat{f}(y) ]_\star.
\ee
The leading term in Eq. \eq{nc-vector} exactly recovers the vector
fields in Eq. \eq{v-field}, i.e.,
\bea \la{c-vector}
\widehat{V}_a[\widehat{f}](y) &=& - i [\widehat{D}_a(y), \widehat{f}(y) ]_\star
= - \theta^{\mu \nu} \frac{\partial D_a(y)}{\partial y^\nu}
\frac{\partial f(y)}{\partial y^\mu} + \cdots \xx & = &
V_a[f](y) + {\cal O}(\theta^3).
\eea

Because the star product in Eq. \eq{d-star} is associative, one can show the
following properties:
\bea \la{leibniz}
&& \widehat{V}_a[\widehat{f} \star \widehat{g}] =
\widehat{V}_a[\widehat{f}] \star \widehat{g} + \widehat{f} \star
\widehat{V}_a[\widehat{g}], \xx
&& \widehat{V}_{-i[\widehat{D}_a,\widehat{D}_a]_\star} =
[\widehat{V}_a, \widehat{V}_b]_\star.
\eea
The above property implies that we can identify the adjoint
operation ${\rm Der}(\CA_\theta) \equiv \{ \widehat{V}_a | a =
1,\cdots, 2n \}$ with the (inner) derivations of a noncommutative
$\star$-algebra $\CA_\theta$, and so the generalization of vector
fields $\Gamma(TM) = \{V_a | a = 1,\cdots, 2n \}$ in
Eq. \eq{v-field}. Using Eq. \eq{leibniz}, one can show that
\bea \la{nc-homo1}
&& \widehat{V}_{\widehat{F}_{ab}} = [\widehat{V}_a,
\widehat{V}_b]_\star, \\
\la{nc-homo2}
&& \widehat{V}_{\widehat{D}_a \widehat{F}_{bc}} = [\widehat{V}_a,
[\widehat{V}_b, \widehat{V}_c]_\star ]_\star,
\eea
which may be compared with Eqs. \eq{map-fvv} and \eq{map-vvv}. We can use
the map in Eq. \eq{nc-homo2} in exactly the same way as in the Poisson algebra
case to translate the Jacobi identity in Eq. \eq{nc-jacobi} and the equations of motion
in Eq. \eq{nc-eom} into some relations between the generalized vector fields $\widehat{V}_a$
defined by Eq. \eq{nc-vector} \ct{hsy-jhep}:
\bea \la{ncmap-jacobi}
&& \widehat{D}_{[a} \widehat{F}_{bc]} = 0 \;\;
\Leftrightarrow \;\; [\widehat{V}_{[a}, [\widehat{V}_b, \widehat{V}_{c]}]_\star ]_\star = 0, \\
\la{ncmap-eom}
&& \widehat{D}^a \widehat{F}_{ab} = 0 \;\;
\Leftrightarrow \;\; [\widehat{V}^a, [\widehat{V}_a, \widehat{V}_b]_\star
]_\star = 0.
\eea

We will consider the system of the derivations of noncommutative
$\star$-algebra $\CA_\theta$ defined by Eqs. \eq{ncmap-jacobi} and
\eq{ncmap-eom} as the quantization of the system given by
Eqs. \eq{map-jacobi} and \eq{map-eom} and, thus, as {\it quantization of Einstein
gravity} in the sense of Eq. \eq{q-bracket}. To support the claim, we
will take the correspondence in Eq. \eq{ncft-matrix} to show \ct{hsy-epjc}
that any large-$N$ gauge theory can be mapped to a noncommutative
$U(1)$ gauge theory like as Eq. \eq{nc-matrix}. Because the large-$N$
gauge theory is believed to provide a theory of quantum geometries
as evidenced by the AdS/CFT correspondence and various matrix models
in string theory, we think it could be reasonable evidence for our claim.

\subsection{Noncommutative Electromagnetism as a Large-$N$ Gauge Theory}

Let us consider $U(N  \to \infty)$ Yang-Mills theory in $d$ dimensions:
\begin{equation} \label{n=4}
S_M = - \frac{1}{G_s} \int d^d z {\rm Tr}
\left(\frac{1}{4} F_{\mu\nu} F^{\mu\nu}  +
\frac{1}{2} D_\mu \Phi^a D^\mu \Phi^a - \frac{1}{4} [\Phi^a, \Phi^b]^2
\right),
\end{equation}
where $G_s \equiv 2\pi g_s/(2\pi\kappa)^{\frac{4-d}{2}}$ and
$\Phi^a \; (a = 1, \cdots, 2n)$ are adjoint scalar fields in $U(N)$.
Here, the $d$-dimensional commutative spacetime $\mathbb{R}^d_C$ will be
taken with either a Lorentzian or a Euclidean signature. Note that, if
$d=4$ and $n=3$, the action in Eq. \eq{n=4} is exactly the bosonic part of
the 4-dimensional ${\cal N} =4$ supersymmetric $U(N)$ Yang-Mills theory,
which is the large-$N$ gauge theory of the AdS/CFT correspondence.

Suppose that a vacuum of the theory, Eq. \eq{n=4}, is given by
\be \la{matrix-vacuum}
\langle \Phi^a \rangle_{\rm vac} = \frac{1}{\kappa} y^a, \qquad
\langle A_\mu \rangle_{\rm vac} = 0.
\ee
We will assume that the vacuum expectation values $y^a \in U(N)$
in the $N \to \infty$ limit satisfy the algebra
\be \la{vacuum-algebra}
[y^a, y^b] = i \theta^{ab} \mathbf{1}_{N \times N},
\ee
where $\theta^{ab}$ is a constant matrix of rank $2n$. If so, the
vacuum in Eq. \eq{matrix-vacuum} is definitely a solution to the theory
in Eq. \eq{n=4} and the large-$N$ matrices $y^a$ can be mapped
to noncommutative fields according to the correspondence
in Eq. \eq{ncft-matrix}. The adjoint scalar fields in vacuum
then satisfy the noncommutative Moyal algebra defined by
Eq. \eq{nc-spacetime} or equivalently
\be \la{nc-vacuum}
[y^a, y^b]_\star = i \theta^{ab}.
\ee

Now, let us expand the large-$N$ matrices in the action in Eq. \eq{n=4} around
the vacuum in Eq. \eq{matrix-vacuum}:
\bea \la{mat-fluct}
&& \Phi^a(z,y) = \frac{1}{\kappa} \big( y^a + \theta^{ab}
\widehat{A}_b(z,y) \big), \\
&& D_\mu (z,y) = \p_\mu - i \widehat{A}_\mu (z,y),
\eea
where we have assumed that the fluctuations $\widehat{A}_{M} (X) \equiv
(\widehat{A}_\mu, \widehat{A}_a) (z,y), \; M = 1, \cdots, d+2n$,
also depend on the vacuum moduli in Eq. \eq{matrix-vacuum}. Therefore,
let us introduce $D=d+2n$-dimensional coordinates $X^M = (z^\mu,
y^a)$, which consist of $d$-dimensional commutative ones denoted by
$z^\mu \; (\mu=1, \cdots, d)$ and $2n$-dimensional noncommutative
ones denoted by $y^a \; (a = 1, \cdots, 2n)$, satisfying the
relation in Eq. \eq{nc-vacuum}. Likewise, $D$-dimensional gauge fields
$\widehat{A}_M(X)$ are also introduced in a similar way \cite{i-notat}:
\bea \la{decomp-cov}
D_M (X) &=& \p_M - i \widehat{A}_M(X) \xx &\equiv& (D_\mu =
\p_\mu - i \widehat{A}_\mu, D_a = -i\kappa B_{ab}
\Phi^b)(z,y).
\eea

According to the correspondence in Eq. \eq{ncft-matrix}, we will replace
the matrix commutator in the action in Eq. \eq{n=4} by the star commutator,
i.e.,
\be \la{map-commutator}
[\clubsuit, \spadesuit]_{N \times N} \quad \to \quad [\clubsuit,
\spadesuit]_\star.
\ee
It is then straightforward to calculate each component in the matrix
action in Eq. \eq{n=4}
\bea \la{nc-comp}
&& F_{\mu\nu} = i [D_\mu, D_\nu]_\star = \partial_\mu
\widehat{A}_\nu - \partial_\nu \widehat{A}_\mu -
i[\widehat{A}_\mu,
\widehat{A}_\nu]_\star := \widehat{F}_{\mu\nu}, \xx
&& D_\mu \Phi^a = i \frac{\theta^{ab}}{\kappa} [D_\mu, D_b]_\star =
\frac{\theta^{ab}}{\kappa} \Big( \partial_\mu \widehat{A}_b -
\partial_b \widehat{A}_\mu - i[\widehat{A}_\mu,
\widehat{A}_b]_\star \Big) := \frac{\theta^{ab}}{\kappa} \widehat{F}_{\mu b},
\\
&& [\Phi^a, \Phi^b] = - \frac{1}{\kappa^2} \theta^{ac} \theta^{bd}
[D_c, D_d]_\star = \frac{i}{\kappa^2} \theta^{ac}
\theta^{bd} \Big(-B_{cd} + \partial_c \widehat{A}_d - \partial_d
\widehat{A}_c - i[\widehat{A}_c, \widehat{A}_d]_\star \Big) \xx
&& \qquad \quad := - \frac{i}{\kappa^2} \Big(\theta (\widehat{F} -
B) \theta \Big)^{ab}, \nonumber
\eea
where we defined $[\partial_\mu, \widehat{f}]_\star = \partial_\mu
\widehat{f}$ and $B = \half B_{ab} dy^a \wedge dy^b$ with rank$(B)= 2n$.
It is important to notice that large-$N$ matrices on the vacuum
in Eq. \eq{matrix-vacuum} are now represented by their master fields, which are
higher-dimensional noncommutative $U(1)$ gauge fields in
Eq. \eq{decomp-cov} whose field strength is given by
\be \la{D-field}
\widehat{F}_{MN} = \p_M \widehat{A}_N - \p_N \widehat{A}_M
- i [\widehat{A}_M, \widehat{A}_N]_\star.
\ee
Collecting all the results in Eq. \eq{nc-comp} and using the trace
in Eq. \eq{trace}, the action in Eq. \eq{n=4} can be recast into the simple form \ct{hsy-epjc}
\begin{equation} \label{higher-action}
\widehat{S}_B = - \frac{1}{4 g^2_{YM}} \int d^D X \sqrt{-G} G^{MP}
G^{NQ} (\widehat{F}_{MN} - B_{MN}) \star (\widehat{F}_{PQ} -
B_{PQ}),
\end{equation}
where we have assumed a constant metric on $\mathbb{R}^D = \mathbb{R}^d_C \times
\mathbb{R}^{2n}_{NC}$ with the form
\bea \la{flat-metric}
ds^2 &=& G_{MN} dX^M dX^N \xx &=& g_{\mu\nu} dz^\mu dz^\nu +
\widehat{g}_{ab} dy^a dy^b
\eea
and the relations in Eqs. \eq{open}, \eq{coupling1} and \eq{coupling2}
were used. In the end, the $d$-dimensional $U(N)$ Yang-Mills theory
in Eq. \eq{n=4} has been transformed into a $D$-dimensional
noncommutative $U(1)$ gauge theory.

Depending on the choice of the base space $\mathbb{R}_C^d$, one can get a
series of matrix models from the large-$N$ gauge theory in Eq. \eq{n=4}:
for instance, the IKKT matrix model for $d=0$ \ct{ikkt}, the BFSS matrix model for $d=1$ \ct{bfss}
and the matrix string theory for $d=2$ \ct{mst}.
The most interesting case is $d=4$ and $n=3$, which is equal to the
bosonic part of the 4-dimensional ${\cal N} =4$ supersymmetric $U(N)$
Yang-Mills theory in the AdS/CFT duality
\ct{ads-cft} and is equivalent to the 10-dimensional noncommutative
$U(1)$ gauge theory on $\mathbb{R}^{4}_C \times \mathbb{R}^{6}_{NC}$. Note that
all these matrix models or large-$N$ gauge theories are
nonperturbative formulations of string or M theories. Therefore, it
should be reasonable to expect that the $d$-dimensional $U(N  \to
\infty)$ gauge theory in Eq. \eq{n=4} and so the $D$-dimensional noncommutative
$U(1)$ gauge theory in Eq. \eq{higher-action} describe a theory of quantum
gravity according to the large-$N$ duality or AdS/CFT correspondence.

We will give further evidences why the matrix action in Eq. \eq{n=4}
contains a variety of quantum geometries and how smooth Riemannian
geometries can be emergent from the action in Eq. \eq{higher-action} in
a commutative limit. First, apply the adjoint operation in Eq. \eq{nc-vector}
to the $D$-dimensional noncommutative gauge fields $D_A(X) = (D_\mu,
D_a)(z,y)$ (after switching the index $M \to A$ to distinguish them from
the local coordinate indices $M, N, \cdots$) to obtain
\bea \la{D-vector}
\widehat{V}_A [\widehat{f}](X) &=& [D_A, \widehat{f}]_\star (z,y)
\xx &\equiv& V_A^M(z,y) \p_M f(z,y) + \CO(\theta^3),
\eea
where $V_A^\mu = \delta^\mu_A$ because the star product acts only on
$y$-coordinates and we define $[\p_\mu,
\widehat{f}(X)]_\star = \frac{\p \widehat{f}(X)}{\p z^\mu}$.
More explicitly, the $D$-dimensional noncommutative $U(1)$ gauge
fields at leading order appear as the usual vector fields
(frames on a tangent bundle) on a $D$-dimensional manifold $M$ given by
\be \la{vec-d}
V_A (X) = (\p_\mu + A_\mu^a \p_a,  D_a^b \p_b)
\ee
or with matrix notation
\be \la{vec-matrix}
V_A^M(X) = \left(
  \begin{array}{cc}
    \delta^\nu_\mu & A_\mu^a  \\
    0 &  D_a^b \\
  \end{array}
\right),
\ee
where
\be \la{com-vec}
A_\mu^a \equiv - \theta^{ab} \frac{\p \widehat{A}_\mu}{\p y^b},
\qquad  D_a^b \equiv \delta^b_a - \theta^{bc} \frac{\p \widehat{A}_a}{\p
y^c}.
\ee
One can easily check that $V_A$'s in Eq. \eq{vec-d} take values in
the Lie algebra of volume-preserving vector fields, i.e., $\p_M
V_A^M = 0$. One can also determine the dual basis $V^A = V^A_M dX^M
\in \Gamma(T^*M)$, i.e., $\langle V^A, V_B \rangle = \delta^A_B$, which is given by
\be \la{form-d}
V^A (X) = \bigl(dz^\mu,  V^a_b(dy^b- A_\mu^b dz^\mu)\bigr)
\ee
or with matrix notation
\be \la{form-matrix}
V_M^A (X) = \left(
  \begin{array}{cc}
    \delta^\nu_\mu & - V_b^a A_\mu^b  \\
    0 &  V_b^a \\
  \end{array}
\right),
\ee
where $V_a^c D_c^b = \delta_a^b$.

From the previous analysis in Section II.C (which corresponds to the
$d=0$ case), we know that the vector fields $V_A$ determined by
gauge fields are related to the orthonormal frames (vielbeins) $E_A$
by $V_A = \lambda E_A$ and $E^A = \lambda V^A$, where the
conformal factor $\lambda$ will be determined later. (This situation
is very reminiscent of the string frame ($V_A$) and the Einstein
frame ($E_A$) in string theory.) Hence, the $D$-dimensional metric
can be determined explicitly, by using the dual basis \eq{form-d} up to a
conformal factor \ct{ward}:
\bea \la{ward-metric}
ds^2 &=& \eta_{AB} E^A \otimes E^B \xx &=& \lambda^2
\eta_{AB} V^A \otimes V^B =
\lambda^2 \eta_{AB} V^A_M V^B_N \; dX^M \otimes dX^N \xx
&=& \lambda^2 \Bigl(\eta_{\mu\nu} dz^\mu dz^\nu +
\delta_{ab} V^a_c V^b_d (dy^c - \mathbf{A}^c)(dy^d - \mathbf{A}^d)
\Bigr)
\eea
where $\mathbf{A}^a = A_\mu^a dz^\mu$.

The conformal factor $\lambda^2$ in the metric in Eq. \eq{ward-metric} can
be determined in exactly the same way as in the Section II.C. Choose a
$D$-dimensional volume form with a matching parameter $\lambda \in
C^\infty(M)$ such that
\be \la{D-volume-form}
\nu = \lambda^2 V^1 \wedge \cdots \wedge V^D
\ee
and
\be \la{D-volume}
\lambda^2 = \nu (V_1, \cdots, V_D).
\ee
Then, the vector fields $V_A$ are volume preserving with respect to a
$D$-dimensional volume form $\nu = \lambda^{(2-D)} \nu_g$, where
\be \la{E-volume-form}
\nu_g = E^1 \wedge \cdots \wedge E^D
\ee
and the vector fields $E_A$ are volume preserving with respect to
another volume form $\widetilde{\nu} = \lambda^{(3-D)}
\nu_g$. Because $\p_M V_A^M = 0$ or $\CL_{V_A} \nu = 0$, we can choose the invariant volume
by turning off all fluctuations in Eq. \eq{D-volume-form} as
\be \la{D-flat-volume}
\nu = dz^1 \wedge \cdots \wedge dz^d \wedge dy^1
\wedge \cdots \wedge dy^{2n}.
\ee
Then, we finally get
\be \la{D-lambda}
\lambda^2 = \det^{-1} V^a_b.
\ee

One can see that the spacetime geometry described by the metric
in Eq. \eq{ward-metric} is completely determined by noncommutative gauge fields
whose underlying theory is defined by the action in Eq. \eq{n=4} or
in Eq. \eq{higher-action}. One may also confirm the claim in Section I.A that a spin-two graviton
arises as a composite of two spin-one vector fields and such
spin-one vector fields arise from electromagnetic fields living in
the noncommutative spacetime of Eq. \eq{nc-vacuum}. However, one has to remember
that the spacetime geometry in Eq. \eq{ward-metric} is responsible only at
the leading order, i.e., $\CO(\theta)$, of the generalized vector
fields defined by Eq. \eq{D-vector}. All higher derivative terms in the
star product in Eq. \eq{D-vector} are simply ignored. If such higher
derivative terms are included in the star product in Eq. \eq{D-vector}
order by order, they will deform the Einstein gravity order by order
as a response to the noncommutative effects of spacetime. (See
Ref. \ct{hsy-ijmpa} for higher-order corrections to emergent
gravity.) If a probe goes into a deep microscopic world where the
noncommutative effect of spacetime will grow significantly, the
gravity description in Eq. \eq{ward-metric} in terms of smooth geometries
will gradually become crude and coarse. In the deep noncommutative
space, we have to replace Einstein gravity by a more fundamental
theory describing quantum gravity or a noncommutative geometry. We
argued that such a fundamental theory could be implemented by using the
large-$N$ gauge theory in Eq. \eq{n=4} or the higher-dimensional
noncommutative $U(1)$ gauge theory \eq{higher-action}.
First note that
\bea \la{D-dd}
&& [D_A, D_B]_\star = - i (\widehat{F}_{AB} - B_{AB}),\\
\la{D-ddd}
&& [D_A, [D_B, D_C]_\star]_\star =  - i \widehat{D}_A
\widehat{F}_{BC},
\eea
where
\be \la{D-cod}
\widehat{D}_A \widehat{F}_{BC} \equiv \p_A \widehat{F}_{BC} - i
[\widehat{D}_A, \widehat{F}_{BC}]_\star.
\ee
Therefore, the Bianchi identity and the equations of motion for the
action in Eq. \eq{higher-action} can be written as
\bea \la{D-jacobi}
&& \widehat{D}_{[A} \widehat{F}_{BC]} = i [D_{[A}, [D_B,
D_{C]}]_\star]_\star = 0, \\
\la{D-eom}
&& \widehat{D}^A \widehat{F}_{AB} = i [D^A, [D_A, D_B]_\star]_\star
= 0.
\eea
Then, the above equations can be translated into some ``geometric"
equations of generalized vector fields defined in Eq. \eq{D-vector}:
\bea \la{Dmap-jacobi}
&& \widehat{D}_{[A} \widehat{F}_{BC]} = 0 \qquad \Leftrightarrow
\qquad [\widehat{V}_{[A}, [\widehat{V}_B, \widehat{V}_{C]}]_\star]_\star = 0, \\
\la{Dmap-eom}
&& \widehat{D}^A \widehat{F}_{AB} = 0 \qquad \Leftrightarrow
\qquad [\widehat{V}^A, [\widehat{V}_A, \widehat{V}_B]_\star]_\star = 0.
\eea
It may be useful to introduce a noncommutative version of the
structure equation, Eq. \eq{lie-v}:
\be \la{lie-ncv}
[\widehat{V}_A, \widehat{V}_B]_\star = -
{\widehat{\mathfrak{F}}_{AB}}^{\quad C}
\widehat{V}_C,
\ee
with the ordering prescription that the structure coefficients
${\widehat{\mathfrak{F}}_{AB}}^{\quad C} \in \CA_\theta$ are always
coming to the left-hand side. Equations \eq{Dmap-jacobi} and
\eq{Dmap-eom} can be rewritten using the structure equation, Eq. \eq{lie-ncv}, as
\bea \la{Dstr-jacobi}
&& \widehat{D}_{[A} \widehat{F}_{BC]} = 0  \qquad \Leftrightarrow
\qquad \widehat{V}_{[A} {\widehat{\mathfrak{F}}_{BC]}}^{\quad D}  -
{\widehat{\mathfrak{F}}_{[BC}}^{\quad E} \star {\widehat{\mathfrak{F}}_{A]E}}^{\quad D} = 0, \\
\la{Dstr-eom}
&& \widehat{D}^A \widehat{F}_{AB} = 0 \qquad \Leftrightarrow
\qquad \eta^{AB} \Bigl(\widehat{V}_A {\widehat{\mathfrak{F}}_{BC}}^{\quad D} -
{\widehat{\mathfrak{F}}_{BC}}^{\quad E}
\star {\widehat{\mathfrak{F}}_{AE}}^{\quad D} \Bigr) = 0.
\eea

We take a commutative limit $|\theta| \to 0$ (in the same sense as
$\hbar \to 0$ in quantum mechanics), and we keep only the leading term
in Eq. \eq{D-vector} for the generalized vector fields
$\widehat{V}_A$. In this limit, we will recover the Einstein gravity
for the emergent metric in Eq. \eq{ward-metric} where Eqs. \eq{Dstr-jacobi} and
\eq{Dstr-eom} reduce to the first Bianchi identity for Riemann tensors and
the Einstein equations, respectively, as we checked in the previous
section. The Einstein gravity is relevant only in this limit. If
$|\theta|$ is finite (in the same sense as $\hbar \to 1$ in quantum
mechanics), we have to rely on Eqs. \eq{Dstr-jacobi} and \eq{Dstr-eom} instead:
What is going on here? In order to answer the question, it is necessary to solve
Eqs. \eq{Dstr-jacobi} and \eq{Dstr-eom} first. Of course, it
will be, in general, very difficult to solve the equations. Instead,
one may introduce linear algebraic conditions of $D$-dimensional
field strengths $\widehat{F}_{AB}$ as a higher-dimensional analogue
of $4$-dimensional self-duality equations such that the Yang-Mills
equations in the action in Eq. \eq{higher-action} follow automatically.
These are of the type \ct{cdfn}
\be \la{high-sd}
\half T_{ABCD}\widehat{F}_{CD} = \chi \widehat{F}_{AB}
\ee
with a constant 4-form tensor $T_{ABCD}$. The relation in Eq. \eq{high-sd}
clearly implies via the Bianchi identity in Eq. \eq{Dmap-jacobi} that the
equations of motion, Eq. \eq{Dmap-eom}, are satisfied provided $\chi$ is
nonzero. For $D > 4$, the 4-form tensor $T_{ABCD}$ cannot be
invariant under $SO(D)$ transformations and Eq. \eq{high-sd} breaks the rotational symmetry
to a subgroup $H \subset SO(D)$. Thus, the resulting first-order equations can be classified
by the unbroken symmetry $H$ under which $T_{ABCD}$ remains invariant
\ct{cdfn}. It was also shown \ct{bpark} that the first-order linear equations above
are closely related to supersymmetric states, i.e., BPS states in higher-dimensional Yang-Mills theories.

Note that
\be \la{D-homo}
\widehat{V}_{-i [D_A, D_B]_\star} = \widehat{V}_{\widehat{F}_{AB}} =
[\widehat{V}_A, \widehat{V}_B]_\star.
\ee
Using the homomorphism in Eq. \eq{D-homo}, one can translate the
generalized self-duality equation, Eq. \eq{high-sd}, into the structure
equation between vector fields,
\be \la{D-sd}
\half T_{ABCD} \widehat{F}_{CD} = \chi \widehat{F}_{AB} \quad \Leftrightarrow \quad
\half T_{ABCD} [\widehat{V}_C, \widehat{V}_D]_\star =
\chi  [\widehat{V}_A, \widehat{V}_B]_\star.
\ee
Therefore, a $D$-dimensional noncommutative gauge field configuration
satisfying the first-order system defined by the left-hand side of
Eq. \eq{D-sd} is isomorphic to a $D$-dimensional emergent ``quantum"
geometry defined by the right-hand side of Eq. \eq{D-sd} whose metric in
the commutative limit is given by Eq. \eq{ward-metric}. For example, in four
dimensions where $T_{ABCD} = \varepsilon_{ABCD}$ and $\chi =
\pm 1$, Eq. \eq{D-sd} goes to Eq. \eq{iso-map} describing gravitational instantons
in the commutative limit. Hence, it would not be absurd for someone
to claim that self-dual noncommutative electromagnetism in four
dimensions is equivalent to self-dual quantum gravity \ct{hsy-epl,hsy-inst}.
Indeed, it was argued in Ref. \ct{hsy-epjc} that
the emergent geometry arising from the self-dual system in Eq. \eq{D-sd} is
closely related to the bubbling geometry in the AdS space found in Ref. \ct{bubbling}.

\subsection{Background-independent Quantum Gravity}

According to Einstein, gravity is the dynamics of spacetime
geometry. Therefore, as emphasized by Elvang and Polchinski
\ct{polchinski}, the emergence of gravity necessarily requires the
emergence of spacetime itself. That is, spacetime is not given {\it
a priori}, but should be derived from fundamental ingredients in
quantum gravity theory, say, ``spacetime atoms". However, for
consistency, the entire spacetime including a flat spacetime must be
emergent. In other words, the emergent gravity should necessarily be
``background independent," where any spacetime structure is not {\it
a priori} assumed, but is defined by the theory. Furthermore, if
spacetime is emergent, then all fields supported on this spacetime
must be emergent, too. The question is how everything, including
spacetime, gauge fields and matter fields, could be emergent
collectively. We know emergent phenomena in condensed matters arise
due to a very coherent condensation in vacuum. Thus, in order to
realize all these emergent phenomena, the emergent spacetime needs
to be derived from an extremely coherent vacuum, which is the lesson
we learned from condensed matter. This turns out to be the case
if a flat spacetime is emergent from a noncommutative algebra such
as quantum harmonic oscillators.

We will carefully recapitulate the emergent gravity derived from the
action in Eq. \eq{n=4} to throw the universe into a fresh perspective and
to elucidate how the emergent gravity based on the noncommutative
geometry achieves background independence. Of course, real
physics is necessarily background dependent because a physical
phenomenon occurs in a particular background with specific initial
conditions. Background independence here means that, although
physical events occur in a particular (spacetime and material)
background, an underlying theory itself describing such a physical
event should presuppose neither any kind of spacetime nor material backgrounds.
The background in itself should also arise from a solution of the underlying theory.

The $U(N)$ gauge theory in Eq. \eq{n=4} is defined by a collection of $N
\times N$ matrices $(A_\mu, \Phi^a)(z)$ on a $d$-dimensional
flat spacetime $\mathbb{R}^d_C$. Note that the $d$-dimensional flat
spacetime $\mathbb{R}^d_C$ already exists from the beginning independently
of $U(N)$ gauge fields and that the theory says nothing about its origin.
It just serves as a playground for the players $(A_\mu, \Phi^a)$.

We showed that the $d$-dimensional matrix theory in Eq. \eq{n=4} in the $N
\to \infty$ limit could be mapped to the $D=d+2n$-dimensional
noncommutative $U(1)$ gauge theory. The resulting higher-dimensional
$U(1)$ gauge theory has been transformed to a theory of
higher-dimensional gravity describing a dynamical spacetime geometry
according to the isomorphism between the noncommutative
$\star$-algebra $\CA_\theta$ and the algebra ${\rm Der}(\CA_\theta)$
of vector fields. Look at the metric in Eq. \eq{ward-metric}. Definitely,
the extra $2n$-dimensional spacetime is emergent and takes part in
the spacetime geometry. It was not a preexisting spacetime background
in the action in Eq. \eq{n=4}. Instead the theory says that it originated
from the vacuum in Eq. \eq{matrix-vacuum}. One can easily check this fact by turning off
all fluctuations in the metric in Eq. \eq{ward-metric}. The $D$-dimensional
flat spacetime comes from the vacuum configuration in Eq. \eq{matrix-vacuum}
whose vector field is given by $V_A^{({\rm vac})} = (\partial_\mu, \partial_a)$
according to Eq. \eq{D-vector}. Furthermore, the vacuum is a solution of the theory in Eq. \eq{n=4}.
Therefore, the underlying theory in Eq. \eq{n=4} by itself entirely
describes the emergence of the $2n$-dimensional space and its dynamical fluctuations.

Also, note that the original $d$-dimensional spacetime is now
dynamical, not a playground any more, although the original flat
spacetime part $\mathbb{R}^d_C$ was assumed {\it a priori} at the outset.
One can see that the existence of nontrivial gauge field
fluctuations $A_\mu(z)$ causes the curving of $\mathbb{R}^d_C$. Therefore,
the large-$N$ gauge theory in Eq. \eq{n=4} almost provides a background-independent
description of spacetime geometry, except the original background $\mathbb{R}^d_C$.

Now, a question is how to achieve a complete background independence
about the emergent geometry. The answer is simple. We may completely
remove the spacetime $\mathbb{R}^d_C$ from the action in Eq. \eq{n=4} and start
with a theory without spacetime from the beginning.
How to do this operation is well-known in matrix models. This change of
dimensionality appears in matrix theory as the `matrix
T-duality' (see Sec. VI.A in Ref. \ct{taylor}) defined by
\be \la{matrix-t}
iD_\mu = i\partial_\mu + A_\mu \rightleftarrows \Phi^a.
\ee
Applying the matrix T-duality in Eq. \eq{matrix-t} to the action
in Eq. \eq{n=4}, on one hand, one can arrive at the 0-dimensional
IKKT matrix model \ct{ikkt} in the case of the Euclidean signature
\be \la{ikkt}
S_{IKKT} = - \frac{2\pi}{g_s \kappa^2} \Tr \left(
\frac{1}{4}[X^M, X^N][X_M, X_N] \right),
\ee
where $X^M = \kappa \Phi^M$, or the 1-dimensional BFSS matrix model
\ct{bfss} in the case of the Lorentzian signature
\begin{equation} \label{bfss}
S_{BFSS} = - \frac{1}{G_s} \int dt {\rm Tr}
\left(\frac{1}{2} D_0 \Phi^a D^0 \Phi^a - \frac{1}{4} [\Phi^a, \Phi^b]^2
\right).
\end{equation}
On the other hand, one can also go up to $D$-dimensional pure $U(N)$
Yang-Mills theory given by
\be \la{pure-ym}
S_C = -\frac{1}{4 g^2_{YM}} \int d^D X \Tr F_{MN} F^{MN}.
\ee
Note that the $B$-field has completely disappeared; i.e., the
spacetime is commutative. In fact, the T-duality between the theories
defined by Eqs. \eq{higher-action} and \eq{pure-ym} is an analogue of the
Morita equivalence on a noncommutative torus stating that
the noncommutative $U(1)$ gauge theory with rational $\theta = M/N$ is
equivalent to an ordinary $U(N)$ gauge theory \ct{SW-ncft}.

Let us focus on the IKKT matrix model in Eq. \eq{ikkt} because it is
completely background independent because it is 0-dimensional. In
order to define the action in Eq. \eq{ikkt}, it is not necessary to assume
the prior existence of any spacetime structure. There are only a bunch of $N
\times N$ Hermitian matrices $X^M \; (M = 1, \cdots, D)$ that are
subject to a couple of algebraic relations given by
\bea \la{matrix-bianchi}
&& [X^M, [X^N, X^P]] + [X^N, [X^P, X^M]] + [X^P, [X^M, X^N]] = 0, \\
\la{matrix-eom}
&& [X_M, [X^M, X^N]] = 0.
\eea

Suppose that a vacuum of the theory in Eq. \eq{ikkt} in the $N \to
\infty$ limit is given by
\be \la{ikkt-vac}
[X^M, X^N] = i \theta^{MN} \mathbf{1}_{N \times N} = \left(
                                                       \begin{array}{cc}
                                                         0 &  0 \\
                                                         0 & i \theta^{ab} \\
                                                       \end{array}
                                                     \right) \mathbf{1}_{N \times
                                                     N},
\ee
where $\theta^{ab}$ is a constant matrix of rank $2n$. In exactly
the same way as the case for Eq. \eq{n=4}, one can map the $N \times N$
matrices $X^M = (X^\mu, X^a)$ into noncommutative fields according
to the correspondence in Eq. \eq{ncft-matrix}:
\be \la{map-m-ncf}
(X^\mu, X^a)_{N \times N} \quad \mapsto \quad
\kappa \Big( \widehat{\Phi}^\mu(y), \frac{i}{\kappa} \theta^{ab}
\widehat{D}_b(y) \Big) \in \CA_\theta,
\ee
where $\widehat{D}_a(y) \in \CA_\theta$ is given by Eq. \eq{nc-photon}.
It is then straightforward to get a $2n$-dimensional noncommutative
$U(1)$ gauge theory from the matrix action in Eq. \eq{ikkt}:
\be \la{2n-u1}
\widehat{S} = \frac{1}{g_{YM}^2} \int d^{2n} y \sqrt{\widehat{g}}
\Big( \frac{1}{4} \widehat{g}^{ac} \widehat{g}^{bd} (\widehat{F}_{ab} - B_{ab})
\star (\widehat{F}_{cd} - B_{cd}) + \frac{1}{2} \widehat{g}^{ab}
\widehat{D}_a \widehat{\Phi}^\mu \star \widehat{D}_b \widehat{\Phi}^\mu
- \frac{1}{4}[\widehat{\Phi}^\mu, \widehat{\Phi}^\nu]_\star^2 \Big),
\ee
where $g_{YM}^2$ and the metric $\widehat{g}^{ab}$ are defined by
Eqs. \eq{open}, \eq{coupling1} and \eq{coupling2}. If $\theta^{MN}$ in
Eq. \eq{ikkt-vac} is a constant matrix of rank $D = d + 2n$ instead, we
will get a $D$-dimensional noncommutative $U(1)$ gauge theory whose
action is basically the same as Eq. \eq{higher-action} except that it
comes with the Euclidean signature and a constant $B$-field of rank $D$.

In summary, we have scanned both $U(N)$ Yang-Mills theories and
noncommutative $U(1)$ gauge theories in various dimensions and
different $B$-field backgrounds by applying the matrix T-duality
in Eq. \eq{matrix-t} and the correspondence in Eq. \eq{ncft-matrix}.
From the derivation of Eq. \eq{higher-action}, one may notice that
the rank of the $B$-field is equal to the dimension of the emergent space,
which is also equal to the number of adjoint scalar fields $\Phi^a
\in U(N)$. Therefore, the matrix theory in Eq. \eq{n=4} can be defined
in different dimensions by changing the rank of the $B$-field if the
dimension $D$ is fixed, e.g., $D = 10$. On the other hand, we can
change the dimensionality of the theory in Eq. \eq{2n-u1} by changing the
rank of $\theta$ in Eq. \eq{ikkt-vac}. In this way, we can connect every
$U(N)$ Yang-Mills theory and noncommutative $U(1)$ gauge theory
in various dimensions by changing the $B$-field background and
applying the matrix T-duality \eq{matrix-t} and the correspondence
in Eq. \eq{ncft-matrix}. It is really remarkable!

However, there is also a caveat. One can change the dimensionality of
the matrix model by any integer number by using the matrix T-duality
in Eq. \eq{matrix-t} while the rank of the $B$-field can be changed only by
an even number because it is supposed to be symplectic. Hence, it is
not obvious what kind of background can explain a noncommutative
field theory with an odd number of adjoint Higgs fields. A plausible
guess is that either the vacuum is described by a noncommutative
space induced by a Poisson structure, e.g., of Lie algebra type,
i.e., $[X^a, X^b] = i {f^{ab}}_c X^c$, or there is a 3-form
$C_{\mu\nu\rho}$ that reduces to the 2-form $B$ in Eq. \eq{open} by
a circle compactification, so may be of M-theory origin. We will
briefly discuss the Lie algebra case later, but, unfortunately, we
don't know much about how to construct a corresponding
noncommutative field theory with the 3-form background. We leave it
as a future problem.

Some critical aspect of quantum geometry may be encountered with the
following question. What is the emergent geometry derived from the
noncommutative $U(1)$ gauge theory in Eq. \eq{2n-u1} ? One may naively
apply the map in Eq. \eq{D-vector} to the noncommutative fields $\Big(
\widehat{\Phi}^\mu(y), \widehat{D}_a(y) \Big) \in \CA_\theta$.
The fields $\widehat{D}_a(y)$ have no problem because they are exactly
the same as Eq. \eq{nc-photon}. However, the fields $\widehat{\Phi}^\mu(y)$
leads to a bizarre circumstance. From the map in Eq. \eq{D-vector}, we may define
\bea \la{fake}
- i [\widehat{\Phi}^\mu (y), \widehat{f} (y)]_\star &=&
- \theta^{ab} \frac{\p \Phi^\mu (y)}{\partial y^b} \partial_a f(y) + \cdots \xx
&``\equiv"& V^{\mu a} (y) \partial_a f(y) + \cdots.
\eea
We immediately get into trouble if we remember that the fields
$\widehat{\Phi}^\mu(y)$ are purely fluctuations and so the `fake'
vector fields $V^\mu = V^{\mu a} (y) \partial_a$ are not invertible,
in general. For example, they tend to vanish at $|y| \to \infty$.
Recall that a Riemannian metric should be nondegenerate, i.e.,
invertible everywhere. This is not the case for $V^{\mu}$. Therefore,
the fields $\widehat{\Phi}^\mu(y)$ are not yet full-fledged as a
classical geometry although they could define a ``bubbling quantum
geometry". This notable difference between $\widehat{\Phi}^\mu(y)$
and $\widehat{D}_a(y)$ is due to the fact that $\widehat{D}_a(y)$
define fluctuations around the uniform vacuum condensation
in Eq. \eq{matrix-vacuum} while $\widehat{\Phi}^\mu(y)$ define pure
fluctuations around ``nothing", say, without any coherent
condensation in vacuum. Therefore we get a very important picture from the above analysis:

{\it In order to describe a classical geometry from a background
independent theory such as Eq. \eq{ikkt}, it is necessary to have a
nontrivial vacuum defined by a ``coherent" condensation of gauge
fields, e.g., the vacuum defined by Eq. \eq{matrix-vacuum}.}

Here, ``coherent" means that a spacetime vacuum is defined by the
Heisenberg algebra such as quantum harmonic oscillators as in
Eq. \eq{vacuum-spacetime}. Its physical significance will be discussed
later.

Also, note that a symplectic structure $B_{ab} \equiv
(\theta^{-1})_{ab}$ is nowhere vanishing, which can be regarded as a
background field strength of noncommutative gauge fields $A_a^{(0)}
\equiv \langle \widehat{A}^{(0)}_a \rangle_{\rm vac} = - B_{ab} y^b$. In
terms of physicist's language, this means that there is an
(inhomogeneous in general) condensation of gauge fields in vacuum,
i.e.,
\be \la{vacuum-condense}
\langle B_{ab}(x) \rangle_{{\rm vac}} = \theta^{-1}_{ab}(x).
\ee
For a constant symplectic structure for simplicity, rewriting the covariant vectors in
Eq. \eq{nc-photon} as (actually to invoke a renowned Goldstone
boson $\varphi = \langle \varphi \rangle + h$) \cite{foot-zee}
\be \la{vacuum-coordinate}
\widehat{D}_a(y) = - \langle \widehat{A}_a^{(0)} \rangle_{{\rm vac}} +
\widehat{A}_a(y)
\ee
would be suggestive. This naturally suggests some sort of spontaneous symmetry breaking
\ct{hsy-ijmpa} in which $y^a$ are vacuum expectation values of $\widehat{D}_a(y)$,
specifying the background in Eq. \eq{vacuum-condense} as usual, and
$\widehat{A}_b(y)$ are fluctuating dynamical coordinates (fields).
We thus arrived at another important point:

{\it The origin of spacetime with a symplectic or a Poisson structure such as
Eq. \eq{p-bracket} or Eq. \eq{vacuum-spacetime} comes from the coherent
condensation of gauge fields in vacuum.}

\subsection{General Noncommutative Spacetime}

So far, we have mostly considered noncommutative spaces defined by a
canonical symplectic structure. Here, we will explain how it is
possible to generalize emergent gravity to a general noncommutative
spacetime, for example, to the case with a nonconstant symplectic
structure or a generic Poisson structure. General results have been beyond
our reach up to now. Thus, we will be brief about this subject. Readers
may skip this part and might attack the emergent gravity for general cases after
a deeper understanding about the simple cases has been realized.

The question is how to generalize the emergent gravity picture to
the case of a nontrivial vacuum, e.g., Eq. \eq{vacuum-condense},
describing an inhomogeneous condensate of gauge fields. In this case, the Poisson
structure $\Theta^{ab}(x) = (\frac{1}{B})^{ab}(x)$ is not constant,
so the corresponding noncommutative field theory is defined by a nontrivial star-product
\be \la{general-nc}
[Y^a, Y^b]_{\star^\prime} = i \Theta^{ab}(Y)
\ee
where $Y^a$ denote vacuum coordinates, which are designed with the
capital letters to distinguish them from $y^a$ for the constant
vacuum in Eq. \eq{vacuum-spacetime}. The star product $[\widehat{f},
\widehat{g}]_{\star'}$ for $\widehat{f},
\widehat{g} \in {\cal A}_\Theta$ can be perturbatively computed via the
deformation quantization \ct{kontsevich}. There are excellent
earlier works \ct{cornal} especially relevant for the analysis of
the DBI action as a generalized geometry though a concrete
formulation of noncommutative field theories for a general
noncommutative spacetime is still out of reach.

We will mostly focus on the commutative limit so that
\bea \la{general-c-limit}
-i [\widehat{f}, \widehat{g}]_{\star^\prime} &=& \Theta^{ab}(Y)
\frac{\p f(Y)}{\p Y^a} \frac{\p g(Y)}{\p Y^b} + \cdots \xx
&\equiv& \{f,g \}_\Theta + \cdots
\eea
for $\widehat{f}, \widehat{g} \in {\cal A}_\Theta$. Using the
Poisson bracket in Eq. \eq{general-c-limit}, we can similarly realize the
Lie algebra homomophism $C^\infty (M) \to TM: f \mapsto X_f$ between
a Hamiltonian function $f$ and the corresponding Hamiltonian vector
field $X_f$. To be specific, for any given function $f
\in C^\infty (M)$, we can always assign a Hamiltonian vector field $X_f$
defined by $X_f (g) = \{g,f\}_{\Theta}$ with any fixed function $g
\in C^\infty (M)$. Then, the Lie algebra homomophism
\be \la{lie-homo-gen}
X_{\{f,g\}_{\Theta}} = [X_f, X_g]
\ee
still holds as long as the Jacobi identity for the Poisson bracket
$\{f,g\}_{\Theta}(x)$ holds or, equivalently, the Schouten-Nijenhuis
bracket in Eq. \eq{sn-bracket} for the Poisson structure $\Theta^{ab}$ vanishes.

As we discussed in Eq. \eq{star-equiv}, there is a natural automorphism
$D(\hbar)$ that acts on star-products \ct{kontsevich}. In the
commutative limit where $D(\hbar) \approx 1$, we can deduce the
following relation from Eq. \eq{star-equiv}:
\be \la{poisson-equiv}
{\{f,g\}_{\Theta}} = {\{f,g\}_{\theta}}.
\ee
Let us explain what Eq. \eq{poisson-equiv} means. For $f=Y^a(y)$ and
$g=Y^b(y)$, Eq. \eq{poisson-equiv} implies that
\be \la{two-theta}
\Theta^{ab}(Y) = \theta^{cd} \frac{\p Y^a}{\p y^c} \frac{\p Y^b}{\p y^d},
\ee
whose statement is, of course, equivalent to the Darboux
transformation in Eq. \eq{darboux-tr}. Also, notice that
Eq. \eq{poisson-equiv} defines diffeomorphisms between vector fields
$X_f^\prime (g) \equiv
\{g,f\}_{\Theta}$ and $X_f (g) \equiv \{g,f\}_{\theta}$ such that
\be \la{poisson-diffeo}
{X^\prime_f}^a = \frac{\p Y^a}{\p y^b} X^b_f.
\ee
Indeed, the automorphism in Eq. \eq{star-equiv} corresponds to a global
statement that the two star-products involved are cohomologically
equivalent in the sense that they generate the same Hochschild
cohomology \ct{kontsevich}.

In order to understand the origin of the nontrivial star product
in Eq. \eq{general-nc}, let us look at the background independent action in Eq. \eq{ikkt}.
As we pointed out, a particular vacuum such as the one in Eq. \eq{vacuum-condense}
should be defined by the theory itself as a solution of the
equations of motion, Eq. \eq{matrix-eom}. Of course, there are infinitely
many solutions. The constant background in Eq. \eq{ikkt-vac} is just one of
them, so let us consider another background
\be \la{inst-vac}
[X^M, X^N] = \left(
\begin{array}{cc}
0 &  0 \\
0 & i (\theta - \theta \widehat{F} \theta)^{ab} \\
 \end{array}
  \right) \mathbf{1}_{N \times N}.
\ee
Using the property in Eq. \eq{poisson-equiv}, one can infer that the above
background can be made equivalent to Eq. \eq{general-nc} by using the
identification $X^a := Y^a = y^a + \theta^{ab} \widehat{A}_b(y)$ and
$\Theta^{ab}(y) = (\theta - \theta \widehat{F}(y) \theta)^{ab}$. If
$\widehat{F}_{ab}(y)$ simply satisfy Eq. \eq{high-sd}, which provides
a very ample class of solutions, the background in Eq. \eq{inst-vac} is a consistent solution
of the theory in Eq. \eq{ikkt}. For example,
the vacuum in Eq. \eq{inst-vac} in four dimensions $(n=2)$ corresponds to
the noncommutative instanton background. In this case, the vacuum
manifold determined by background gauge fields is a hyper-K\"ahler manifold.

Therefore, we may understand that the nontrivial star product
in Eq. \eq{general-nc} results from an inhomogeneous condensation of gauge
fields on the constant vacuum in Eq. \eq{ikkt-vac}. This observation can be
applied to the identity in Eq. \eq{mirror} in a very interesting way. Let
us decompose the nontrivial B-field in Eq. \eq{vacuum-condense} as
\be \la{back-b-bar}
B_{ab}(x) = (\bar{B} + \bar{F} (x))_{ab},
\ee
where $\bar{B}_{ab}= \big({\theta}^{-1}\big)_{ab}$ describes a
constant background such as the one in Eq. \eq{vacuum-spacetime} while $
\bar{F}(x) = d \bar{A}(x)$ describes an inhomogeneous condensate of gauge fields.
Then, the left-hand side of Eq. \eq{mirror} is of the form $g + \kappa
(\bar{B} + \mathcal{F})$, where $\mathcal{F} = d \mathcal{A}$ with
$\mathcal{A}(x) = \bar{A}(x) + A(x)$. It should be completely
conceivable that it can be mapped to a noncommutative gauge theory
of the gauge field $\mathcal{A}(x)$ in the constant $\bar{B}$-field
background according to the Seiberg-Witten equivalence
\ct{SW-ncft}. Let us denote the corresponding noncommutative gauge field as
$\widehat{\mathcal{A}}_a \equiv \widehat{\bar{A}}_a +
\widehat{A}_a$. The only notable point is that the gauge field
$\widehat{\mathcal{A}}_a$ contains an inhomogeneous background part
$\widehat{\bar{A}}_a$. This situation is, of course, analogous to an
instanton (or soliton) background in gauge theory, as we remarked
before.

Thus, everything will go parallel with the constant case. We will
consider a general situation in the context of the action
in Eq. \eq{higher-action}, where background gauge fields are given by
$\widehat{\bar{A}}_\mu (z,y)$ as well as $\widehat{\bar{A}}_b (z,y)$,
which also depend on the commutative coordinates $z^\mu$. Let us
introduce the following covariant coordinates:
\bea \la{cov-coord-gen}
\widehat{X}^a(z, y) &=& y^a + \theta^{ab} \widehat{\mathcal{A}}_b (z, y) = y^a +
\theta^{ab} \widehat{\bar{A}}_b (z,y) + \theta^{ab} \widehat{A}_b (z, y) \xx
& \equiv & Y^a (z,y) + \theta^{ab} \widehat{A}_b (z, y),
\eea
where we identified the vacuum coordinates $Y^a$ in Eq. \eq{general-nc}
because we have to recover them after turning off the fluctuation
$\widehat{A}_a$. Also, introduce the covariant derivatives
\bea \la{cov-gder}
\widehat{D}_\mu (z, y) &=& \p_\mu - i \widehat{\mathcal{A}}_\mu (z, y) =
\p_\mu - i \widehat{\bar{A}}_\mu (z,y) - i \widehat{A}_\mu (z, y)
\xx &\equiv & \widehat{\bar{D}}_\mu (z, y) - i \widehat{A}_\mu (z, y).
\eea
Then, the covariant derivatives in Eq. \eq{decomp-cov} can be defined in
exactly the same way:
\be \la{cov-der-gen}
\widehat{D}_A = \p_A - i \widehat{\mathcal{A}}_A(z, y) = (\widehat{D}_\mu, - i \bar{B}_{ab}
\widehat{X}^b)(z, y),
\ee
where $\p_A = (\p_\mu, - i \bar{B}_{ab} y^b)$. Now, the
noncommutative fields $\widehat{D}_A$ in Eq. \eq{cov-der-gen} can be
mapped to vector fields using Eq. \eq{D-vector}.

Because the results in Section III.B can be applied to arbitrary
noncommutative gauge fields in a constant $B$-field, the same
formulae can be applied to the present case with the understanding
that the vector fields $V_A$ in Eq. \eq{D-vector} refer to total gauge
fields including the inhomogeneous background. This means that the
vector fields $V_A = \lambda E_A \in \Gamma(TM)$ reduce to
$\bar{V}_A =
\bar{\lambda} \bar{E}_A$ after turning off the fluctuations,
where $\bar{V}_A$ is determined by the background $(\p_\mu - i
\widehat{\bar{A}}_\mu (z,y), - i
\bar{B}_{ab} Y^b(z,y))$ and $\bar{\lambda}$ satisfies the relation
\be \la{back-lambda}
{\bar{\lambda}}^2 = \nu(\bar{V}_1, \cdots, \bar{V}_D).
\ee
Therefore, the $D$-dimensional metric is precisely given by
Eq. \eq{ward-metric} with $\mathbf{A}^a = \mathcal{A}_\mu^a dz^\mu$,
and the metric for the background is given by
\bea \la{D-back-metric}
ds^2 &=& \eta_{AB} \bar{E}^A \otimes \bar{E}^B \xx &=&
{\bar{\lambda}}^2
\eta_{AB} \bar{V}^A \otimes \bar{V}^B =
{\bar{\lambda}}^2 \eta_{AB} \bar{V}^A_M \bar{V}^B_N
\; dX^M \otimes dX^N.
\eea
Here, we have implicitly assumed that the background $\bar{V}_A$
satisfies Eqs. \eq{Dmap-jacobi} and \eq{Dmap-eom}. In four dimensions,
for instance, we know that the metric in Eq. \eq{D-back-metric} describes
Ricci-flat K\"ahler manifolds if $\bar{V}_A$ satisfies the self-duality
equation, Eq. \eq{iso-map}.

Now, let us look at the picture of the right-hand side of
Eq. \eq{mirror}. After applying the Darboux transform, Eq. \eq{darboux-tr},
only for the symplectic structure, Eq. \eq{back-b-bar}, and
leaving the fluctuations intact, the right-hand side becomes of the
form $G_{ab}(y) + \kappa (\bar{B}_{ab} + \mathfrak{F}_{ab}(y))$, where
\be \la{f-darboux}
\mathfrak{F}_{ab}(y) = \frac{\p x^\alpha}{\p y^a} \frac{\p x^\beta}{\p y^b}
F_{\alpha\beta}(x) \equiv \p_a \mathfrak{A}_b (y) - \p_b
\mathfrak{A}_a (y),
\ee
and the metric $G_{ab}(y)$ is given by Eq. \eq{gauge-metric}. Note that
in this picture, the gauge fields $\mathfrak{A}_a (y)$ are regarded
as fluctuations propagating in the background $G_{ab}(y)$ and
$\bar{B}_{ab}$. Therefore, it would be reasonable to interpret the
right-hand side of Eq. \eq{mirror} as a noncommutative gauge theory
of the gauge field $\mathfrak{A}_a (y)$ defined by the canonical
noncommutative space in Eq. \eq{vacuum-spacetime}, but in a curved space
described by the metric $G_{ab}(y)$.

Although the formulation of noncommutative field theory in a generic
curved spacetime is still a challenging problem, there is no
obstacle to formulating emergent gravity if one is confined to the
commutative limit. Because the inhomogeneous condensate of gauge
fields in the vacuum \eq{back-b-bar} now appears as an explicit
background metric, the metric in Eq. \eq{ward-metric} in this picture will
be replaced by
\bea \la{D-darboux-metric}
ds^2 &=& h_{AB} E^A \otimes E^B \xx &=& \Lambda^2 h_{AB} V^A
\otimes V^B =  \Lambda^2 h_{AB}
V^A_M V^B_N \; dX^M \otimes dX^N,
\eea
where $h_{AB}$ is the metric in the space spanned by noncoordinate
bases $V_A = \Lambda E_A$ \ct{cho-freund}. Because the metric
in Eq. \eq{D-darboux-metric} has the Riemannian volume form $\nu_g =
\sqrt{-h}  E^1 \wedge \cdots \wedge E^D$ instead of Eq. \eq{E-volume-form},
the volume form $\nu = \Lambda^{(2-D)}\nu_g$ in Eq. \eq{D-volume-form}
will be given by
\be \la{D-curved-volume-form}
\nu =  \sqrt{-h} \Lambda^2 V^1 \wedge
\cdots \wedge V^D.
\ee
Thus, the function $\Lambda$ in Eq. \eq{D-darboux-metric} will be
determined by the condition
\be \la{D-volume-curved}
\sqrt{-h} \Lambda^2 = \nu (V_1, \cdots, V_D).
\ee
Because the anholonomic basis $V^A$ in Eq. \eq{D-darboux-metric} will
become flat when fluctuations are turned off, i.e.,
$\mathfrak{F}_{ab} = 0$, the background metric in this picture is
simply given by
\be \la{D-back}
ds^2 = \overline{\Lambda}^2 h_{MN} \; dX^M \otimes dX^N,
\ee
where $\overline{\Lambda}^2 = 1/\sqrt{-h}$.

As usual, the torsion-free condition, Eq. \eq{torsion-free}, for the metric in Eq. \eq{D-darboux-metric}
will be imposed to get the relation in Eq. \eq{spin-structure} in which $\omega_{ABC} = h_{BD}{{\omega_A}^D}_C$
and $f_{ABC} = h_{CD}{f_{AB}}^D$, where the indices $A,B,
\cdots$ are raised and lowered using the metric $h_{AB}$.
Because $h_{AB}$ is not a flat metric, ${{\omega_A}^B}_C$ in
Eq. \eq{D-spin1} or Eq. \eq{D-spin2} will actually be the Levi-Civita
connection in noncoordinate bases rather than a spin connection, but
we will keep the notation for convenience. Also, the condition that
the metric in Eq. \eq{D-darboux-metric} be covariantly constant, i.e.,
$\nabla_C \Bigl( h_{AB} E^A \otimes E^B \Bigr) = 0$, leads to the
relation \ct{cho-freund}
\be \la{metric-condition-curved}
\omega_{ABC} = \half \big(E_A h_{BC} - E_B h_{CA} + E_C h_{AB} \big)
+ \half \big(f_{ABC} - f_{BCA} + f_{CAB} \big).
\ee
The curvature tensors have exactly the same form as Eq. \eq{D-riemann}.
All the calculation in Section II.C can be repeated in the same way
even for this case although the details will be much more
complicated and have not been performed so far.
By comparing the two metrics, Eqs. \eq{D-back-metric} and
\eq{D-back}, we finally get the following relations \ct{hsy-jhep}:
\be \la{2-metric}
h_{MN} = \eta_{AB} \bar{V}^A_M \bar{V}^B_N, \qquad
\overline{\Lambda}^2 = {\bar{\lambda}}^2  = \frac{1}{\sqrt{-h}},
\ee
which is, of course, consistent with our earlier observation.

One may wonder whether the emergent gravity for symplectic
structures can be smoothly taken over to the case where a symplectic
structure is not available. It was shown in Ref. \ct{hsy-siva} that emergent
gravity can nicely be generalized to a Poisson manifold $(M, \pi)$.
A Poisson manifold $M$ is a differentiable manifold $M$ equipped
with a bivector field (not necessarily nondegenerate) $\pi = \pi^{\mu\nu} \p_\mu \wedge \p_\nu \in
\Gamma(\Lambda^2 TM)$ which defines an $\mathbb{R}$-bilinear antisymmetric operation
$\{ \cdot , \cdot \}_\pi :C^\infty(M) \times C^\infty(M) \to C^\infty(M)$ by
\be \la{poisson-bracket}
(f,g) \mapsto \{f,g\}_\pi = \langle \pi, df
\otimes dg \rangle = \pi^{\mu\nu}(x) \p_\mu f(x) \p_\nu g(x).
\ee
The bracket, called the Poisson bracket, satisfies
\bea \la{p-leibniz}
&&  1) \; {\rm Leibniz \; rule}: \{f,gh\}_\pi = g
\{f,h\}_\pi + \{f,g\}_\pi h, \\
\la{p-jacobi}
&& 2) \; {\rm Jacobi \; identity}: \{f,\{g,h\}_\pi\}_\pi +
\{g,\{h,f\}_\pi\}_\pi + \{h,\{f,g\}_\pi\}_\pi = 0,
\eea
$\forall f,g,h \in C^\infty(M)$. Poisson manifolds appear as a
natural generalization of symplectic manifolds where the Poisson
structure reduces to a symplectic structure if $\pi$ is nongenerate
\ct{mechanics}.

One can show that the Jacobi identity in Eq. \eq{p-jacobi} for the bracket
$\{ \cdot , \cdot \}_\pi$ is equivalent to the condition that the
Schouten-Nijenhuis bracket \ct{vaisman} for the Poisson tensor $\pi$
vanishes, i.e.,
\be \la{sn-bracket}
 [\pi, \pi]_{SN} \equiv \Big( \pi^{\lambda \mu} \frac{\partial
 \pi^{\nu\rho}}{\partial x^\lambda} +
\pi^{\lambda\nu} \frac{\partial \pi^{\rho\mu}}{\partial x^\lambda} + \pi^{\lambda\rho}
\frac{\partial \pi^{\mu\nu}}{\partial x^\lambda} \Big) \frac{\partial}{\partial x^\mu}
\wedge \frac{\partial}{\partial x^\nu} \wedge \frac{\partial}{\partial
x^\rho} = 0.
\ee
Like the Darboux theorem in symplectic geometry, the Poisson
geometry also enjoys a similar property known as the splitting
theorem proven by Weinstein \ct{weinstein}. The splitting theorem
states that a $d$-dimensional Poisson manifold is locally equivalent
to the product of $\mathbb{R}^{2n}$ equipped with the canonical symplectic
structure with $\mathbb{R}^{d-2n}$ equipped with a Poisson structure of
rank zero at the origin. That is, the Poisson manifold $(M, \pi)$ is
locally isomorphic (in a neighborhood of $x \in M$) to the direct
product $S \times N$ of a symplectic manifold $(S, \sum_{i=1}^n dq^i
\wedge dp_i)$ with a Poisson manifold $(N_x, \{\cdot, \cdot\}_N)$
whose Poisson tensor vanishes at $x$.

A well-known example of a Poisson manifold is four-sphere where no
symplectic structure is available. If $M$ is a compact symplectic
manifold, the second de Rham cohomology group $H^2(M)$ is nontrivial,
so the only $n$-sphere that admits a symplectic form is the
2-sphere. For example, let ${\bf S}^4 =
\{(u, v, t) \in \mathbb{C} \times \mathbb{C} \times \mathbb{R}: |u|^2 + |v|^2 =
t(2-t) \}$. Then, the bivector field $\pi = u v
\partial_u \wedge \partial_v - u v^* \partial_u \wedge
\partial_{v^*} - u^* v \partial_{u^*} \wedge \partial_v
+ u^* v^* \partial_{u^*} \wedge \partial_{v^*}$ is a Poisson tensor,
that is, $[\pi, \pi]_{SN} =0$, and $\pi \wedge \pi = 4 |u|^2 |v|^2
\partial_u \wedge \partial_{v} \wedge
\partial_{u^*} \wedge \partial_{v^*}$.
Therefore, the Poisson tensor $\pi$ vanishes on a subspace of either
$u = 0$ or $v =0$, so the Poisson structure becomes degenerate there.
In this case, we have to rely on a Poisson structure to formulate an emergent gravity \ct{hsy-siva}.

The Poisson tensor $\pi$ of a Poisson manifold $M$ induces a bundle
map $\pi^\sharp: T^* M \to TM$ by
\be \la{anchor}
A \mapsto \pi^\sharp(A) = \pi^{\mu\nu}(x) A_\mu(x)
\frac{\partial}{\partial x^\nu}
\ee
for $A = A_\mu(x) dx^\mu \in T_x^* M$, which is called the anchor
map of $\pi$ \ct{vaisman}. See also Section VI. The rank of the
Poisson structure at a point $x \in M$ is defined as the rank of the
anchor map at that point. If the rank equals the dimension of the
manifold at each point, the Poisson structure reduces to a
symplectic structure, which is also called nondegenerate. The
nondegenerate Poisson structure uniquely determines the symplectic
structure defined by a 2-form $\omega = \half \omega_{\mu\nu}(x) d
x^\mu \wedge d x^\nu = \pi^{-1}$, and the condition in Eq. \eq{sn-bracket} is
equivalent to the statement that the 2-form $\omega$ is closed,
$d\omega = 0$. In this case, the anchor map $\pi^\sharp: T^* M \to
TM$ becomes a bundle isomorphism, as we discussed in Section I.

To define a Hamiltonian vector field $\pi^\sharp(df)$ of a smooth
function $f \in C^\infty(M)$, what one really needs is a Poisson
structure that reduces to a symplectic structure for the
nondegenerate case. A Hamiltonian vector field $X_f = -
\pi^\sharp(df)$ for a smooth function $f \in C^\infty(M)$ is defined
by the anchor map in Eq. \eq{anchor} as follows:
\be \la{ham-vec}
X_f(g) = - \langle \pi, df \otimes dg \rangle  =
\{g, f\}_\pi = \pi^{\mu\nu}(x)
\frac{\partial f}{\partial x^\nu} \frac{\partial g}{\partial x^\mu}.
\ee
Given a smooth Poisson manifold $(M, \pi)$, the map $f \mapsto X_f =
- \pi^\sharp(df)$ is a homomorphism  \ct{vaisman} from the Lie
algebra $C^\infty(M)$ of smooth functions under the Poisson bracket
to the Lie algebra of smooth vector fields under the Lie bracket. In
other words, the Lie algebra homomorphism in Eq. \eq{poisson-lie} is still
true even for any Poisson manifold.

As we just noticed, it is enough to have a Poisson structure to
achieve the map $C^\infty(M) \to \Gamma(TM): f \mapsto X_f = -
\pi^\sharp(df)$ such as Eq. \eq{duality}. As we discussed earlier,
any Poisson manifold can be quantized via deformation quantization
\ct{kontsevich}:
\be \la{pnc-space}
\{ x^\mu, x^\nu \}_\pi = \pi^{\mu\nu}(x) \;\; \to \;\;
[\widehat{x}^\mu, \widehat{x}^\nu]_{\widetilde{\star}} = i
\kappa \, \widehat{\pi}^{\mu\nu}(\widehat{x}),
\ee
where we introduced a deformation parameter $\kappa$ of
$(\rm{length})^2$ and $\widehat{\pi}^{ab}(\widehat{x}) \in
\CA_\pi$ are assumed to be dimensionless operators.
Therefore, the anchor map in Eq. \eq{ham-vec} can be lifted to a
noncommutative manifold as in Eq. \eq{nc-vector},
\be \la{nc-anchor}
\widehat{V}_a [\widehat{f}](x) \equiv - i
[\widehat{D}_a(x), \widehat{f}(x)]_{\widetilde{\star}},
\ee
for any noncommutative field $\widehat{D}_a(x) \in \CA_\pi$
(dropping the hat in the coordinates $\widehat{x}^\mu \in \CA_\pi$
for simple notation). Then, everything will go exactly parallel with
the symplectic case if we define emergent quantum gravity from a
gauge theory defined on the noncommutative space in Eq. \eq{pnc-space} with
the generalized vector fields in Eq. \eq{nc-anchor}. It was studied in
Ref. \ct{hsy-siva} how a fuzzy Poisson manifold can be derived from a mass deformed
matrix model, from which the picture of emergent gravity was checked.

\section{EMERGENT MATTER}

We have stressed that quantum gravity should be background
independent where no kind of spacetime structure is assumed.
Only morphisms between objects need to be postulate. An
underlying theory, for example, only has matrices (as objects) that
are subject to some algebraic relations such as the Jacobi identity
and the equations of motion (as morphisms). However, we can derive a
spacetime geometry from these algebraic relations between objects by
mapping the matrix algebra to a Poisson or noncommutative
$\star$-algebra $\CA_\theta$ and then deriving the algebra ${\rm
Der}(\CA_\theta)$ of $\CA_\theta$. We observed that such an operator
algebra, e.g., $\star$-algebra, can be defined by using noncommutative
gauge fields and that a smooth geometry emerges from them in a macroscopic
world. Depending on the choice of an algebraic relation, we get a
different geometry. In this scheme, the geometry is a derived
concept defined by the algebra \ct{connes}. In a deep noncommutative
space, a smooth geometry is doomed; instead, an algebra between
objects becomes more fundamental. Ergo, the motto of emergent
gravity is that an algebra defines a geometry. One has to specify an
underlying algebra to talk about a corresponding geometry.

As a recitation, the emergence of gravity necessarily requires the
emergence of spacetime itself. If spacetime is emergent, then all
fields supported on this spacetime must be emergent too. Somehow,
matter fields and other non-Abelian gauge fields for weak and strong
forces must be emergent together with spacetime. How is this possible?
How are matter fields describing quarks and leptons to be defined in the
context of emergent geometry?

We may start with a naive reasoning. First, note that translations in
noncommutative directions are an inner automorphism of the
noncommutative $\star$-algebra ${\cal A}_\theta$ generated by the
coordinates in Eq. \eq{vacuum-spacetime}:
\be \la{nc*-algebra}
e^{-i k^a B_{ab}y^b} \star \widehat{f}(y) \star e^{i k^a B_{ab}y^b}
= \widehat{f}(y + k)
\ee
for any $\widehat{f}(y)\in {\cal A}_\theta$. The inner automorphism
in Eq. \eq{nc*-algebra} is nontrivial only in the case of a noncommutative
algebra \ct{nc-review}; that is, commutative algebras do not possess
any inner automorphism, so all ``points" in noncommutative space
are indistinguishable, i.e., unitarily equivalent while all points
in commutative space are distinguishable, i.e., unitarily
inequivalent. As a result, one loses the meaning of ``point" in
noncommutative space. Hence, the concept of ``particle" becomes
ambiguous, too. Thus, before matter fields, first we may address the question:
What is a particle in noncommutative spacetime?

When a space becomes noncommutative, there is a Hilbert space $\CH$
associated with the space such as Eq. \eq{fock}, so a point or a
particle may be replaced by a state in $\CH$. Then, the most natural
concept of a particle in noncommutative space may be a localized
state in $\CH$. However, because the Hilbert space $\CH$ is a complex
vector space as usual, such a localized state will tend to be
dissipative due to a linear superposition between nearby states.
Therefore, the most natural and pertinent concept of a particle in
noncommutative space may be a {\it stable localized state} in $\CH$.
This means \ct{hsy-jhep} that a particle may be realized as a
topological object in the noncommutative $\star$-algebra $\CA_\theta$.

As illustrated by quantum mechanics, noncommutative algebras admit a
much greater variety of algebraic and topological structures
compared to commutative ones. Likewise, when spacetime at
a fundamental level is replaced by a noncommutative algebra, algebraic
and topological structures in the noncommutative spacetime actually
become extremely rich and coherent \ct{connes}, which would, we
guess, be responsible for emergent properties such as
diffeomorphisms, gauge symmetries and matter fields.

This line of thought is our naive reasoning about how to realize a
particle or matter field in noncommutative spacetime. We think this
idea should direct us to a reasonable track, but an involved math is often
abstruse. Therefore, we will try to get more insights from physics.

\subsection{Feynman's View on Electrodynamics}

In a very charming paper \ct{feynman}, Dyson explains the Feynman's
view about the electrodynamics of a charged particle. Feynman starts
with an assumption that a particle exists with position $x^i$ and
velocity $\dot{x}_i$ satisfying commutation relations
\be \la{feynman-comm}
[x^i, x^k] = 0, \qquad m [x^i, \dot{x}_k] = i \hbar\delta^i_k.
\ee
Then, he asks a question: What is the most general form of forces
appearing in Newton's equation $m \ddot{x}_i = F_i(x,\dot{x},t)$
consistent with the commutation relation in Eq. \eq{feynman-comm}?
Remarkably, he ends up with the electromagnetic force
\be \la{force-law-time}
m \frac{d\mathbf{v}}{dt} = e \big( \mathbf{E} + \mathbf{v} \times
\mathbf{B} \big).
\ee
In a sense, Feynman's result is a no-go theorem for the
consistent interaction of particles in quantum mechanics. It turns
out that the conditions in Eq. \eq{feynman-comm} are restrictive enough
that only the electromagnetic force in Eq. \eq{force-law-time} is
compatible with them.

We here reproduce his argument with a puny refinement. We will start
with the Feynman's assumption, together with the Hamilton's equation
\be \la{hamilton-eq}
\frac{df}{dt} = \frac{i}{\hbar} [H, f] + \frac{\p f}{\p t},
\ee
where $f = f(x,p,t), \; H = H(x,p,t) \in \CA_\hbar$ and $\dot{x}_i
\equiv \dot{x}_i(x,p)$. However, we will not assume Newton's equation $m
\ddot{x}_i = F_i(x,\dot{x},t)$. To be precise, we replaced
Newton's equation by Eq. \eq{hamilton-eq}, i.e.,
\be \la{newton-eq}
m \frac{d\dot{x}_i}{dt} = \frac{i m}{\hbar} [H, \dot{x}_i] \equiv
F_i(x,p,t).
\ee
First, consider the following commutator:
\bea \la{comm-1}
[H, [x^i, \dot{x}_k]] &=& [x^i, [H, \dot{x}_k]] - [\dot{x}_k, [H,
x^i]]
\xx &=& -i
\hbar \Big(\frac{1}{m}[x^i, F_k] +  [\dot{x}_i, \dot{x}_k] \Big) = 0.
\eea
The Jacobi identity $[x^l, [\dot{x}_i, \dot{x}_k]] + [\dot{x}_i,
[\dot{x}_k, x^l]] + [\dot{x}_k, [x^l, \dot{x}_i]] = [x^l,
[\dot{x}_i, \dot{x}_k]] = 0$ with Eq. \eq{comm-1} implies
\be \la{comm-2}
[x^l, [x^i, F_k]] = 0.
\ee
Equation \eq{comm-1} also implies $[x^i, F_k] + [x^k, F_i] = 0$, so we may write
\be \la{comm-3}
[x^i, F_k] = - \frac{i \hbar}{m} \varepsilon^{ikl} B_l.
\ee
Equation \eq{comm-3} is the definition of the field $B_l =
B_l(x,p,t) \in \CA_\hbar$, but Eq. \eq{comm-2} says
\be \la{comm-4}
[x^l, B_m] = 0,
\ee
which means that $B_m$ is a function of $x$ and $t$ only, i.e., $B_m
= B_m(x,t)$. Then, we can solve Eq. \eq{comm-3} with
\be \la{comm-5}
F_i(x,p,t) = E_i(x,t) +  \varepsilon^{ikl} \big\langle \dot{x}_k
B_l(x,t) \big\rangle,
\ee
where $E_i(x,t) \in \CA_\hbar $ is an arbitrary function that also
depends on $x$ and $t$ only and the symbol $\langle \cdots \rangle$
denotes the Weyl-ordering, i.e., the complete symmetrization of
operator products.

Combining Eqs. \eq{comm-1} and \eq{comm-3} leads to
\be \la{comm-6}
B_l = - \frac{im^2}{2 \hbar} \varepsilon^{lik} [\dot{x}_i,
\dot{x}_k].
\ee
Another Jacobi identity $\varepsilon^{ijk}[\dot{x}_i, [\dot{x}_j,
\dot{x}_k]] = 0$ then implies
\be \la{comm-7}
[\dot{x}_i, B_i] = - \frac{i \hbar}{m} \Big\langle \frac{\p B_i}{\p
x^i}
\Big\rangle = 0.
\ee
Taking the total derivative of Eq. \eq{comm-6} with respect to time gives
\bea \la{comm-8}
\Big\langle \dot{x}_i \frac{\p B_l}{\p x^i} \Big\rangle +
\frac{\p B_l}{\p t} &=& \frac{m^2}{2 \hbar^2} \varepsilon^{lik} [H,
[\dot{x}_i, \dot{x}_k]] \xx &=& \frac{im}{\hbar} \varepsilon^{lik}
[\dot{x}_k, F_i]
\xx &=& \frac{im}{\hbar} \Big( - \varepsilon^{lik} [\dot{x}_i,E_k] -
[\dot{x}_i, \dot{x}_l] B_i - \dot{x}_l [\dot{x}_i, B_i] +
 \big\langle \dot{x}_i [\dot{x}_i, B_l] \big\rangle \Big) \xx &=& - \varepsilon^{lik}
\Big\langle \frac{\p E_k}{\p x^i}
\Big\rangle + \frac{1}{m}\varepsilon^{lik} B_i B_k
+ \Big\langle \dot{x}_i \frac{\p B_l}{\p x^i} \Big\rangle \xx &=& -
\varepsilon^{lik} \Big\langle \frac{\p E_k}{\p x^i}
\Big\rangle + \Big\langle \dot{x}_i \frac{\p B_l}{\p x^i}
\Big\rangle.
\eea
From Eq. \eq{comm-8}, we finally get
\be \la{comm-9}
\frac{\p B_l}{\p t} + \varepsilon^{lik} \Big\langle \frac{\p E_k}{\p x^i}
\Big\rangle = 0.
\ee

We arrived at the force in Eq. \eq{comm-5} (by using the definition
in Eq. \eq{newton-eq}), where the fields $E_i(x,t)$ and $B_i(x,t)$ should
satisfy Eqs. \eq{comm-7} and \eq{comm-9}. We immediately recognize that
they are electromagnetic fields. Therefore, we get a remarkable
result \ct{feynman} that the Lorentz force in Eq. \eq{force-law-time} is
the only consistent interaction with a quantum particle satisfying the
commutation relations in Eq. \eq{feynman-comm}. Remember that we have
only assumed the commutation relation \eq{feynman-comm} and have only
used the Hamilton's equation in Eq. \eq{hamilton-eq} and the Jacobi
identity to find a consistent interaction of quantum particles. However,
we could get only the electromagnetic force in Eq. \eq{force-law-time}.
What a surprise (at least to us)!

Feynman's observation raises a curious question. We know that, beside the electromagnetic force,
other interactions, weak and strong forces, exist in Nature. Thus, the question
is how to incorporate the weak and the strong forces into Feynman's scheme.
Because he started only with very natural axioms,
there seems to be no room to relax his postulates to include the
weak and the strong forces except by introducing extra dimensions.
Surprisingly, it works with extra dimensions!

Consider a particle motion defined on $\mathbb{R}^3 \times F$ with an
internal space $F$ whose coordinates are $\{x^i: i=1,2,3 \} \in
\mathbb{R}^3$ and $\{Q^I: I = 1,\cdots, n^2-1 \} \in F$.
The dynamics of the particle carrying an internal charge in $F$
\ct{crlee,feynman-gen} is defined by a symplectic structure on $T^*
\mathbf{R}^3 \times F$ whose commutation relations are given by
\bea \la{na-comm-1}
&& [x^i, x^k] = 0, \qquad m [x^i, \dot{x}_k] = i \hbar \delta^i_k, \\
\la{na-comm-2}
&& [Q^I, Q^J] = i \hbar {f^{IJ}}_K Q^K, \\
\la{na-comm-3}
&& [x^i, Q^I] = 0.
\eea
Note that the internal space $F$ is a Poisson manifold $(F, \pi)$
whose Poisson structure is given by $\pi = \frac{1}{2} \pi^{IJ} \p_I
\wedge \p_J = \frac{1}{2} {f^{IJ}}_K Q^K \p_I \wedge \p_J$
and defines the $SU(n)$ Lie algebra in Eq. \eq{na-comm-2}. That is, by
Eq. \eq{sn-bracket},
\be \la{sn-jacobi}
[\pi, \pi]_{SN} = 0 \qquad \Leftrightarrow
\qquad {f^{JK}}_L  {f^{LI}}_M  + {f^{KI}}_L  {f^{LJ}}_M  +
{f^{IJ}}_L {f^{LK}}_M = 0.
\ee
Also, the internal coordinates $Q^I$ are assumed to obey Wong's equation \ct{wong}
\be \la{wong-eq}
\dot{Q}^I + {f^I}_{JK} A_i^J(x,t) Q^K \dot{x}_i = 0.
\ee
Wong's equation just says that the internal charge $Q^I$ is
parallel-transported along the trajectory of a particle under the
influence of the non-Abelian gauge field $A_i^I$.

The geometrical meaning of Wong's equation, Eq. \eq{wong-eq}, can be seen as follows:
Taking the total derivative of Eq. \eq{na-comm-3} with respect to time gives
\be \la{wong1}
[\dot{x}_i, Q^I] = - [x^i, \dot{Q}^I] = \frac{i \hbar}{m} {f^I}_{JK}
A^J_i(x,t) Q^K.
\ee
This property can be used to show the formula for any field
$\phi(x,t) = \phi^I(x,t) Q^I \in \CA_\hbar \times \CG$:
\bea \la{wong2}
[\dot{x}_i, \phi^I(x,t) Q^I] &=& [\dot{x}_i, \phi^I(x,t)] Q^I +
\phi^I(x,t)[\dot{x}_i, Q^I] \xx
&=& - \frac{i\hbar}{m} \big( \p_i \phi^I + {f^I}_{JK} A_i^J \phi^K
\big)Q^I
\xx
&=& - \frac{i\hbar}{m} \big( \p_i \phi - \frac{i}{\hbar} [A_i, \phi]
\big)
\equiv - \frac{i\hbar}{m} D_i \phi.
\eea
Recall that $p_i = m \dot{x}_i + A_i(x,t)$ are translation
generators along $\mathbb{R}^3$, and remember the geometrical meaning of the
Wong's equation, Eq. \eq{wong-eq}, stated above.

Now repeat Feynman's question: What is the most general
interaction of a quantum particle carrying an internal charge
satisfying Eq. \eq{wong-eq} and the commutation relations
in Eqs. \eq{na-comm-1}-\eq{na-comm-3}? The calculation follows almost the
same line \ct{crlee} as that for the electromagnetic force except that the
fields $E_i(x,t) = E_i^I(x,t) Q^I \in \CA_\hbar
\times \mathfrak{g}$ and $B_i(x,t) = B_i^I(x,t) Q^I
\in \CA_\hbar \times \mathfrak{g}$ now carry internal charges in the Lie algebra
$\mathfrak{g}$; thus, Wong's equation, Eq. \eq{wong-eq}, has to be taken into
account. We will not echo the derivation because it is almost
straightforward with a careful Weyl-ordering. That may be a good
exercise for graduate students.

The resulting force exerted on a quantum particle moving in $\mathbb{R}^3
\times F$ is the generalized non-Abelian Lorentz force \ct{crlee}
\be \la{na-lorentz-force}
F_i = E_i + \varepsilon^{ikl} \dot{x}_k B_l,
\ee
where the fields $E_i(x,t) = E_i^I(x,t)Q^I$ and $B_i(x,t) =
B_i^I(x,t) Q^I$ satisfy
\be \la{ym-bianchi}
\p_i B_i - \frac{i}{\hbar} [A_i, B_i] = 0, \qquad \frac{\p B_i}{\p t} +
\varepsilon^{ikl} \Big( \p_k E_l - \frac{i}{\hbar} [A_k, E_l] \Big) = 0.
\ee
The equations in Eq. \eq{ym-bianchi}, of course, can be summarized with the
Lorentz covariant form (in the temporal gauge, $A_0 = 0$) as
\be \la{ym-lbianchi}
\varepsilon^{\mu\nu\rho\sigma} D_\nu F_{\rho\sigma} = 0,
\ee
and the Bianchi identity, Eq. \eq{ym-lbianchi}, can be solved by
introducing non-Abelian gauge fields $A_\mu$ such that
\be \la{ym-fields}
F_{\mu\nu} = \p_\mu A_\nu - \p_\nu A_\mu - \frac{i}{\hbar} [A_\mu,
A_\nu].
\ee
One can check the expression in Eq. \eq{ym-fields}. For example,
one can get
\be \la{na-b}
\varepsilon^{ikl}B_l^I = \p_i A^I_k - \p_k A^I_i + {f^I}_{JK} A_i^J
A_k^K
\ee
by using the Jacobi identity
\be \la{lie-jacobi}
[Q^I, [\dot{x}_i, \dot{x}_k]] + [\dot{x}_i, [\dot{x}_k, Q^I]] +
[\dot{x}_k, [Q^I, \dot{x}_i]] = 0
\ee
together with Eqs. \eq{comm-6} and \eq{wong1} where $B_l = B^I_l Q^I$.

\subsection{Symplectic Geometry Again}

The previous argument by Feynman clearly implies that the fundamental interactions
such as electromagnetic, weak and strong forces can be understood as a symplectic
or Poisson geometry of a particle phase space. Feynman starts with a very natural assumption
about the Poisson structure of a particle interacting with external
forces. In the case of a free particle, Eq. \eq{nc*-algebra} is the
well-known Heisenberg algebra: $[x^i,x^k] = 0, \; [x^i, p_k] = i
\hbar \delta^i_k$. (Note that Feynman and Dyson intentionally use $m\dot{x}_i$
instead of $p_i$.) If some external fields are turned on, then the
particle momentum $m \dot{x}_i$ is no longer equal to $p_i$, but is
shifted by $p_i - A_i$, where $A_i$ are arbitrary external fields.
We easily see that, if the external fields $A_i$ depend only on
$x$ and $t$, i.e., $A_i = A_i(x,t)$, and satisfy Wong's equation, Eq. \eq{wong-eq},
in the non-Abelian cases (to preserve the localizability in Eq. \eq{na-comm-3}),
the commutation relations, Eq. \eq{nc*-algebra}, remain intact, but that is
not the whole story. We have repeatedly used the Jacobi identity of
the algebra $\CA_\hbar$ or $\CA_{\hbar} \times \CG$, which
originally comes from the Poisson algebra in Eq. \eq{p-poisson} of a particle phase
space $P$ or $P \times F$. Recall that the Jacobi identity of Poisson bracket is not
automatically guaranteed. The Schouten-Nijenhuis bracket for the
Poisson tensor should vanish \ct{vaisman}. See Eq. \eq{sn-bracket}.
Therefore, the external fields $A_i$ cannot be completely arbitrary.
They should not ruin the underlying Poisson structure. We know that,
if the Poisson structure is nondegenerate, this condition is
equivalent to the statement that the symplectic 2-form uniquely
determined by the Poisson structure must be closed. See the discussion below Eq. \eq{anchor}.
This is precisely the condition for gauge fields Feynman found.
In gauge theory, it is called the Bianchi identity, e.g., $dF =0$ or $DF=0$.

There is an another beautiful observation \ct{sternberg} (orginally
due to Jean-Marie Souriau) realizing Feynman's idea. Let $(P,
\omega)$ be a symplectic manifold. One can properly choose local
canonical coordinates $y^a
\equiv (x^1, p_1, \cdots, x^n, p_n)$ in $P$ such that the symplectic
structure $\omega$ can be written in the form
\be \la{symplectic-phase}
\omega = \sum_{i = 1}^n dx^i \wedge dp_i.
\ee
Then, $\omega \in \Gamma(\Lambda^2 T^*P)$ can be thought of as a bundle map
$\omega: TP \to T^*P$. Because $\omega$ is nondegenerate at any point
$y \in P$, we can invert this map to obtain the map $\eta \equiv
\omega^{-1}: T^* P \to TP$. This cosymplectic structure $\eta  = \frac{\partial}{\partial x^i} \wedge
\frac{\partial}{\partial p_i} \in \omega \in \Gamma(\Lambda^2 TP)$ is called the Poisson structure of $P$
and defines a Poisson bracket $\{\cdot,\cdot\}_\hbar$. See Section III.D.
In a local chart with coordinates $y^a$, we have
\be \la{poisson-phase}
\{f, g\}_\hbar = \sum_{a,b=1}^{2n} \eta^{ab} \frac{\p f}{\p y^a}
\frac{\p g}{\p y^b}.
\ee

Let $H: P \to \mathbb{R}$ be a smooth function on a Poisson manifold $(P,
\eta)$. The vector field $X_H$ defined by $\iota_{X_H} \omega = dH$
is called a Hamiltonian vector field with the energy function $H$.
We define a dynamical flow by using the differential equation
\ct{mechanics}
\be \la{hamilton-peq}
\frac{df}{dt} = X_H (f) + \frac{\p f}{\p t} = \{f, H\}_\hbar + \frac{\p f}{\p t}.
\ee
A solution to the above equation is a function $f$ such that for any
path $\gamma: [0,1] \to M$, we have
\be \la{hamilton-path}
\frac{df(\gamma(t))}{dt} = \{f, H\}_\hbar(\gamma(t)) + \frac{\p f(\gamma(t))}{\p t}.
\ee

The dynamics of a charged particle in an external static magnetic
field is described by the Hamiltonian
\be \la{particle-hamiltonian}
H = \frac{1}{2m}\big(\mathbf{p}- e \mathbf{A} \big)^2,
\ee
which is obtained from the free Hamiltonian $H_0 =
\frac{\mathbf{p}^2}{2m}$ with the replacement
\be \la{minimal-coupling}
\mathbf{p}  \to \mathbf{p}- e \mathbf{A}.
\ee
Here, the electric charge of an electron is $q_e = - e$, and $e$ is a
coupling constant identified with $g_{YM}$. The symplectic structure
in Eq. \eq{symplectic-phase} leads to the Hamiltonian vector field $X_H$ given by
\be \la{ham-vec-x}
X_H = \frac{\p H}{\p p_i}\frac{\p}{\p x^i} - \frac{\p H}{\p
x^i}\frac{\p}{\p p_i}.
\ee
Then, the Hamilton's equation, Eq. \eq{hamilton-peq}, reduces to the
well-known Lorentz force law
\be \la{force-law}
m \frac{d\mathbf{v}}{dt} = e \mathbf{v} \times \mathbf{B}.
\ee

The observation in Ref. \ct{sternberg} (and of Jean-Marie Souriau) is that the Lorentz force
law in Eq. \eq{force-law} can be derived by keeping the Hamiltonian $H = H_0$, but instead shifting
the symplectic structure
\be \la{soriau}
\omega \to \omega' = \omega - e B,
\ee
where $B = \half B_{ik} (x) dx^i \wedge dx^k$. In this case, the
Hamiltonian vector field $X_H$ is defined by $\iota_{X_H} \omega' =
dH_0$ and given by
\be \la{ham-vec-f}
X_H = \frac{\p H_0}{\p p_i}\frac{\p}{\p x^i} -
\Big( \frac{\p H_0}{\p x^i} - e B_{ik} \frac{\p H_0}{\p p_k} \Big)
\frac{\p}{\p p_i}.
\ee
Then, one can easily check that the Hamilton's equation,
Eq. \eq{hamilton-peq}, with the vector field in Eq. \eq{ham-vec-f} reproduces the
Lorentz force law in Eq. \eq{force-law}. Actually, one can show that the
symplectic structure $\omega'$ in Eq. \eq{soriau} introduces a
noncommutative phase space \ct{nc-review} such that the momentum
space becomes noncommutative, i.e., $[p'_i, p'_j] = - i \hbar eB_{ij}$.

If a particle is interacting with electromagnetic fields, the
influence of the magnetic field $B=dA$ is described by `minimal
coupling', Eq. \eq{minimal-coupling}, and the new momenta $\mathbf{p}' =
-i\hbar (\nabla - i \frac{e}{\hbar} \mathbf{A})$ are covariant under
$U(1)$ gauge transformations. Let us point out that the minimal
coupling can be understood as the Darboux transformation, Eq. \eq{darboux-tr},
between $\omega$ and $\omega'$.
Consider the coordinate transformation $y^a \mapsto x^a(y) = (X^1,
P_1, \cdots, X^n, P_n)(x,p)$ such that
\be \la{darboux-phase}
\sum_{i = 1}^n dx^i \wedge dp_i =
\sum_{i = 1}^n dX^i \wedge dP_i - \frac{e}{2} \sum_{i,j=1}^n B_{ij} (X) dX^i \wedge dX^j,
\ee
with the Hamiltonian being unchanged, i.e., $H_0 =
\frac{\mathbf{P}^2}{2m}$. The condition in Eq. \eq{darboux-phase} is
equivalent to the following equations:
\bea \la{darboux-123}
&& \frac{\p x^i}{\p X^j} \frac{\p p_i}{\p X^k} - \frac{\p x^i}{\p
X^k} \frac{\p p_i}{\p X^j} = - e B_{jk}, \xx && \frac{\p x^i}{\p
X^j} \frac{\p p_i}{\p P_k} - \frac{\p x^i}{\p P_j} \frac{\p p_i}{\p
X^k} = \delta_j^k, \\
&& \frac{\p x^i}{\p P_j} \frac{\p p_i}{\p P_k} - \frac{\p x^i}{\p
P_k} \frac{\p p_i}{\p P_j} = 0. \nonumber
\eea
The above equations are solved by
\be \la{darboux-sol}
x^i = X^i, \qquad p_i = P_i + e A_i(X).
\ee

In summary the dynamics of a charged particle in an electromagnetic
field has two equivalent descriptions \ct{hsy-jhep}:
\be \la{particle-equivalence}
\Big(H=\frac{(\mathbf{p}- e \mathbf{A})^2}{2m}, \omega \Big)(x,p)
\quad \cong \quad \Big(H_0 = \frac{\mathbf{P}^2}{2m}, \omega'= \omega - eB \Big)(X,P).
\ee
The equivalence in Eq. \eq{particle-equivalence} can easily be generalized
to a time-dependent background $A^\mu = (A^0, \mathbf{A})(x,t)$ with
the Hamiltonian $H = \frac{1}{2m}\big(\mathbf{p}- e \mathbf{A}
\big)^2 + eA^0$. The Hamilton's equation, Eq. \eq{hamilton-peq}, in this case
is given by Eq. \eq{force-law-time}. The equivalence
in Eq. \eq{particle-equivalence} now means that the Lorentz force law,
Eq. \eq{force-law-time}, can be obtained by using the Hamiltonian vector field
in Eq. \eq{ham-vec-f} with the Hamiltonian $H_0 =
\frac{\mathbf{p}^2}{2m} + eA^0$ and noticing that the time dependence
of the external fields now appears as the explicit $t$-dependence of
momenta $p_i =p_i(t)$. Indeed, the electric field $\mathbf{E}$
appears as the combination $\mathbf{E} = - \nabla A^0 +
\frac{1}{e} \frac{\p \mathbf{p}}{\p t}$, but note that the coordinates
$(x^i, p_i)$ in Eq. \eq{ham-vec-f} correspond to $(X^i, P_i)$ in the
notation of Eq. \eq{darboux-phase} and so $\frac{\p \mathbf{p}}{\p t} = - e
\frac{\p \mathbf{A}}{\p t}$ by Eq. \eq{darboux-sol}.

Feynman's approach transparently shows that electromagnetism is an inevitable structure
in quantum particle dynamics and that we need an internal space (extra dimensions) to
introduce non-Abelian forces. Furthermore, as emphasized by Dyson
\ct{feynman}, Feynman's formulation also shows that nonrelativistic
Newtonian mechanics and relativistic Maxwell's equations
coexist peacefully. This is due to the underlying symplectic
geometry as Souriau and Sternberg showed \ct{sternberg}.
We know that the Lorentz force, Eq. \eq{force-law-time}, is generated by the minimal coupling $p_\mu \to
\mathfrak{P}_\mu \equiv p_\mu - e A_\mu$ and that the minimal coupling can be encoded
into the deformation of symplectic structure, which can be
summarized as the relativistic form \ct{symplectic-tech}: $\omega = - d \xi \to \omega' = \omega -
e F = - d \big( \xi + e A \big) $, where $\xi = \mathfrak{P}_\mu
dx^\mu$ and $A = A_\mu (x) dx^\mu$. Therefore, the Maxwell equation
$dF=0$ is simply interpreted as the closedness of the symplectic
structure, and the minimal coupling is the Darboux transformation
in Eq. \eq{darboux-tr} from the deformed symplectic structure $\omega' = \omega -
e F$, as was shown in Eq. \eq{darboux-phase}. In this symplectic
formulation of particle dynamics, the gauge symmetry defined by $A
\to A + d\lambda$ is actually symplectomorphisms, i.e.,
diffeomorphisms generated by Hamiltonian vector fields $X_\lambda$
satisfying $\CL_{X_\lambda} \omega =0$. In this sense, the gauge
symmetry is derived from the symplectic or Poisson geometry, so
one may regard the underlying symplectic or Poisson structure as a
more fundamental structure of particle dynamics. Also, one may notice
a great similarity between the symplectic geometries of particles
and spacetime geometry (gravity).

A symplectic formulation of the equations of motion of a particle
was generalized to a Yang-Mills field by Sternberg in
\ct{sternberg} and Weinstein in \ct{weinstein-lmp}.
Let $\pi:P \to M$ be a principal $G$-bundle, and let $F$ be a
Hamiltonian $G$-space. This means that $F$ is a symplectic manifold
with symplectic form $\Omega$ such that $G$ acts on $F$ as a group of
symplectic diffeomorphisms so that there is a homomorphism of the
Lie algebra $\mathfrak{g}$ of $G$ into the algebra of Hamiltonian
vector fields and that we are given a lifting of this homomorphism
to a homomorphism of $\mathfrak{g}$ into the Lie algebra of
functions on $F$, where the Lie algebra structure is given by Poisson
bracket. Thus, to each $\xi \in \mathfrak{g}$, we get a function
$f_\xi$ on $F$ and a Hamiltonian vector field $\xi_F$ on $F$ so that
$\iota_{\xi_F} \Omega = - df_\xi$.

Let $E \subset T^* M \times P$ be the pull-back of $P$ by the
canonical projection $\widetilde{\pi}: T^*M \to M$, i.e., the following diagram commutes:
$$
\xymatrix@+0.5cm{ E
\ar[r]^-{Pr_2}  \ar[d]_{Pr_1}\ & P
\ar[d]^{\pi} \cr
T^* M \ar[r]^-{\widetilde{\pi}} & M \cr } \
$$
Sternberg shows \ct{sternberg} how a connection on $P$ can be used
to put a symplectic structure on the associated bundle $E \times_G F
\to T^* M$ with fiber $F$. Given a Hamiltonian function $H: T^* M
\to \mathbb{R}$, one may pull it back to $E \times_G F$ and thereby obtain a
Hamiltonian flow that represents the motion of a classical particle
under the influence of the field for which the given connection is a
Yang-Mills field. That is, every connection on a principal bundle
$P$ induces a Poisson structure on the associated bundle $E \times_G
F$. The resulting symplectic mechanics of a particle in a Yang-Mills
field is actually equivalent to Feynman's approach in Section IV.A.
More details in terms of the local formula will be discussed elsewhere.

\subsection{Emergent Matters from Stable Geometries}

Now let us pose our original problem about what matter is in
emergent geometry. We speculated that particles or matter fields may
be realized as a topological object in a noncommutative
$\star$-algebra $\CA_\theta$ and, thus, as a stable localized state in a
Hilbert space $\CH$, e.g., the Fock space \eq{fock}.
If so, we can assign the concept of positions and
velocities (as collective variables) to these localized states such
that they satisfy some well-defined (quantum) Poisson algebra, e.g.,
Eqs. \eq{na-comm-1}-\eq{na-comm-3}, which are inherited from the original
noncommutative $\star$-algebra $\CA_\theta$. Here, we will suggest a
plausible picture based on the Fermi-surface scenario in Refs. \ct{horava,volovik},
but we will not insist on our proposal.

Particles are by definition characterized by their positions and
momenta, besides their intrinsic charges, e.g., spin, isospin and an
electric charge. They should be replaced by matter fields in
relativistic quantum theory in order to incorporate pair creation
and pair annihilation. Moreover, in a noncommutative space such as
Eq. \eq{vacuum-spacetime}, the very notion of a point is replaced by a
state in the Hilbert space of Eq. \eq{fock}; thus, the concept of
particles (and matter fields, too) becomes ambiguous, so the
following question should be meaningful and addressed: What is the
most natural notion of a particle or a corresponding matter field in
the noncommutative $\star$-algebra of Eq. \eq{matrix-basis}?
We suggested in Ref. \ct{hsy-jhep} that it should be a K-theory object
in the sense of Ref. \ct{horava}.

We consider the $U(N)$ Yang-Mills theory described by the
action in Eq. \eq{n=4} defined on a $d$-dimensional Minkowski spacetime
$\mathbb{R}_C^d$. As we explained in Section III.C, the theory in Eq. \eq{n=4} can
be related to both $U(N)$ Yang-Mills theories and noncommutative
$U(1)$ gauge theories in various dimensions and different $B$-field
backgrounds by applying the matrix T-duality in Eq. \eq{matrix-t} and the
correspondence in Eq. \eq{ncft-matrix}. Thereby, we will assume that the
$U(N)$ Yang-Mills theory in Eq. \eq{n=4} has been obtained from the BFSS
matrix model in Eq. \eq{bfss} by using the $(d-1)$-fold matrix T-duality
in Eq. \eq{matrix-t}. In particular, it will be important to remember that
the $U(N \to \infty)$ gauge theory in Eq. \eq{n=4} in the Moyal background
in Eq. \eq{matrix-vacuum} can be mapped to the $D=d+2n$-dimensional noncommutative
$U(1)$ gauge theory in Eq. \eq{higher-action}.

Motivated by this fact, we will specify our problem as follows: We
want to classify a stable class of ``time-independent solutions" in
the action in Eq. \eq{higher-action} satisfying the asymptotic boundary
condition in Eq. \eq{matrix-vacuum}. For such kind of solutions, we may
simply forget about time and work in the temporal gauge, $A_0 = 0$.
Therefore, we will consider the $U(N)$ gauge-Higgs system
$(\mathbf{A}, \Phi^a)(\mathbf{x})$ as a map from $\mathbb{R}_{C}^p$ to
$GL(N, \mathbb{C})$, where $p \equiv d-1$ and $z^\mu = (t, \mathbf{x})$. As
long as we require the fields in the theory to approach the common
limit in Eq. \eq{matrix-vacuum} (which does not depend on $\mathbf{x}$) as
$\mathbf{x} \to \infty$ in any direction, we can think of $\mathbb{R}^p$ as
having the topology of a sphere $\mathbf{S}^p = \mathbb{R}^p \cup
\{\infty\}$, with the point at infinity being included as an ordinary point.

Note that the matrices $\Phi^a (\mathbf{x}) \; (a = 1, \cdots,2n)$
are nondegenerate along $\mathbf{S}^p$ because we have assumed
Eq. \eq{matrix-vacuum}. Therefore, $\Phi^a$ defines a well-defined map
\ct{horava}
\be \la{homotopy}
\Phi^a: \mathbf{S}^p \to GL(N,\mathbb{C})
\ee
from $\mathbf{S}^p$ to the group of nondegenerate complex $N
\times N$ matrices. If this map represents a nontrivial class in the
$p$th homotopy group $\pi_p(GL(N,\mathbb{C}))$, the solution
in Eq. \eq{homotopy} will be stable under small perturbations, and the
corresponding nontrivial element of $\pi_p(GL(N,\mathbb{C}))$ represents a
topological invariant. Note that the map
in Eq. \eq{homotopy} is contractible to the group of maps from $\mathbf{S}^p$ to
$U(N)$ \ct{k-theory2}.

If we think of $GL(N, \mathbb{C})$ as an endomorphism from $\mathbb{C}^N$ to
itself, $\mathbb{C}^N$ is already big enough to embed $\mathbf{S}^p$ into
it if $N > p/2$. This leads to a remarkable point that there is the
so-called stable regime at $N > p/2$, where $\pi_p(GL(N,\mathbb{C}))$ is
independent of $N$. In this stable regime, the homotopy groups of
$GL(N,\mathbb{C})$ or $U(N)$ define a generalized cohomology theory, known
as K-theory \ct{karoubi,k-theory1,k-theory2,k-theory3}. In K-theory, which
also involves vector bundles and gauge fields, any smooth manifold
$X$ is assigned an Abelian group $K(X)$. Aside from a deep relation
to D-brane charges and RR fields in string theory
\ct{k-theory1,k-theory2}, the K-theory is also deeply connected with the theory
of Dirac operators, the index theorem, Riemannian geometry,
noncommutative geometry, etc. \ct{connes}.

The matrix action in Eq. \eq{n=4} describes a $U(N \to \infty)$ vector
(Chan-Paton) bundle supported on $\mathbb{R}_{C}^{d}$. The homotopy map
in Eq. \eq{homotopy} is to classify stable solutions of the $U(N)$ Chan-Paton bundle
that cannot be dissipated by small perturbations. However, the
topological classification should be defined up to pair creations
and pair annihilations because there is no way to suppress such
quantum effects. This is the reason \ct{k-theory2,k-theory3} $K(X)$ is the right answer
for classifying the topological class of excitations in the $U(N)$ gauge-Higgs system.
For $X$ noncompact, $K(X)$ is to be interpreted as compact K-theory
\ct{karoubi}. For example, for $X = \mathbb{R}^d$, this group is given by
\be \la{k-theory}
K(\mathbb{R}^d) = \pi_{d-1}(GL(N, \mathbb{C})),
\ee
with $N$ in the stable regime. The corresponding groups are known to
exhibit Bott periodicity such that $K(\mathbb{R}^d) = \mathbf{Z}$ for
even $d$ and $K(\mathbb{R}^d) = 0$ for odd $d$.

With the above understanding, let us find an explicit construction
of a topologically non-trivial excitation. It is well-known
\ct{k-theory3} that this can be done using an elegant construction
due to Atiyah, Bott and Shapiro (ABS) \ct{abs}. The construction
uses the gamma matrices of the Lorentz group $SO(p,1)$ for $X =
\mathbb{R}_{C}^{d}$ to construct explicit generators of the K-theory group
in Eq. \eq{k-theory}, where $p = d - 1$. Let $X$ be even dimensional so
that $K(X) = \mathbf{Z}$, and $S_\pm$ be two irreducible spinor
representations of $Spin(d)$ Lorentz group, and $p_\mu = (\omega,
\mathbf{p}), \; \mu = 0, 1, \cdots, p,$ be the momenta along $X$. We define the
gamma matrices $\Gamma^\mu: S_+
\to S_-$ of $SO(p,1)$ obeying $\{ \Gamma^\mu, \Gamma^\nu \} = 2 \eta^{\mu\nu}$.
Also, we introduce an operator $\mathcal{D}: \mathcal{H} \times S_+ \to \mathcal{H} \times S_-$
\ct{horava} such that
\be \la{dirac-op}
\mathcal{D} = \Gamma^\mu p_\mu + \cdots,
\ee
which is regarded as a linear operator acting on a Hilbert space
$\mathcal{H}$, as well as the spinor vector space $S_\pm$. Here, the
Hilbert space $\CH$ is possibly much smaller than the Fock space
in Eq. \eq{fock}, because the Dirac operator in Eq. \eq{dirac-op} acts on collective
(coarse-grained) modes of the solution in Eq. \eq{homotopy}.

The ABS construction implies \ct{horava,volovik} that the Dirac
operator in Eq. \eq{dirac-op} is a generator of $\pi_p(U(N))$ as a
nontrivial topology in momentum space $(\mathbf{p}, \omega)$ and
acts on a low lying excitation near the vacuum in Eq. \eq{matrix-vacuum}
which carries K-theory charges and so is stable. Such modes are
described by using coarse-grained fermions $\chi^A (\omega, \mathbf{p}, \theta)$,
with $\theta$ denoting possible collective coordinates of
the solution in Eq. \eq{homotopy} \ct{horava}. The ABS construction determines the range
$\widetilde{N}$ of the index $A$ carried by the coarse-grained
fermions $\chi^A$ to be $\widetilde{N} = 2^{[p/2]} n
\leq N$ complex components. The precise form of the fermion $\chi^A$
depends on its K-theory charge, whose explicit representation on
$\mathcal{H} \times S_\pm$ will be given later. Feynman's
approach \ct{feynman} in Section IV.A will provide a clear-cut
picture to see what the multiplicity $n$ means. At low energies, the
dispersion relation of the fermion $\chi^A$ is given by the
relativistic Dirac equation
\be \la{dirac-eq}
i \Gamma^\mu \p_\mu \chi + \cdots =0,
\ee
with possible gauge interactions and higher order corrections in
higher energies. Thus, we get a spinor of the Lorentz group $SO(p,1)$
from the ABS construction as a topological solution in momentum
space \ct{horava}. For example, in four dimensions, i.e., $p=3$,
$\chi^A$ has two complex components when $n=1$, so it describes a chiral Weyl fermion.

Although the emergence of $(p+1)$-dimensional spinors is just a
consequence due to the fact that the ABS construction uses the
Clifford algebra to construct explicit generators of
$\pi_{p}(U(N))$, its physical origin is mysterious and difficult to
understand. However, we believe that the coherent spactime vacuum
in Eq. \eq{vacuum-spacetime} would be the crux for the origin of the
fermionic nature of particles and the mysterious connection between
the Clifford module and K-theory \ct{abs}. An important
future problem would be to clearly understand this issue.

Now, let us address the problem to determine the multiplicity $n$ of
the coarse-grained fermions $\chi^{\alpha a}$, where we decomposed
the index $A = (\alpha a)$ with $\alpha$ the spinor index of the
$SO(d)$ Lorentz group and $a = 1, \cdots, n$ an internal index of an
$n$-dimensional representation of some compact symmetry $G$. In
order to understand this problem, we will identify the
noncommutative $\star$-algebra ${\cal A}_\theta$ with $GL(N,
\mathbb{C})$ by using the relation in Eq. \eq{ncft-matrix}. Under this
correspondence, the $U(N \to \infty)$ gauge theory in Eq. \eq{n=4} in the
Moyal background in Eq. \eq{matrix-vacuum} can be mapped to the
$D$-dimensional noncommutative $U(1)$ gauge theory
in Eq. \eq{higher-action} defined on $\mathbb{R}_C^d \times \mathbb{R}_{NC}^{2n}$
where $D=d+2n$. Then, the K-theory in Eq. \eq{k-theory} for any sufficiently
large $N$ can be identified with the K-theory $K(\CA_\theta)$ for
the noncommutative $\star$-algebra ${\cal A}_\theta$ \ct{k-theory2}.

As we showed in Section III.B, the generic fluctuation in
Eq. \eq{D-vector} will deform the background spacetime lattice
defined by the Fock space in Eq. \eq{fock}, which generates gravitational
fields given by the metric in Eq. \eq{ward-metric}. For simplicity, we will
consider low-energy excitations around the solution in Eq. \eq{homotopy}
whose K-theory class is given by $K(\CA_\theta)$. In this case, the solution
in Eq. \eq{homotopy} would be a sufficiently localized state described by a
compact (bounded self-adjoint) operator in ${\cal A}_\theta$. This
means that it does not appreciably disturb the ambient gravitational
field. Therefore, we may reduce the problem to quantum particle
dynamics on $X \times F$ \ct{hsy-jhep}, where $X = \mathbb{R}_C^d$ and $F$
is an internal space describing collective modes of the solution
in Eq. \eq{homotopy}. It is natural to identify the coordinates of $F$
with an internal charge of $G$ carried by the fermion
$\chi^{\alpha a}$. To be specific, the (collective) coordinates of
$F$ will take values in the Lie algebra $\mathfrak{g}$ of $G$, such
as the isospins or colors, and will be denoted by $Q^I \; (I = 1,
\cdots, n^2-1)$. In the end, we essentially revisit Feynman's
problem, which we addressed in Section IV.A.

The quantum particle dynamics on $X \times F$ naturally requires the
introduction of non-Abelian gauge fields in the representation of the Lie
algebra in Eq. \eq{na-comm-2}, and the dynamics of the particle carrying an
internal charge in $F$ will be defined by a symplectic structure on
$T^* X \times F$. Note that $\mathbb{R}_{NC}^{2n}$ already has its
symplectic structure $B = \half B_{ab} dy^a \wedge dy^b$, originated from the noncommutative
space in Eq. \eq{nc-vacuum}. Also, note that the action in Eq. \eq{higher-action}
has only $U(1)$ gauge fields on $\mathbb{R}_C^d \times \mathbb{R}_{NC}^{2n}$, so
the problem is how to get the Lie algebra generators in
Eq. \eq{na-comm-2} from the space $\mathbb{R}_{NC}^{2n}$ and how to get the
non-Abelian gauge fields $A_\mu^I(z) \in
\mathfrak{g}$ on $X$ from the $U(1)$ gauge fields on $\mathbb{R}_C^d \times
\mathbb{R}_{NC}^{2n}$. Here, it is enough to consider only the transverse gauge
fields $A_\mu^I(z)$ as low-lying excitations because the solution
in Eq. \eq{homotopy} is actually coming from the longitudinal gauge field
$\widehat{A}_a(z, y)$ in Eq. \eq{decomp-cov}.

The problem is solved \ct{hsy-jhep} by noting that the
$n$-dimensional harmonic oscillator in quantum mechanics can realize
$SU(n)$ symmetries (see the Chapter 14 in Ref. \ct{georgi}). The
generators of the $SU(n)$ symmetry on the Fock space in Eq. \eq{fock} are
given by
\be \la{sun}
Q^I = a^\dagger_i T^I_{ik} a_k,
\ee
where the creation and the annihilation operators are given by
Eq. \eq{vacuum-spacetime} and the $T^I$'s are constant $n \times n$
matrices satisfying $[T^I, T^J] = i {f^{IJ}}_K T^K$ with the same
structure constants as Eq. \eq{na-comm-2} . It is easy to check that the
$Q^I$'s satisfy the $SU(n)$ Lie algebra \eq{na-comm-2}. We introduce
the number operator $Q^0 \equiv a^\dagger_i a_i$ and identify it with a
$U(1)$ generator. The operator $\mathfrak{C} =
\sum_I Q^I Q^I$ is the quadratic Casimir operator of the $SU(n)$ Lie
algebra and commutes with all $Q^I$'s. Thus, one may identify
$\mathfrak{C}$ with an additional $U(1)$ generator.

Let $\rho(\CH)$ be a representation of the Lie algebra
in Eq. \eq{na-comm-2}  in a Hilbert space $\CH$. We take an $n$-dimensional
representation in $\CH = L^2(\mathbb{C}^n)$, a square integrable Hilbert
space. Because the solution in Eq. \eq{homotopy} is described by a compact operator
in $\CA_\theta$, its representation space $\CH = L^2(\mathbb{C}^n)$ will be much smaller
(with finite basis in generic cases) than the original Fock space in Eq. \eq{fock}.
Thus, let us expand the $U(1)$ gauge field $\widehat{A}_\mu(z,y)$ in
Eq. \eq{decomp-cov} with the $SU(n)$ basis in Eq. \eq{sun}:
\bea \la{sun-expansion}
\widehat{A}_\mu(z,y) &=& \sum_{n=0}^\infty \sum_{I_i \in \rho(\CH)}  A^{I_1 \cdots
I_n}_\mu (z, \rho, \lambda_n)
\; Q^{I_1} \cdots Q^{I_n} \xx
&=& A_\mu(z) + A^I_\mu (z, \rho, \lambda_1)\; Q^I + A^{IJ}_\mu (z,
\rho, \lambda_2)\; Q^I Q^J + \cdots,
\eea
where $\rho$ and $\lambda_n$ are eigenvalues of $Q^0$ and
$\mathfrak{C}$, respectively, in the representation $\rho(\CH)$. The
expansion in Eq. \eq{sun-expansion} is formal, but it is assumed that each
term in Eq. \eq{sun-expansion} belongs to the irreducible
representation of $\rho(\CH)$. Through the expansion in Eq. \eq{sun-expansion},
we get $SU(n)$ gauge fields $A^I_\mu(z)$, as well as $U(1)$ gauge fields $A_\mu(z)$,
as low-lying excitations \ct{hsy-jhep}.

Note that the coarse-grained fermion $\chi$ in Eq. \eq{dirac-eq}
behaves as a stable relativistic particle in the spacetime $X =
\mathbb{R}_C^d$. When these fermionic excitations are given, there will
also be bosonic excitations arising from changing the position along
$X$ of the internal charge $F$. According to Feynman's picture,
especially Wong's equation, Eq. \eq{wong-eq}, the gauge fields in
Eq. \eq{sun-expansion} represent collective modes for the position
change in $X = \mathbb{R}_C^d$ of the charge $F$ \ct{volovik}.
See Eq. \eq{wong2} for the geometrical interpretation of
Wong's equation, Eq. \eq{wong-eq}. Thus, they can be regarded as collective modes
in the vicinity of an internal charge living in $F$ and
interact with the fermions in Eq. \eq{dirac-eq}.

Therefore, we think of the Dirac operator, Eq. \eq{dirac-op}, as an
operator $\mathcal{D}: \mathcal{H} \times S_+ \to \mathcal{H} \times
S_-$, where $\mathcal{H} = L^2(\mathbb{C}^n)$, and we introduce a minimal
coupling with the $U(1)$ and $SU(n)$ gauge fields in
Eq. \eq{sun-expansion} by the replacement $p_\mu \to p_\mu - e A_\mu
- A^I_\mu Q^I$. Then, the Dirac equation, Eq. \eq{dirac-eq}, becomes
\be \la{dirac-eq-gauge}
i \Gamma^\mu (\p_\mu -i e A_\mu - i A^I_\mu Q^I) \chi + \cdots =0.
\ee
Here, we see that the coarse-grained fermion $\chi$ in the homotopy
class $\pi_p (U(N))$ is in the fundamental representation of
$SU(n)$, so we identify the multiplicity $n$ in the ABS construction
in Eq. \eq{dirac-eq} with the number of colors in $SU(n)$ \ct{hsy-jhep}.

The most interesting case in Eq. \eq{higher-action} is of $d=4$ and
$n=3$, that is, 10-dimensional noncommutative $U(1)$ gauge theory on
$\mathbb{R}_C^4 \times \mathbb{R}_{NC}^6$. In this case, Eq. \eq{dirac-eq-gauge} is
the 4-dimensional Dirac equation, where $\chi$ is a quark, an $SU(3)$
multiplet of chiral Weyl fermions, which couples with gluons
$A^I_\mu(z)$, $SU(3)$ gauge fields for the color charge $Q^I$, as
well as photons $A_\mu(z)$, $U(1)$ gauge fields for the electric
charge $e$. One may consider a similar ABS construction in the
vector space $\mathbb{C}^2 \times \mathbb{C} \subset \mathbb{C}^3$, i.e., by breaking the
$SU(3)$ symmetry into $SU(2) \times U(1)$, where $\chi$ would be a
lepton, an $SU(2)$ doublet of chiral Weyl fermions coupling with
$SU(2)$ gauge fields. In this case, $Q^I \;(I=1,2,3)$ in Eq. \eq{sun}
are the famous Schwinger representation of the $SU(2)$ Lie algebra.

To conclude, we may go back to our starting point. Our starting
point was the $d$-dimensional $U(N)$ Yang-Mills theory defined by
the action in Eq. \eq{n=4} or equivalently $D$-dimensional noncommutative
$U(1)$ gauge theory defined by the action in Eq. \eq{higher-action}. We
observed that the theory in Eq. \eq{n=4} allows topologically stable
solutions as long as the homotopy group in Eq. \eq{homotopy} is nontrivial,
and we argued that a matter field, such as leptons and quarks, simply
arises from such a stable solution and that non-Abelian gauge fields
correspond to collective zero-modes of the stable localized
solution. Although we intended to interpret such excitations as
particles and gauge fields and to ignore their gravitational effects,
we have to remember that these are originally a part of spacetime
geometry according to the map in Eq. \eq{D-vector}. Consequently, we get a
remarkable picture, if any, that matter fields, such as leptons and
quarks, simply arise as a stable localized geometry, which is a
topological object in the defining algebra (noncommutative
$\star$-algebra) of quantum gravity.

\section{ANATOMY OF SPACETIME}

It is in order to discuss the most beautiful aspects of emergent
gravity. Remarkably, the emergent gravity reveals a novel picture
about the origin of spacetime, dubbed as emergent spacetime, which
is radically different from any previous physical theory all of
which describe what happens in a given spacetime. Thus, we may take it
for granted that emergent gravity leads to many results that are radically
different from Einstein gravity.

\subsection{Emergent Time in Emergent Gravity}

We have intentionally postponed posing the formidable issue how
``Time" emerges, together with space, and how it is entangled with
space to unfold into a single entity, spacetime, and take the
shape of Lorentz covariance. Now we are ready to address this
formidable issue.

Let $(M, \pi)$ be a Poisson manifold. We previously defined the
anchor map $\pi^\sharp : T^* M \to TM$ in Eq. \eq{anchor} for a general
Poisson bivector $\pi \in \Gamma(\Lambda^2 TM)$ and the Hamiltonian vector field
in Eq. \eq{ham-vec} by
\be \la{hamilton-vec}
X_H  \equiv - \pi^\sharp (dH)  = \{\cdot , H\}_\pi = \pi^{\mu\nu} (x)
\frac{\p H}{\p x^\nu} \frac{\p }{\p x^\mu},
\ee
where $H: M \to \mathbb{R}$ is a smooth function on the Poisson manifold
$(M, \pi)$. If the Poisson tensor $\pi$ is nondegenerate so that
$\pi^{-1} \equiv \omega \in \Gamma(\Lambda^2 T^* M)$ is a symplectic
structure on $M$, the anchor map $\pi^\sharp : T^* M \to TM$ defines
a bundle isomorphism because $\pi^\sharp$ is nondegenerate everywhere.
We will speak of the flow $\phi_t$ of a vector field $X$ on $M$ when
referring to a 1-parameter group of diffeomorphisms generated by $X$.

Any Poisson manifold $(M,\pi)$ always admits a Hamiltonian dynamical
system on $M$ defined by a Hamiltonian vector field $X_H$ and it is
described by
\be \la{hamilton-poisson}
\frac{df}{dt} = X_H (f) + \frac{\p f}{\p t} = \{f,H\}_\pi + \frac{\p f}{\p t}
\ee
for any $f \in C^\infty (\mathbb{R} \times M)$. If $\phi_t$ is a flow
generated by a Hamiltonian vector field $X_H$, the following
identity holds \ct{mechanics}:
\bea \la{flow-hamilton}
\frac{d}{dt} (f \circ \phi_t) &=& \frac{d}{dt} \phi_t^* f
= \phi_t^* \CL_{X_H} f + \phi_t^* \frac{\p f}{\p t} \xx &=&
\phi_t^*
\{f, H\}_\pi + \frac{\p f}{\p t} \circ \phi_t \xx &=& \Big( \{f,
H\}_\pi +
\frac{\p f}{\p t} \Big) \circ \phi_t.
\eea
Thus, we can get $f(x,t) = g(\phi_t(x))$, where $g(x) \equiv f(x,0)$.
If $\pi = \eta = \frac{\p}{\p x^i} \wedge \frac{\p}{\p p_i}$, we
precisely reproduce Eqs. \eq{hamilton-peq} and \eq{hamilton-path} from
Eqs. \eq{hamilton-poisson} and \eq{flow-hamilton}, respectively.
In this case, the evolution of a particle system is described by the
dynamical flow in Eq. \eq{flow-hamilton} generated by the Hamiltonian
vector field in Eq. \eq{hamilton-vec} for a given Hamiltonian $H$.

Introduce an extended Poisson tensor on $\mathbb{R} \times M$
\ct{mechanics}
\be \la{ext-poisson}
\widetilde{\pi} = \pi + \frac{\partial}{\partial t}
\wedge \frac{\partial}{\partial H}
\ee
and a generalized Hamiltonian vector field
\be \la{gen-hamilton-vec}
\widetilde{X}_H  \equiv - \widetilde{\pi}^\sharp (dH)  = \{\cdot , H\}_{\widetilde{\pi}} =
\pi^{\mu\nu} (x) \frac{\p H}{\p x^\nu} \frac{\p }{\p x^\mu} + \frac{\p }{\p
t}.
\ee
We can then rewrite Hamilton's equation, Eq. \eq{hamilton-poisson},
compactly in the form
\be \la{gen-hamilton-poisson}
\frac{df}{dt} = \widetilde{X}_H (f) = \{f, H\}_{\widetilde{\pi}}
= \{f,H\}_{\pi} + \frac{\p f}{\p t}.
\ee
Similarly, we can extend the symplectic structure $\omega =
\pi^{-1}$ to the product manifold $\mathbb{R} \times M$ by considering a
new symplectic structure $\widetilde{\omega} =
\pi_2^* \omega$, where $\pi_2: \mathbb{R} \times M \to M$ is the projection
such that $\pi_2(t,x) = x$. Define $\omega_H = \widetilde{\omega} +
dH \wedge dt$. Then, the pair $(\mathbb{R} \times M, \omega_H)$ is called a
contact manifold \ct{mechanics}.

Suppose that observables $f \in C^\infty(M)$ do not depend on time
explicitly, i.e., $\frac{\partial f}{\partial t} =0$. Look at Eq. \eq{flow-hamilton}.
We understand that the time evolution of the
system in this case is determined by simply calculating the Poisson
bracket with a Hamiltonian function $H$. In other words, in the case
of $\frac{\partial f}{\partial t} =0$, the time evolution is just
the inner automorphism of the Poisson algebra $(M, \{\cdot, \cdot \}_\pi$) \cite{hei-alg}.
Therefore, time in Hamilton's equation, Eq. \eq{hamilton-poisson}, is basically
an affine parameter to trace the history of a particle, and it is operationally
defined by the Hamiltonian. That is, time in Hamiltonian dynamics is
intrinsically the histories of the particles themselves. However, we have to
notice that, only when the symplectic structure is fixed for a given
Hamiltonian, the evolution of the system is completely determined by
the evolution equation in Eq. \eq{flow-hamilton}. In this case, the dynamics
of the system can be formulated in terms of an evolution with a
single time parameter. In other words, we have a globally
well-defined time for the evolution of the system. This is the usual
situation we consider in classical mechanics.

If observables $f \in C^\infty(M)$ including the Hamiltonian
$H$, explicitly depend on time, i.e., $\frac{\partial f}{\partial t}
\neq 0$, the time evolution of the system is not completely
determined by the inner automorphism of the Poisson algebra only, so
the time evolution partially becomes an outer automorphism. However,
as we remarked above, we can extend an underlying Poisson structure
as in Eq. \eq{ext-poisson} or introduce a contact manifold $(\mathbb{R}
\times M, \omega_H)$ by extending an underlying symplectic structure. The time
evolution of a particle system is again defined by an inner
automorphism of the extended Poisson algebra $(\mathbb{R} \times M, \{\cdot,
\cdot \}_{\widetilde{\pi}}$). In this case, time should be
regarded as a dynamical variable whose conjugate momentum is given
by the Hamiltonian $H$, as indicated by the Poisson structure
in Eq. \eq{ext-poisson}. Thus, the time should be defined locally in this case.
Let us clarify this situation.

Consider a dynamical evolution described by a change of a
symplectic structure from $\omega$ to $\omega_t = \omega + t
(\omega' - \omega)$ for all $0 \leq t \leq 1$, where $\omega' -
\omega = - e dA$. The Moser lemma, Eq. \eq{time-flow}, says that
there always exists a one-parameter family of diffeomorphisms
generated by a smooth time-dependent vector field $X_t$ satisfying
$\iota_{X_t} \omega_t = e A$. Although the vector field $X_t$
defines a dynamical one-parameter flow, the vector field $X_t$ is, in
general, not even locally Hamiltonian because $dA
\neq 0$. The evolution of the system in this case is locally
described by the flow $\phi_t$ of $X_t$ starting at $F_0$ = identity,
but it is no longer a (locally) Hamiltonian flow. In this case, we fail
to have the property $\CL_{X_t} f = \{f, H\}_\pi$ in
Eq. \eq{flow-hamilton}, so we have no global Hamiltonian flow.
That is, there is no well-defined or global time for the particle
system. In other words, the time flow $\phi_t$ of $X_t$ is defined on a
local chart and describes only a local evolution of the system.

We observed the equivalence in Eq. \eq{particle-equivalence} for the
dynamics of a charged particle. Let us consider the above situation
by looking at the left-hand side picture of
Eq. \eq{particle-equivalence} by fixing the symplectic structure, but
instead by changing the Hamiltonian. (Note that the magnetic field in
the Lorentz force, Eq. \eq{force-law}, does not do any work, so there is
no energy flow during the evolution.) At time $t=0$, the system is
described by the free Hamiltonian $H_0$, but it ends up with the
Hamiltonian in Eq. \eq{particle-hamiltonian} at time $t=1$. Therefore, the
dynamics of the system cannot be described with a single time
parameter covering the entire period $0 \leq t \leq 1$.
We can introduce at most a local time during $\delta t <
\epsilon$ on a local patch and smoothly adjust to a neighboring
patch. To say, a clock of the particle will tick each time with a
different rate because the Hamiltonian of the particle is changing
during time evolution. As we already remarked before, we may also
need to quantize the time according to the Poisson structure
in Eq. \eq{ext-poisson} in order to describe a quantum evolution of a
system in terms of an extended inner automorphism such as
that in Eq. \eq{gen-hamilton-poisson}.

Now, we can apply the same philosophy to the case of the Poisson
structure in Eq. \eq{p-bracket} defined on a space itself \ct{hsy-jhep}.
The mathematics is exactly the same. An essential point in defining
the time evolution of a system was that any Poisson manifold
$(M,\pi)$ always admits the Hamiltonian dynamical system
in Eq. \eq{hamilton-poisson} on $M$ defined by the Hamiltonian vector field
$X_H$ given by Eq. \eq{hamilton-vec}. We have faced the same situation with the
$\theta$-bracket, Eq. \eq{p-bracket}, whose time evolution was summarized
in Eq. \eq{time-flow}. Of course, one should avoid a confusion between
the dynamical evolution of a particle system related to the phase
space in Eq. \eq{p-poisson} and the dynamical evolution of spacetime
geometry related to the noncommutative space in Eq. \eq{p-bracket}.

We learn an important lesson from Souriau and Sternberg
\ct{sternberg} that the Hamiltonian dynamics in the presence of
electromagnetic fields can be described by the deformation of
a symplectic structure of a phase space. More precisely, we observed
that the emergent geometry is defined by a one-parameter family of
diffeomorphisms generated by a smooth vector field $X_t$ satisfying
$\iota_{X_t} \omega_t + A = 0$ for the change of a symplectic
structure within the same cohomology class from $\omega$ to
$\omega_t = \omega + t (\omega' - \omega)$ for all $0 \leq t \leq 1$
where $\omega' - \omega = dA$. The vector field $X_t$ is, in general,
not a Hamiltonian flow, so no global time can be assigned to the
evolution of the symplectic structure $\omega_t$. However, if there is
no fluctuation of the symplectic structure, i.e., $F=dA = 0$ or $A = -
dH$, there can be a globally well-defined Hamiltonian flow. In this
case, we can define a global time by introducing a unique Hamiltonian
such that the time evolution is defined by $df/dt = X_H(f) = \{f, H
\}_{\theta= \omega^{-1}}$ everywhere. In particular, when the initial symplectic
structure $\omega$ is constant (homogeneous), a clock will tick
everywhere at the same rate. Note that this situation happens for
the constant background in Eq. \eq{vacuum-spacetime} from which a flat
spacetime emerges as we will discuss soon in some detail. If
$\omega$ is not constant, the time evolution will not be uniform
over space and a clock will tick at different rates at
different places. This is consistent with Einstein gravity because a
nonconstant $\omega$ corresponds to a curved space in our picture,
as we explained in Section III.D.

In the case of a changing symplectic structure, we can apply the
same strategy as we did in the particle case with the Poisson structure $\pi =
\theta$, so we suggest, in general, the concept of ``Time" in emergent
gravity \ct{hsy-jhep} as a contact manifold $(\mathbb{R} \times M,
\omega_H)$, where $(M, \omega)$ is a symplectic manifold and $\omega_H =
\widetilde{\omega} + dH \wedge dt$, with $\widetilde{\omega} = \pi_2^*
\omega$ defined by the projection $\pi_2: \mathbb{R} \times M
\to M, \; \pi_2(t,x) = x$. A question is then how to
recover the (local) Lorentz symmetry in the end. As we pointed out
above, if $(M, \omega)$ is a canonical symplectic manifold, i.e., $M
=\mathbb{R}^{2n}$ and $\omega$ = constant, a $(2n+1)$-dimensional Lorentz
symmetry appears from the contact manifold $(\mathbb{R} \times M,
\omega_H)$. (For a more general case such as our $(3+1)$-dimensional
Lorentzian world and a Poisson spacetime, Eq. \eq{sn-bracket},
we may instead use the Poisson structure in Eqs. \eq{ext-poisson}-\eq{gen-hamilton-poisson},
or we may need an even more general argument, which we don't know yet.)
Once again, the Darboux theorem says that there always
exists a local coordinate system in which the symplectic structure has
a canonical form. See Eq. \eq{darboux-tr}. For the Poisson case,
we can apply Weinstein's splitting theorem instead. Then, it is quite plausible that the
Lorentz symmetry on a local Darboux chart would be recovered in a local way.
Furthermore, Feynman's argument in Section IV.A implies that the gauge symmetry, as well as
the Lorentz symmetry, is just derived from the symplectic structure on the contact
manifold $(\mathbb{R} \times M, \omega_H)$. For example, one can recover
the gauge symmetry along the time direction by defining the
Hamiltonian $H = A_0 + H'$ and the time evolution of a spacetime
geometry by the Hamilton's equation $D_0 f \equiv df/dt + \{A_0, f
\}_{\widetilde{\theta} = \widetilde{\omega}^{-1}} =
\{f, H'\}_{\widetilde{\theta} = \widetilde{\omega}^{-1}}$.
Then, one may interpret Hamilton's equation as an infinitesimal
version of an inner automorphism like \eq{D-vector}, which was,
indeed, used to define the vector field $V_0(X)$ in Eq. \eq{vec-d}.

Our proposal for the emergent time  \ct{hsy-jhep} is based on the
fact that a symplectic manifold $(M, \omega)$ always admits a
Hamiltonian dynamical system on $M$ defined by a Hamiltonian vector
field $X_H$, i.e., $\iota_{X_H} \omega = dH$. The emergent time can
be generalized to the noncommutative space in Eq. \eq{nc-spacetime} by
considering the inner derivation, Eq. (\ref{nc-vector}), instead of the
Poisson bracket $\{f,H\}_\theta$. If time is emergent in this way,
it implies a very interesting consequence. Note that every
symplectic manifold $(M,B)$ is canonically oriented and comes with a
canonical measure, the Liouville measure, $B^n = \frac{1}{n!} B
\wedge \cdots \wedge B$, which is a volume form of the symplectic manifold $(M,B)$
and nowhere vanishing on $M$. Therefore, the symplectic structure $B$
triggered by the vacuum condensate in Eq. \eq{vacuum-condense} not only
causes the emergence of spacetime but also specifies an orientation
of spacetime. Because the time evolution of spacetime is
defined by the Poisson structure $\pi = \theta = B^{-1}$ as in
Eq. \eq{flow-hamilton}, a global time evolution of spacetime manifold will have
a direction that depends on the orientation $B^n$, although a local time
evolution has time reversal symmetry. If gravity is emergent
from the electromagnetism supported on a symplectic manifold as we
have envisaged so far, it may also be possible to explain the ``arrow of
time" in the cosmic evolution of our Universe - the most notoriously
difficult problem in quantum gravity.

\subsection{Cosmological Constant Problem and Dark Energy}

In general relativity, gravitation arises out of the dynamics of
spacetime being curved by the presence of stress-energy, and the
equations of motion for the metric fields of spacetime are
determined by the distribution of matter and energy:
\be \la{einstein-eq}
R_{\mu\nu} - \half g_{\mu\nu} R = \frac{8 \pi G}{c^4} T_{\mu\nu}.
\ee
The Einstein equations, Eq. \eq{einstein-eq}, describe how the geometry of
spacetime on the left-hand side is determined dynamically in harmony with matter fields
on the right-hand side, at first sight. We know that the existence of spacetime leads
to a ``metrical elasticity" of space, i.e., to an inertial force that opposes the curving of space.

However, there is a deep conflict between the spacetime geometry
described by general relativity and the matter fields described by
quantum field theory \cite{jpcs-hsy}. If spacetime is flat, i.e., $g_{\mu\nu} =
\eta_{\mu\nu}$, the left-hand side of Eq. \eq{einstein-eq} identically vanishes,
so the energy-momentum tensor of matter fields should vanish,
i.e., $T_{\mu\nu} = 0$. In other words, a flat spacetime is completely empty with no energy.
Thus, the concept of empty space in Einstein gravity is in an acute contrast to the concept
of vacuum in quantum field theory, where the vacuum is not empty but is full of
quantum fluctuations. As a result, a vacuum is extremely heavy,
and its weight is on the order of the Planck mass, i.e., $\rho_{\rm vac} \sim M_P^4$.

The conflict rises to the surface that gravity and matters respond
differently to the vacuum energy and perplexingly brings about the
notorious cosmological constant problem. Indeed, the clash manifests
itself as a mismatch of symmetry between gravity and matter \ct{tpad}.
To be precise, if we shift a matter Lagrangian $\CL_M$ by
a constant $\Lambda$, that is,
\be \la{shift}
\CL_M \to \CL_M - 2 \Lambda,
\ee
it results in a shift of the matter energy-momentum tensor by
$T_{\mu\nu} \to T_{\mu\nu} - \Lambda g_{\mu\nu}$ in the Einstein
equation, Eq. \eq{einstein-eq}, although the equations of motion for
matter are invariant under the shift in Eq. \eq{shift}. Definitely the
$\Lambda$-term in Eq. \eq{shift} will appear as the cosmological
constant in Einstein gravity, and it affects the spacetime structure.
For example, a flat spacetime is no longer a solution of Eq. \eq{einstein-eq}.

Let us sharpen the problem arising from the conflict between
geometry and matter. In quantum field theory, there is no way to
suppress quantum fluctuations in a vacuum. Fortunately, the vacuum
energy due to the quantum fluctuations, regardless of how large they
are, does not cause any trouble for quantum field theory thanks to the
symmetry in Eq. \eq{shift}. However, general covariance requires that
gravity couple universally to all kinds of energy. Therefore, the
vacuum energy $\rho_{\rm vac} \sim M_P^4$ will induce a highly-curved spacetime whose curvature
scale $R$ would be $\sim M_P^2$ according to Eq. \eq{einstein-eq}. If
so, the quantum field theory framework in the background of quantum
fluctuations must be broken down due to a large back-reaction of
background spacetim, but we know that it is not the case. The
quantum field theory is well-defined, even in the presence of the vacuum
energy $\rho_{\rm vac} \sim M_P^4$, and the background spacetime
still remains flat, as we empirically know. So far, there is no experimental evidence for the
vacuum energy really coupling to gravity, although it is believed that
the vacuum energy is real as experimentally verified by the Casimir effect.

Which side of Eq. \eq{einstein-eq} is the culprit giving rise to the
incompatibility? After consolidating all the suspicions inferred
above, we throw doubt on the left-hand side of
Eq. \eq{einstein-eq}, especially, on the result that a flat spacetime
is free gratis, i.e., costs no energy. We should remark that
such a result is not compatible with the inflation scenario either
because it implies that a huge vacuum energy in a highly nonequilibrium
state is required to generate an extremely large spacetime. Note
that Einstein gravity is not completely background independent because
it assumes the prior existence of a spacetime manifold.
Here, we refer to a background-independent theory in which no spacetime structure
is {\it a priori} assumed, but is defined by the theory.
In particular, the flat spacetime is a geometry of special relativity rather
than general relativity, and so it is assumed to be {\it a priori} given
without reference to its dynamical origin. This reasoning implies that
the negligence about the dynamical origin of a flat spacetime defining
a local inertial frame in general relativity might be the core root of
the incompatibility inherent in Eq. \eq{einstein-eq}.

All in all, one may be tempted to infer that a flat spacetime may
not be free gratis, but a result of Planck energy condensation in a
vacuum. Now, we will show that inference to be true \ct{hsy-cc}.
Surprisingly, the emergent spacetime picture then appears as the
H\'oly Gr\'ail to cure several notorious problems in theoretical
physics; for example, the cosmological constant problem,
the nature of dark energy and the reason gravity
is so weak compared to other forces. After all, our final destination
is to check whether the emergent gravity from noncommutative
geometry is a physically viable theory to correctly explain the
dynamical origin of flat spacetime.

Let us start with a background-independent matrix theory, for
example, Eqs. \eq{nc-matrix} or \eq{ikkt}, where no spacetime structure
is introduced. A specific spacetime background, e.g., a flat
spacetime, has been defined by specifying the vacuum,
Eq. (\ref{matrix-vacuum}), of the theory. Now, look at the metric
in Eq. \eq{emergent-metric} or in Eq. \eq{ward-metric} to trace back
to where the flat spacetime comes from. The flat spacetime is the case
with $V^\mu_a = \delta^\mu_a$, so $\lambda^2 = 1$. The vector field
$V_a = V^\mu_a \p_\mu = \p_a$ in this case comes from the
noncommutative gauge field $A_a^{(0)}
\equiv \langle \widehat{A}^{(0)}_a \rangle_{\rm vac} = - B_{ab} y^b$
in Eq. \eq{vacuum-coordinate} whose field strength is $\langle
\widehat{F}^{(0)}_{ab} \rangle_{\rm vac} = - B_{ab}$, describing a uniform condensation of
gauge fields in vacuum. See Eq. \eq{vacuum-condense}. Therefore, we
see that the flat spacetime is emergent from the vacuum algebra
in Eq. (\ref{nc-spacetime}) induced by a uniform condensation of gauge
fields in vacuum. This is a tangible difference from Einstein
gravity in which the flat spacetime is completely an empty space.

The emergent gravity defined by the action in Eq. \eq{higher-action}, for
example, responds completely differently to the constant shift
in Eq. \eq{shift}. To be specific, let us consider a constant
shift of the background $B_{MN} \to B_{MN} + \delta B_{MN}$. Then,
the action in Eq. \eq{higher-action} in the new background becomes
\begin{equation} \label{shift-action}
\widehat{S}_{B + \delta B} = \widehat{S}_{B} + \frac{1}{2 g^2_{YM}} \int d^D X
\widehat{F}_{MN} \delta B_{MN} - \frac{1}{4 g^2_{YM}} \int d^D X
\Big( \delta B_{MN}^2 - 2 B^{MN} \delta B_{MN} \Big).
\end{equation}
The last term in Eq. \eq{shift-action} is simply a constant; thus,
it will not affect the equations of motion, Eq. \eq{D-eom}. The second
term is a total derivative, so it will vanish if $\int d^DX
\widehat{F}_{MN} =0$. (It is a defining
property \ct{nc-review} in the definition of a star product that
$\int d^D X \widehat{f} \star \widehat{g} = \int d^D X \widehat{f} \cdot \widehat{g}$.
Then, the second term should vanish as far as $\widehat{A}_M \to 0$
at infinity.) If spacetime has a nontrivial boundary, the second
term could be nonvanishing at the boundary, which would change the
theory under the shift. We will not consider a nontrivial spacetime
boundary because the boundary term is not an essential issue here,
though there should be interesting physics at the boundary \cite{h-mis}.
Then we get the result $\widehat{S}_{B + \delta B}
\cong \widehat{S}_{B}$. Indeed, this is the Seiberg-Witten equivalence between
noncommutative field theories defined by the noncommutativity
$\theta'= \frac{1}{B+ \delta B}$ and $\theta = \frac{1}{B}$
\ct{SW-ncft}. Although the vacuum in Eq. \eq{vacuum-condense} readjusts
itself under the shift, the Hilbert spaces ${\cal H}_{\theta'}$ and
${\cal H}_{\theta}$ in Eq. \eq{fock} are completely isomorphic if and only if $\theta$ and
$\theta'$ are nondegenerate constants. Furthermore, the vector fields
in Eq. \eq{D-vector} generated by $B + \delta B$ and $B$ backgrounds are
equally flat as long as they are constant. Consequently two
different constant backgrounds are related by a global Lorentz
transformation. Equation \eq{einstein-tensor} also shows that the
background gauge field does not contribute to the energy-momentum
tensor in Eq. \eq{emergent-eq}.

Therefore, we clearly see that a constant shift of energy density
such as Eq. \eq{shift} is a symmetry of the theory in Eq. \eq{higher-action}
although the action in Eq. \eq{higher-action} defines a theory of gravity
in the sense of emergent gravity. As a consequence, {\it there is no
cosmological constant problem in emergent gravity} \ct{hsy-cc}. Now,
let us estimate the dynamical scale of the vacuum condensation
in Eq. \eq{vacuum-condense}. Because gravity emerges from noncommutative gauge fields,
the parameters $g_{YM}^2$ and $|\theta|$ defining a
noncommutative gauge theory should be related to the Newton constant
$G$ in emergent gravity. A simple dimensional analysis leads to the
relation in Eq. \eq{newton-constant}. This relation immediately leads to
the fact \ct{hsy-jhep} that the energy density of the vacuum
in Eq. \eq{vacuum-condense} is
\be \la{vacuum-energy}
\rho_{\rm vac} \sim |B_{ab}|^2 \sim M_P^4,
\ee
where $M_P = (8\pi G)^{-1/2} \sim 10^{18} GeV$ is the Planck mass.
Therefore, the emergent gravity finally reveals a remarkable picture
that the huge Planck energy $M_P$ is actually the origin of the flat
spacetime. Hence, we conclude that {\it a vacuum energy does not
gravitate differently from Einstein gravity, and a flat spacetime is not
free gratis, but is a result of Planck energy condensation in vacuum} \ct{hsy-cc}.

If the vacuum algebra in Eq. (\ref{nc-spacetime}) describes a flat
spacetime, it can have a very important implication to cosmology.
According to our picture for emergent spacetime, a flat spacetime is
emergent from Planck energy condensation in vacuum; thus, the
time scale for the condensate will be roughly on the order of the Planck time.
We know that there was an epoch of very violent time-varying vacuum,
the so-called cosmic inflation. Therefore, it is natural to expect
that the explosive inflation era that lasted roughly $10^{-33}$ seconds at
the beginning of our Universe corresponds to a dynamical process enormously spreading out
a flat spacetime by the instantaneous condensation of vacuum energy $\rho_{{\rm vac}} \sim M^4_P$ .
Unfortunately, it is not clear how to microscopically describe this
dynamical process by using the matrix action (\ref{ikkt}). Nevertheless,
it is quite obvious that the cosmological inflation should be a
dynamical condensation of the vacuum energy $\rho_{{\rm vac}} \sim M^4_P$
for the generation of (flat) spacetime according to our emergent gravity picture.

In addition, our picture for the emergent spacetime implies that the
global Lorentz symmetry should be a perfect symmetry up to the Planck
energy because the flat spacetime was emergent from Planck
energy condensation in vacuum - the maximum energy in Nature. The
huge vacuum energy $\rho_{{\rm vac}}
\sim |B_{ab}|^2 \sim M^4_P$ was simply used to make a flat
spacetime and, surprisingly, does not gravitate \cite{hsy-cc}!
Then, the gravitational fields generated by the
deformations of the background in Eq. \eq{vacuum-condense} will be very
weak because the spacetime vacuum is very solid with a stiffness of
the Planck scale. Hence the dynamical origin of flat spacetime is
intimately related to the weakness of the gravitational force.
Furthermore, the vacuum algebra in Eq. (\ref{vacuum-spacetime}) describes
an extremely coherent condensation because it is equal to the
Heisenberg algebra of an $n$-dimensional quantum harmonic oscillator.
As a consequence, the noncommutative algebra
(\ref{vacuum-spacetime}) should describe a {\it zero-entropy state}
in spite of the involvement of the Planck energy. This is very mysterious,
but it should be the case because a flat spacetime emergent from the
algebra in Eq. (\ref{vacuum-spacetime}) is completely an empty space from
the viewpoint of Einstein gravity and, so, has no entropy. This
reasoning also implies that the condensation of vacuum energy
$\rho_{{\rm vac}} \sim M^4_P$ happened at most once.

We observed that the dynamical scale of the vacuum condensate is on the order of
the Planck scale. The emergence of spacetime was caused by
Planck energy accumulating in vacuum, but the Planck energy
condensation causes the underlying spacetime to be noncommutative,
which will introduce an uncertainty relation between microscopic
spacetimes. Therefore, a further accumulation of energy over the
noncommutative spacetime will be subject to UV/IR mixing \cite{uv-ir}.
UV/IR mixing in noncommutative spacetime then
implies that any UV fluctuations on the order of the Planck scale $L_P$ will be
necessarily paired with IR fluctuations of a typical scale $L_H$.
These vacuum fluctuations around the flat spacetime will add a tiny
energy $\delta \rho$ to the vacuum in Eq. \eq{vacuum-energy} so that the
total energy density is equal to $\rho \sim M_P^4 + \delta
\rho$. A simple dimensional analysis and a symmetry consideration,
e.g., the cosmological principle, lead to the following estimate of the
vacuum fluctuation \ct{tpad}:
\be \la{dark-fluct}
\rho = \rho_{\rm vac} + \delta\rho \sim M_P^4 \Big(1 + \frac{L_P^2}{L_H^2} \Big)
= M_P^4 + \frac{1}{L_P^2 L_H^2}.
\ee
It might be remarked that, though the second term in Eq. \eq{dark-fluct} is nearly constant
within a Hubble patch, it is not completely constant over the entire spacetime
while the first term is a true constant
because the vacuum fluctuation $\delta \rho$ has a finite size of $L_H$,
so it will act as a source of spacetime curvature of the order of $1/L_H^2$.
Because the first term in $\rho$ does not gravitate, the second term
$\delta \rho$ will, thus, be a leading contributor to the deformation of the
global spacetime curvature, leading possibly to a de Sitter phase.
Interestingly, this energy of vacuum fluctuations, $\delta \rho \sim
\frac{1}{L_P^2 L_H^2}$, is in good agreement with the observed value
of current dark energy \ct{tpad,hsy-cc} if $L_H$ is identified with the size of the cosmic
horizon of our universe.

As we reasoned above, the existence of the energy $\delta \rho$ in
Eq. \eq{dark-fluct} seems to be a generic property of emergent gravity based
on a noncommutative spacetime. Therefore, the emergent spacetime would
leave the vestige of this energy everywhere. Readers may remember
that we discussed some strange energy in Section II.C, so let us go
back to Eq. \eq{dark-energy}. Although we have taken the Euclidean
signature to get the result in Eq. \eq{dark-energy}, we will simply assume
that it can be analytically continued to the Lorentzian signature.
The Wick rotation will be defined by $y^4 = iy^0$. Under this Wick
rotation, $\delta_{ab} \to \eta_{ab}=(-+++)$ and $\varepsilon^{1234}
= 1 \to - \varepsilon^{0123} = -1$. Then, we get $\Psi_a^{(E)} = i
\Psi_a^{(L)}$ according to the definition in Eq. \eq{3-form}.
It is then given by \ct{hsy-jhep}
\be \la{dark-energy-4d}
T_{\mu\nu}^{(L)} = \frac{1}{16 \pi G_4 \lambda^2} \Big( \rho_\mu
\rho_\nu + \Psi_\mu \Psi_\nu - \frac{1}{2} g_{\mu\nu}
(\rho_\lambda^2 + \Psi_\lambda^2) \Big),
\ee
where $\rho_\mu = 2 \p_\mu \lambda$ and $\Psi_\mu = E_\mu^a \Psi_a$.

The Raychaudhuri equation \ct{raychauduri,hawking-ellis} represents the
evolution equations of the expansion, shear and rotation of flow
lines along the flow generated by a vector field in a background
spacetime. Here, we introduce an affine parameter $\tau$ labeling
points on the curve of the flow. Given a timelike unit vector field
$u^\mu$, i.e., $u^\mu u_\mu = -1$, the Raychaudhuri equation in four
dimensions is given by
\be \la{raychaudhuri}
\dot{\Theta} -  {\dot{u}^\mu}_{;\mu} + \Sigma_{\mu\nu} \Sigma^{\mu\nu} -
\Omega_{\mu\nu} \Omega^{\mu\nu} + \frac{1}{3} \Theta^2 = - R_{\mu\nu} u^\mu u^\nu.
\ee
$\Theta ={u^\mu}_{;\nu}$ represents the expansion/contraction of
volume and $\dot{\Theta} = \frac{d \Theta}{d \tau}$ while
$\dot{u}^\mu = {u^\mu}_{;\nu} u^\nu$ represents the acceleration due
to nongravitational forces, e.g., the Lorentz force.
$\Sigma_{\mu\nu}$ and $\Omega_{\mu\nu}$ are the shear tensor and the
vorticity tensor, respectively, which are all orthogonal to $u^\mu$,
i.e., $\Sigma_{\mu\nu} u^\nu = \Omega_{\mu\nu} u^\nu =0$. The
Einstein equation, Eq. \eq{emergent-eq}, can be rewritten as
\be \la{einstein-eq-energy}
R_{\mu\nu} = 8 \pi G \big( T_{\mu\nu} - \half g_{\mu\nu}
{T_\lambda}^\lambda \big),
\ee
where $T_{\mu\nu} = E_\mu^a E_\nu^b T_{ab}$. One can see from
Eq. \eq{einstein-eq-energy} that the right-hand side of
Eq. \eq{raychaudhuri} is given by
\be \la{ray-scalar}
 - R_{\mu\nu} u^\mu u^\nu = - \frac{1}{2\lambda^2} u^\mu u^\nu
 (\rho_\mu \rho_\nu  + \Psi_\mu \Psi_\nu),
\ee
where we have considered the energy-momentum tensor, Eq. \eq{dark-energy-4d},
only for simplicity.

Suppose that all the terms on the left-hand side of Eq. \eq{raychaudhuri}, except
the expansion evolution $\dot{\Theta}$, vanish or become negligible.
In this case, the Raychaudhuri equation reduces to
\be \la{reduce-raychaudhuri}
\dot{\Theta} = - \frac{1}{2\lambda^2} u^\mu u^\nu
(\rho_\mu \rho_\nu + \Psi_\mu \Psi_\nu).
\ee
Note that the Ricci scalar is given by $R = \frac{1}{2\lambda^2}
g^{\mu\nu} (\rho_\mu \rho_\nu + \Psi_\mu \Psi_\nu)$. Therefore, $R <
0$ when $\rho_\mu$ and $\Psi_\mu$ are timelike while $R > 0$ when
they are spacelike. Remember that our metric signature is $(-+++)$,
so, for timelike perturbations, $\dot{\Theta} < 0$, which means
that the volume of a three-dimensional spacelike hypersurface
orthogonal to $u_\mu$ decreases. However, if spacelike perturbations
are dominant, the volume of the three-dimensional spacelike
hypersurface can expand. For example, consider the scalar
perturbations in Eq. \eq{dec-emde}, i.e.,
\be \la{so31-invariant}
\langle \rho_a \rho_b \rangle = \frac{1}{4} \eta_{ab} \rho_c^2, \quad
\langle \Psi_a \Psi_b  \rangle = \frac{1}{4} \eta_{ab} \Psi_c^2.
\ee
For spacelike purturbations, Eq. \eq{reduce-raychaudhuri} becomes
\be \la{raychaudhuri-desitter}
\dot{\Theta} = \frac{R}{4} > 0.
\ee
The perturbation in Eq. \eq{so31-invariant} does not violate the energy
condition because $u^\mu u^\nu T_{\mu\nu}^{(L)} = \frac{R}{32 \pi G}
> 0$ according to Eq. \eq{dark-energy-4d}. This means that the Liouville energy-momentum
tensor in Eq. \eq{dark-energy-4d} can act as a source of gravitational
repulsion and exert a negative pressure causing an expansion of
the universe, possibly leading to a de Sitter phase \ct{hawking-ellis}.
As was pointed out in Eq. \eq{em-cc}, it can behave like a cosmological
constant, i.e., $\rho = - p$, in a constant (or almost constant)
curvature spacetime. Another important property is that the
Liouville energy in Eq. \eq{dark-energy-4d} is vanishing for the flat spacetime,
so it should be small if spacetime is not so curved.

To be more quantitative, let us consider the fluctuation
in Eq. \eq{so31-invariant} and look at the energy density $u^\mu u^\nu
T_{\mu\nu}^{(L)}$ along the flow represented by a timelike unit
vector $u^\mu$ as in Eq. \eq{ray-scalar}. Note that the Riemannian
volume is given by $\nu_g = \lambda^2 \nu = \sqrt{-g} d^4 y$. Also,
it was shown in Ref. \ct{hsy-jhep} that $\Psi_\mu$ is the Hodge-dual to
the 3-form $H$. Thus, $u^\mu \rho_\mu$ and $u^\mu
\Psi_\mu$ refer to the volume change of a three-dimensional spacelike
hypersurface orthogonal to $u^\mu$. Assume that the radius of the
three-dimensional hypersurface is $L_H(\tau)$ at time $\tau$, where
$\tau$ is an affine parameter labeling the curve of the flow. Then,
it is reasonable to expect that $u^\mu \rho_\mu = 2 u^\mu \p_\mu
\lambda \approx 2 \lambda/ L_H(\tau) \approx u^\mu \Psi_\mu $ because the Ricci scalar $R \sim
\frac{1}{L_H^2}$. After all, we approximately get \ct{hsy-jhep}
\be \la{dark-energy-cal}
u^\mu u^\nu T_{\mu\nu}^{(L)} \sim \frac{1}{8 \pi G L_H^2} =
\frac{1}{L_P^2 L_H^2}.
\ee
If we identify the radius $L_H$ with the size of cosmic horizon, the
energy density in Eq. \eq{dark-energy-cal} reproduces the dark energy
$\delta \rho$ in Eq. \eq{dark-fluct} up to a factor.

\section{CONCLUSION}

We suggested that the quantum gravity must be defined by quantizing
spacetime itself by the Newton constant $G$. This quantization
scheme is very different from the conventional one in which
quantization basically quantizes an infinite-dimensional
particle phase space associated with spacetime metric fields in
terms of the Planck constant $\hbar$. Our observation is that the
existence of the Newton constant in Nature can be translated into a
symplectic or Poisson structure of spacetime and that the canonical
quantization of the underlying symplectic or Poisson structure
inevitably leads to spacetime noncommutative geometry. It turns out
that electromagnetism defined on the symplectic or Poisson
spacetime enjoys very beautiful properties: the Darboux theorem and
the Moser lemma. From these theorems, we can formulate the
equivalence principle even for the electromagnetic force such that
there always exists a coordinate transformation to locally eliminate
the electromagnetic force. This equivalence principle can be fully
lifted to a noncommutative spacetime; thus, the so-called ``quantum
equivalence principle" can be identified with a gauge equivalence
between star products. This implies that quantum gravity can
consistently be derived from the quantum equivalence principle and
that matter fields can arise from the quantized spacetime.

If gravity emerges from a field theory, it is necessary to realize
the Newton constant $G$ from the field theory. That is the reason
the field theory should be defined with an intrinsic parameter
of $(\rm{length})^2$, and a noncommutative spacetime elegantly
carries out this mission. The only other example of such a theory
carrying an intrinsic constant of $({\rm length})^2$ is string
theory in which $\alpha'$ plays the role of $G$ or $|\theta|$. A unique
feature of string theory due to the existence of $\alpha'$ is
T-duality \ct{string-book}, which is a symmetry between small and
large distances, symbolically represented by $R
\leftrightarrow \alpha'/ R$. This symmetry implies the existence of a minimum length scale
in spactime and signifies an intrinsic noncommutative spacetime
geometry. The T-duality is a crucial ingredient for various string
dualities and mirror symmetry. For the very similar reason,
gravity in string theory also basically arises in the context of
emergent gravity although many string theorists seem to be reluctant
to accept this interpretation. Recently, Blau and Theisen vividly
summarized this picture in their review article \ct{blau-theisen}:

\begin{quote} {\it There are basically two approaches to formulate a
quantum theory of gravity. The first treats gravity as a fundamental
interaction which it attempts to quantise. In the second approach
gravity is not fundamental but an emergent phenomenon. String theory
falls into the second category. It has the gratifying feature that
not only gravity but also the gauge interactions which are mediated
by a spin one gauge boson are emergent. String theory thus provides
a unifying framework of all elementary particles and their
interactions: it inevitably and automatically includes gravity (in
the form of a massless traceless symmetric second-rank tensor
excitation of the closed string, identified with the graviton) in
addition to gauge forces which arise from massless excitation of the
open or closed string (depending on the perturbative formulation of
the theory)}. \end{quote}

We think the emergent gravity we have discussed so far is very
parallel to string theory in many aspects. We may understand this
wonderful similarity by noticing the following fact \ct{hsy-jhep}:
A Riemannian geometry is defined by a pair $(M,g)$, where the metric
$g$ encodes all geometric information, while a symplectic geometry
is defined by a pair $(M, \omega)$, where the 2-form $\omega$ encodes
all. A basic concept in Riemannian geometry is a distance defined by
the metric. One may identify this distance with a geodesic worldline
of a ``particle" moving in $M$. On the contrary, a basic concept in
symplectic geometry is an area defined by the symplectic structure.
One may regard this area as a minimal worldsheet swept by a
``string" moving in $M$. In this picture, the wiggly string, so a
fluctuating worldsheet, may be interpreted as a deformation of the
symplectic structure in spacetime $M$. Then, we know that a Riemannian
geometry (or gravity) is emergent from wiggly strings or the
deformation of the symplectic structure! Amusingly, the Riemannian
geometry is probed by particles while the symplectic geometry would
be probed by strings.

Hence the emergent gravity we have reviewed in this paper may be
deeply related to string theory. This may be supported by the fact
that many essential aspects of string theory, for example, AdS/CFT
correspondence, open-closed string duality, noncommutative geometry,
mirror symmetry, etc. have also been realized in the context of
emergent noncommutative geometry. Thus, we may moderately claim that
string theory is simply a ``stringy" realization of symplectic or
Poisson spacetime.

There are many important issues that we didn't even touch on.
Although we have speculated that matter fields can emerge from
stable localized geometries defined by noncommutative
$\star$-algebra, we could not understand how particle masses can be
generated from the noncommutative $\star$-algebra, in other words,
how to realize spontaneous electroweak symmetry breaking or the
Higgs mechanism. We believe this problem could be deeply related to
the question of how the extra internal space $F$ for weak and strong
forces in Section IV is dynamically compactified. We don't know this
yet even though we have some vague ideas. Thus, from the background independent
formulation of quantum gravity, the Standard Model is completely
unexplored territory. The emergent spacetime picture
may present a radically new understanding of the Standard Model.

We have no idea how supersymmetry arises from a background-independent quantum
gravity theory or what the role of supersymmetry is in the emergent
geometry and emergent matter. We do not know how to break it,
but this issue should be understood in the near future.

Though we have tried to concretely formulate emergent gravity as
much as possible, a rigorous mathematical formulation of emergent
gravity, especially background-independent quantum gravity, is
highly demanded. We think that the Lie algebroid may be a useful mathematical framework
for emergent gravity. Here, we will introduce the definition of a Lie algebroid \ct{vaisman}
only to appreciate some flavor of its mathematical structure for emergent
quantum gravity. Progress along this line will be published elsewhere.

A Lie algebroid is a triple $(E, [\cdot,
\cdot],
\rho)$ consisting of a smooth vector bundle $E$ over a manifold $M$,
together with a Lie algebra structure $[\cdot, \cdot]$ on the vector
space $\Gamma(E)$ of the smooth global sections of $E$, and a
morphism of vector bundles $\rho: E \to TM$, called the anchor map,
where $TM$ is the tangent bundle of $M$. The anchor map and the bracket
satisfy the Leibniz rule such that
\be \la{lie-algebroid}
[X, fY] = f[X, Y] + (\rho(X)f) \cdot Y
\ee
for all $X, Y \in \Gamma(E)$ and $f \in C^\infty(M)$. Here,
$\rho(X)f$ is the derivative of $f$ along the vector field
$\rho(X)$. The anchor $\rho$ defines a Lie algebra homomorphism from
the Lie algebra of sections of $E$, with Lie bracket $[\cdot,
\cdot]$, into the Lie algebra of vector fields on $M$, i.e.,
\be \la{anchor-map}
\rho \Big( [X,Y] \Big) = [\rho(X), \rho(Y)].
\ee

If $M$ is a Poisson manifold, then the cotangent bundle $T^*M \to M$
is, in a natural way, a Lie algebroid over $M$. The anchor is the
map $\pi^\sharp : T^*M \to TM$ defined by the Poisson bivector
$\pi$. See Eq. \eq{anchor}. The Lie bracket $[\cdot, \cdot]$ of
differential 1-forms satisifes $[df, dg] = d\{f, g\}_\pi$ for any
functions $f, g \in C^\infty(M)$, where $\{f, g\}_\pi = \pi(df, dg)$
is the Poisson bracket defined by $\pi$. When $\pi$ is
nondegenerate, $M$ is a symplectic manifold, and this Lie algebra
structure of $\Gamma(T^*M)$ is isomorphic to that of $\Gamma(TM)$. A
noncommutative generalization, i.e. $\{f, g\}_\pi \to -i
[\widehat{f}, \widehat{g}]_\star$, seems to be possible.

Because background-independent quantum gravity does not assume any
kind of spacetime structure, a natural question is then why
spacetime on large scales is four dimensions. If gravity is emergent
from gauge field interactions, we may notice that electromagnetism
is now only a long-range force in Nature. Weak and strong forces are
short-range forces, so they will affect only microscopic structure
of spacetime. Then, we may infer that only electromagnetism is
responsible for the large-scale structure of spacetime. In this
regard, there is a funny coincidence \ct{hsy-bjp}. If we compare the
number of physical polarizations of photons and gravitons in $D$
dimensions and find the matching condition of the physical
polarizations, we get a cute number: $\maltese (A_\mu) = D-2 =
\frac{D(D-3)}{2} = \maltese(g_{\mu\nu}) \Rightarrow D=1 \; {\rm or} \; D=4$,
where $\maltese$ denotes the number of polarizations. Of course, we
have to throw $D=1$ away because it is not physically meaningful. Does
this unfledged math have some meaning?

\begin{acknowledgments}

This work was supported by the Daejin University Research Grants 2014.
This work was partly done when HSY visited the KEK Theory Center using the International Visitor Program.
He thanks Yoshihisa Kitazawa and Jun Nishimura for warm hospitality and helpful discussions
during his visit and Koji Hashimoto for enlightening discussion at RIKEN theory seminar.
\end{acknowledgments}


\begin{references}
\bibitem{gravmac} Nevertheless, gravitational phenomena are ubiquitous
in our everyday life. The reason is that the gravitational force is
only attractive and always additive. As a result, the standard
gravitational parameter $GM$ for an astronomical body with mass $M$
is not small. For example, $GM_e = 4 \times 10^{14} m^3/s^2$ for the Earth,
where $M_e = 5.96 \times 10^{24} Kg$, which corresponds to $1 \, cm$ compared to the Planck
length $L_{pl} = \sqrt{G} \sim 10^{-33} cm$.
\bibitem{nhabel} Niels Henrik Abel (1802-1829): {\it A problem that seems insurmountable is
just seemingly so because we have not asked the right question. You should always ask the right
question and then you can solve the problem.}
\bibitem{6d-mphys} There is an interesting peculiarity in $D=6$ that the gauge coupling constant $e^2$
in natural units $c=\hbar=1$ carries the physical dimension of $(\mathrm{length})^2$.
Thus, the gauge coupling constant $e^2$ in six dimensions operates like a noncommutative parameter $|\theta|$
or a string length $\alpha'$, i.e., $e^2 \sim |\theta| \sim \alpha'$.
This may imply that a six-dimensional gauge theory intrinsically behaves like a string theory
without assuming any noncommutative spacetime.
\bibitem{dbi} The theory on a D-brane also needs an intrinsic length parameter because it is
always defined with the action $\int_{M} \sqrt{\det(g + 2\pi \alpha' F)}$,
so $\alpha' \equiv l_s^2$ should carry the dimension of
$({\rm length})^2$ for a dimensional reason. Because the parameter
$l_s^2$ disappears from the theory if $F = 0$, it is necessary for
the field strength $F$ to be nowhere vanishing in order for the
parameter $l_s^2$ to have an operational meaning as an intrinsic
length scale of the theory. In this case where $\langle F
\rangle_{r \to \infty} \equiv B$, the two theories will be physically equivalent,
as was shown in Ref. \ct{SW-ncft}.
\bibitem{emg-gauge} However, Eq. \eq{spin-connection} shows that
the $SU(2)$ gauge fields $A_\mu^{(\pm)}$ have their own gauge fields, i.e., the vierbeins $E_a$
whose degrees of freedom match with those of the $U(1)$ gauge fields $D_a$ in the action
in Eq. \eq{semi-matrix} when applying the map in Eq. \eq{v-field}. Therefore, it is sensible that $SU(2)$
gauge theory describing self-dual gravity can be mapped to the $U(1)$ gauge theory described
by the action in Eq. \eq{semi-matrix}, as will be shown later.
\bibitem{cartan-homo} This formula can be easily derived by applying Cartan's homotopy
formula ${\cal L}_X \beta = \iota_X d \beta + d \iota_X \beta$ for
$\beta = \iota_Y \alpha$ and $[{\cal L}_X, \iota_Y] = \iota_{[X,Y]}$ \ct{mechanics}.
\bibitem{sign-prob} There is a perverse sign problem in reversing the map.
We still don't understand how to properly translate the map $\Gamma(TM) \to C^\infty(M)$ to avoid the
sign problem. In a way, one may consider some analytic continuation
such that every module is defined over $\mathbb{C}$, as was suggested in
Ref. \ct{hsy-jhep}. Anyway, the sign issue will be insignificant at this stage.
\bibitem{d-para} Here, $\hbar$ is a formal deformation parameter that
may be identified with either $|\theta|$ or the Planck constant $\hbar$,
depending on the choice of a Poisson manifold $M$ or $P$.
For the Poisson manifold $M$ with the Poisson structure $\theta \in \Gamma(\Lambda^2 TM)$,
it would be convenient to take the rescaling $\theta^{\mu\nu} \to \hbar \, \theta^{\mu\nu}$
because $\hbar$ is regarded as a plain parameter without any indices.
In the end, one may set $\hbar = 1$.
\bibitem{i-notat} Our notation here may be sloppy. $D_a$ in Eq. \eq{decomp-cov}
contains an extra $-i$ compared to Eq. \eq{nc-photon}. This sloppy
notation is for a parallel march with $D_\mu$.
\bibitem{foot-zee} In this respect, quoting a recent comment by Zee \ct{zee} would be interesting:
``The basic equation for the graviton field has the same form $g_{\mu\nu} = \eta_{\mu\nu} + h_{\mu\nu}$. This naturally suggests that $\eta_{\mu\nu} = \langle g_{\mu\nu} \rangle$ and
perhaps some sort of spontaneous symmetry breaking." It turns out
that this pattern is not an accidental happening.
\bibitem{hei-alg} One may imagine that the system in Eq. \eq{hamilton-poisson} is
quantized {\it a la} Eq. \eq{qm-bracket}. Then, the time evolution of a
quantum system is derived from a Heisenberg algebra, i.e., $i\hbar \frac{d \widehat{f}}{d t}
= [\widehat{f}, \widehat{H}]$ for $\widehat{f}, \widehat{H} \in \CA_\hbar$.
This is also in line with our philosophy that a geometry is derived from an algebra.
\bibitem {h-mis} This argument was corrected in Ref. \cite{jpcs-hsy}, but we will not delete
this incorrect argument to illustrate how a prejudice easily ignores important physics.
\end{references}
\end{document}